 \documentclass[utf8,sort&compress]{frontiersinFPHY_FAMS} 

\setcitestyle{square} 
\usepackage{url,hyperref,lineno,microtype,subcaption}
\usepackage[onehalfspacing]{setspace}

\usepackage[labelformat=empty, position=top]{subcaption}

\def\keyFont{\fontsize{8}{11}\helveticabold }
\def\firstAuthorLast{A. Meg\'ias and A. Santos} 
\def\Authors{Alberto Meg\'ias\,$^{1,*}$ and Andr\'es Santos\,$^{1,2}$}
\def\Address{$^{1}$Departamento de F\'isica, Universidad de Extremadura, E-06006 Badajoz, Spain \\
$^{2}$ Instituto de Computaci\'on Cient\'ifica Avanzada (ICCAEx),
Universidad de Extremadura, E-06006 Badajoz, Spain}
\def\corrAuthor{A.~Meg\'ias}

\def\corrEmail{albertom@unex.es}

\newcommand{\st}{{\mathrm{st}}}
\newcommand{\rf}{{\mathrm{ref}}}
\newcommand{\ct}{{\mathrm{cr}}}
\newcommand{\dif}{\mathrm{d}}
\newcommand{\cc}{\mathbf{c}}

\newcommand{\rr}{\mathbf{r}}
\newcommand{\vv}{\mathbf{v}}
\newcommand{\vva}{\mathbf{v}_1}
\newcommand{\vvb}{\mathbf{v}_2}
\newcommand{\vvab}{\mathbf{v}_{12}}

\newcommand{\Tt}{T_\text{tr}}
\newcommand{\Tr}{T_\text{rot}}
\newcommand{\Tn}{T^{\mathrm{wn}}}
\newcommand{\chit}{\chi_\text{tr}}
\newcommand{\chir}{\chi_\text{rot}}
\newcommand{\xit}{\xi_\text{tr}}
\newcommand{\xir}{\xi_\text{rot}}
\newcommand{\nun}{\nu^{\mathrm{wn}}}

\newcommand{\een}{\alpha}
\newcommand{\eet}{\beta}

\newcommand{\FWN}{\mathbf{F}^{\mathrm{wn}}}
\newcommand{\MWN}{\tau^{\mathrm{wn}}}

\begin{document}
\onecolumn
\firstpage{1}

\title[Mpemba effect for inelastic and rough hard disks]{Mpemba-like effect protocol for granular gases of inelastic and rough hard disks}

\author[\firstAuthorLast ]{\Authors} 
\address{} 
\correspondance{} 

\extraAuth{}

\maketitle

\begin{abstract}

\section{}
We study the conditions under which a Mpemba-like effect emerges in granular gases of inelastic and rough hard disks driven by a class of thermostats characterized by the splitting of the noise intensity into translational and rotational counterparts. Thus, granular particles are affected by a stochastic force and a stochastic torque, which inject translational and rotational energy, respectively. We realize that a certain choice of a thermostat of this class can be characterized just by the total intensity and the fraction of noise transferred to the rotational degree of freedom with respect to the translational ones. Firstly, Mpemba effect is characterized by the appearance of a crossing between the temperature curves of the considered samples. Later, an overshoot of the temperature evolution with respect to the steady-state value is observed and the mechanism of Mpemba effect generation is changed. The election of parameters allows to design plausible protocols based on these thermostats for generating the initial states to observe  the  Mpemba-like effect in experiments. In order to obtain explicit results, we use a well-founded Maxwellian approximation for the evolution dynamics and the steady-state quantities. Finally, theoretical results are compared with direct simulation Monte Carlo and molecular dynamics results, and a very good agreement is found.

\tiny
 \keyFont{ \section{Keywords:} granular gases, kinetic theory, Mpemba effect, direct simulation Monte Carlo, molecular dynamics} 
\end{abstract}

\section{Introduction}
\label{sec:1}
Since the Antiquity, the fact that water could start freezing earlier for initially hotter samples was observed and commented by very influential people of different epochs like Aristotle \cite{aristotle_works_1931}, Francis Bacon \cite{Bacon1620}, or Ren\'e Descartes \cite{Descartes1637}. This counterintuitive phenomenon contradicts Isaac Newton's formulation of its well-known cooling's law \cite{Newton1701,Newton1782}, but otherwise it is part of the popular belief in cold countries. The scientific community started to pay attention to this effect since the late 60s of the last century thanks to its accidental rediscovery by a Tanzanian high-school student, Erasto B. Mpemba. Later, he and  Dr.\ Denis Osborne reported their findings \cite{MO69,O79} and since then the effect is usually known as  Mpemba effect (ME).

Whereas the original tested system for ME has been water~\cite{M67,MO69,K69,F71,D71,F74,G74,W77,O79,F79,K80,H81,WOB88,A95,K96,M96,J06,ERS08,K09,VM10,B11,VM12,BT12,ZHMZZZJS14,VK15,S15,BT15,R15,JG15,IC16,GLH19,BKC21}, it is still under discussion and no consensus about its occurrence  has been agreed~\cite{BL16,BH20,ES21}. In fact, the statistical physics community is currently paying attention to Mpemba-like effects that have been described in a huge variety of complex systems in the last decades, such as ideal gases~\cite{ZMHM22}, molecular gases~\cite{SP20,PSP21,MSP22}, gas mixtures~\cite{GKG21}, granular gases~\cite{LVPS17,TLLVPS19,BPRR20,MLTVL21,GG21,BPR21,BPR22}, inertial suspensions~\cite{THS21,T21}, spin glasses~\cite{Betal19}, Ising models~\cite{GMLMS21,TYR21,VD21}, non-Markovian mean-field systems~\cite{YH20,YH22}, carbon nanotube resonators~\cite{GLCG11}, clathrate hydrates~\cite{AKKL16} , active systems~\cite{SL21}, or quantum systems~\cite{CLL21}. The theoretical approach to the fundamentals of the problem has been done via different routes like Markovian statistics~\cite{LR17,KRHV19,CKB21,BGM21,LLHRW22} or  Landau's theory of phase transitions~\cite{HR22}. Recently, in the context of a molecular gas under a nonlinear drag force, new interpretations and definitions of ME from thermal and entropic point of views, as well as a classification of the whole possible phenomenology, have been carried out \cite{MSP22}. In addition, ME has been experimentally observed in colloids~\cite{KB20,KCB22}, proving that it is a real effect present in nature.

The very first time that ME was observed theoretically in granular gaseous rapid flows was in Ref.~\cite{LVPS17}. The considered system was a set of inelastic and smooth hard spheres (with constant coefficient of normal restitution) heated by a stochastic thermostat, the effect arising by initially preparing the system in far from Maxwellian states. The same type of initial preparation was applied to  the case of molecular gases with nonlinear drag \cite{MSP22,SP20}. Essentially, the temperature evolution depends on the whole moment hierarchy of the velocity distribution function (particularly on  the excess kurtosis, or fourth cumulant, and, more weakly, on the sixth cumulant), this dependence giving rise to the possible appearance of ME.

On the other hand, there is no need to consider an initial velocity distribution function (VDF) far from the Maxwell--Boltzmann one if the temperature is coupled to other basic variables that can be fine-tuned in the initial preparation of the system. This occurs in the case of a monocomponent granular gas made by inelastic and rough hard spheres thermostatted by a stochastic force \cite{TLLVPS19}, as well as in driven binary mixtures of either molecular or inelastic  gases \cite{BPRR20,GKG21}. In those systems, one does not need to invoke strong nonGaussianities, since the temperature relaxation essentially depends on the rotational-to-translational temperature ratio (in the case of rough particles) or on the partial component temperatures  (in the case of mixtures). A similar situation applies in the presence of  anisotropy in either the injection of energy \cite{BPR21} or in the velocity flow \cite{THS21}. However, there is still a lack of protocol  defining a possible nearly realistic preparation of the initial states for a granular or molecular gas in homogeneous and isotropic states. Unlike other memory effects, such as Kovac's effect \cite{K63,KAHR79}, ME has not a predefined way to elaborate a protocol.

In this work, we have addressed the latter preparation problem for a specific model of granular gases. We consider a monodisperse granular gas of inelastic and rough hard disks, where inelasticity is parameterized via a constant coefficient of normal restitution, $\een$, and the roughness is accounted for by a coefficient of tangential restitution, $\eet$, assumed to be constant as well. Disks are ``heated'' by a stochastic thermostat which injects energy to both translational and rotational degrees of freedom through a combination of a stochastic force and a stochastic torque, both with properties of a white noise. The relative amount of energy injected to the rotational degree of freedom, relative to that injected to the translational degrees of freedom, can be freely chosen. Therefore, we will denote this thermostat as \emph{splitting thermostat} (ST). The quantity coupled to the temperature that will monitor the possible occurrence of ME will be the rotational-to-translational temperature ratio, as in Ref.~\cite{TLLVPS19}, where, however, a stochastic torque was absent. This double energy-injection based on ST allows us to fix the initial conditions of the  variables that play a role in the evolution process, namely the temperature and its coupling. A side effect of providing energy to the rotational degree of freedom is that it favors the possibility of an overshoot of the temperature with respect to its steady-state value. This might cause ME, even in the absence of  a crossing between the temperatures of the two samples \cite{MSP22}. Therefore, the protocol must be adapted to this specific phenomenon.

It is worth saying that our theoretical approach is based on a Maxwellian approximation (MA), that is, we assume that both transient and steady-state VDFs are close to a two-temperature Maxwellian. This approach is founded on previous works for the case of zero stochastic torque  \cite{VS15} and on preliminary results for the system at hand \cite{paperIII}. Moreover, the two-dimensional characterization of the physical system is thought to be plausible for hopefully being reproduced in some experimental setup.
As will be seen, the reliability of our theoretical approach is confirmed by computer simulations via the direct simulation Monte Carlo (DSMC) method and event-driven molecular dynamics (EDMD).

The paper is structured as follows. In section \ref{sec:2}, the model system for a granular gas of inelastic and rough hard disks thermostatted by stochastic force and torque is introduced. Also, explicit evolution equations and expressions for the steady-state  dynamic variables are shown under the MA, and the theoretical results are compared with DSMC and EDMD. Section \ref{sec:3} collects the definition and necessary conditions for ME to occur taking into account the emergence or not of overshoot during evolution. Subsequently, and based on the analysis of this section, two different protocols are presented for observing ME in cases without and with overshoot, respectively. This discussion is accompanied by its proper comparison with simulation results. Finally, concluding remarks are presented in section \ref{sec:4}.

\section{The Model}
\label{sec:2}
We consider a set of mechanically identical inelastic and rough hard disks of mass $m$, diameter $\sigma$, and reduced moment of inertia $\kappa\equiv 4I/m\sigma^2$ ($I$ being the moment of inertia). The translational velocities lie on the $xy$ plane, i.e., $\mathbf{v}=v_x\widehat{\mathbf{x}} +v_y\widehat{\mathbf{y}}$, while the angular velocities point along the orthogonal $z$ axis, $\boldsymbol{\omega}=\omega \widehat{\mathbf{z}}$. Inelasticity and roughness are characterized by the coefficients of normal and tangential restitution, $\alpha$ and $\beta$, respectively, which are assumed to be constant and defined as \cite{G19}
\begin{equation}
    \widehat{\boldsymbol{\sigma}}\cdot\vvab^\prime=-\alpha\widehat{\boldsymbol{\sigma}}\cdot\vvab, \quad \widehat{\boldsymbol{\sigma}}_\perp\cdot\vvab^\prime=-\beta\widehat{\boldsymbol{\sigma}}_\perp\cdot\vvab,
\end{equation}
where $\widehat{\boldsymbol{\sigma}}\equiv(\rr_2-\rr_1)/|\rr_2-\rr_1|$ is the unit intercenter vector along the line of centers from particle 1 to particle 2, $\widehat{\boldsymbol{\sigma}}_\perp \equiv \widehat{\boldsymbol{\sigma}}\times \widehat{\mathbf{z}}$ is orthogonal to $\widehat{\boldsymbol{\sigma}}$,
$\vvab\equiv \vva-\vvb$ is the relative velocity between particles 1 and 2, and primed quantities account for their postcollisional values. Because of their definitions, the ranges of the coefficients of restitution are $0\leq \alpha\leq 1$ and $-1\leq \beta\leq 1$,  $\alpha=1$ corresponding to  elastic collisions, $\beta=-1$ describing a perfectly smooth disks, and $\beta=1$ standing for completely rough disks. In fact, total kinetic energy is only conserved if $\alpha=|\beta|=1$ \cite{G19,MS19,MS19b,S18}.

\begin{figure}[h!]
    \centering
    \begin{minipage}[t]{.45\linewidth}
    	\textbf{A}\\
    \includegraphics[width=\linewidth]{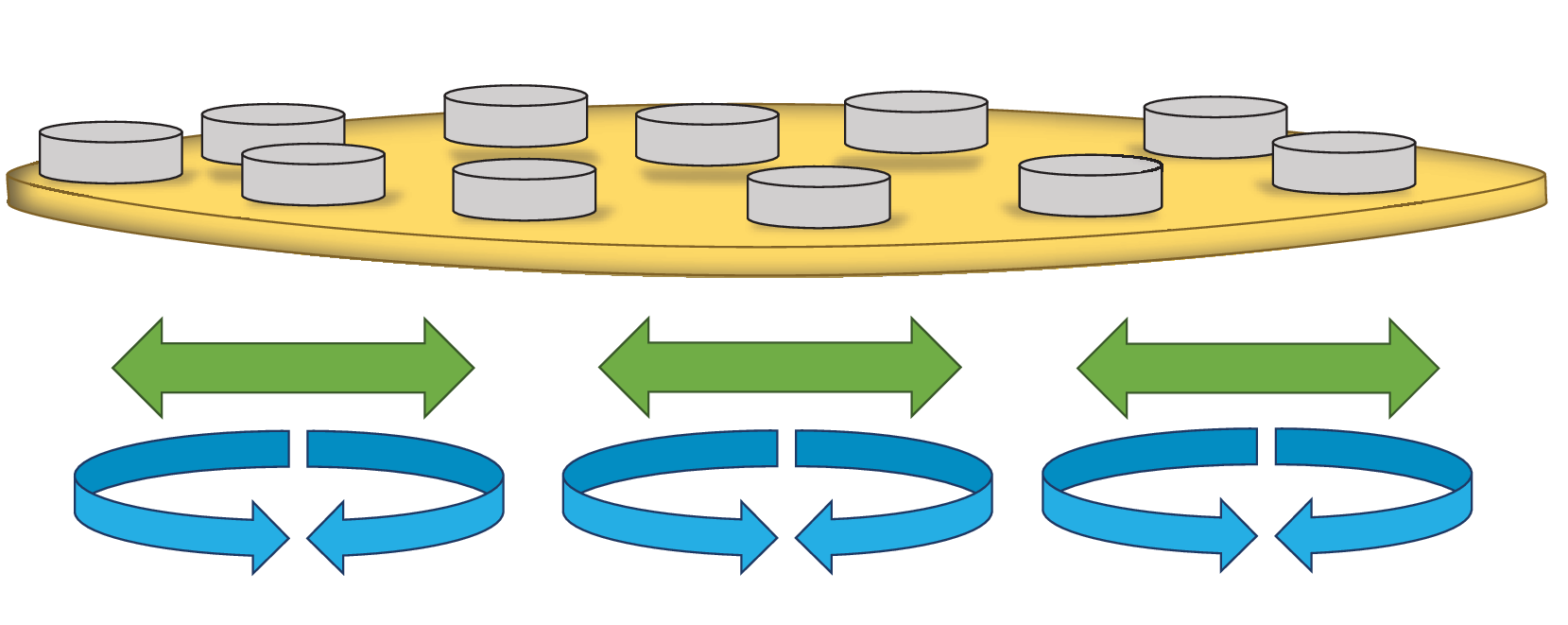}
    \end{minipage}
    \begin{minipage}[t]{0.45\linewidth}
    	\textbf{B}\\
    \includegraphics[width=\linewidth]{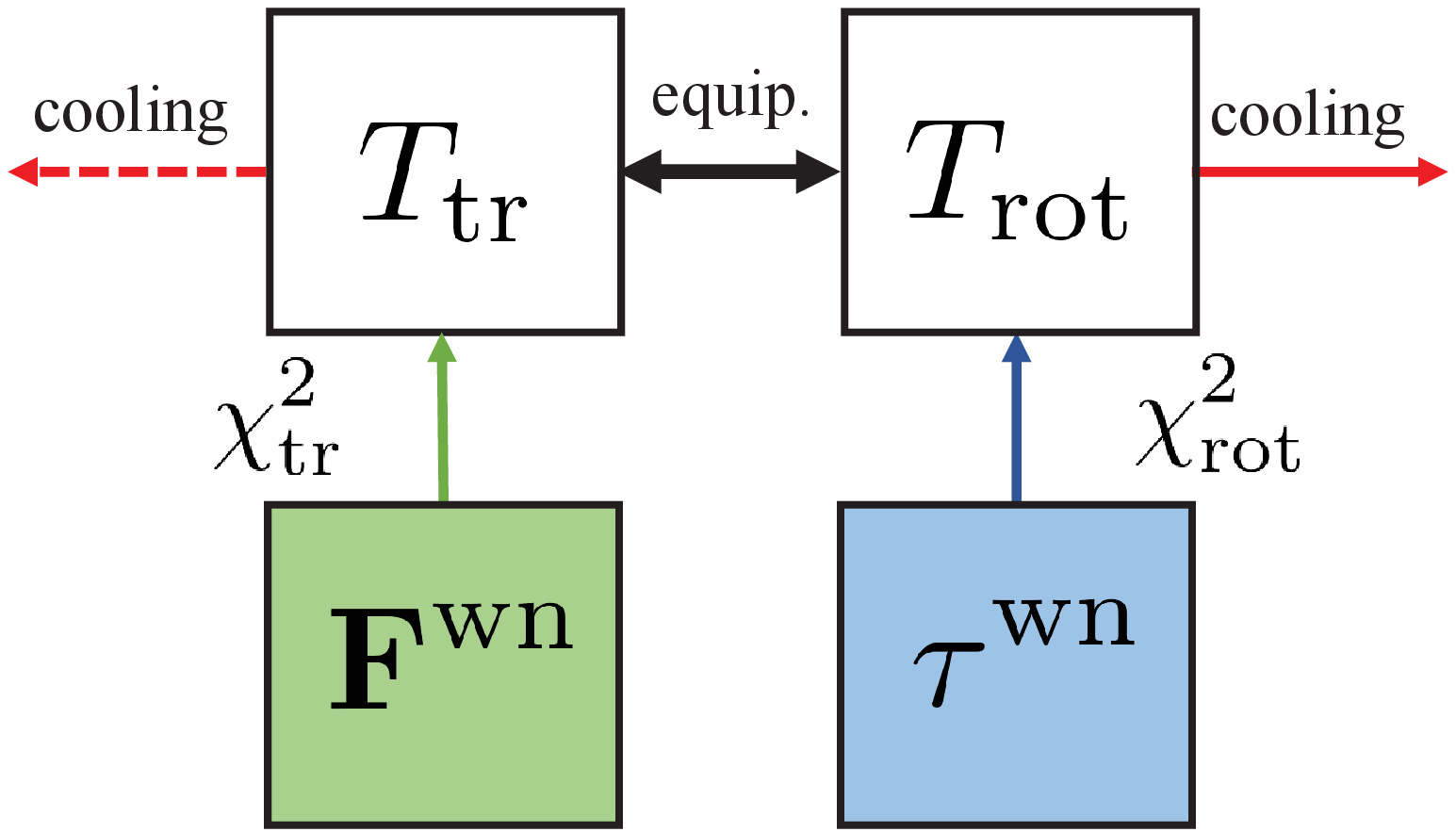}
    \end{minipage}
    \caption{\textbf{(A)} Illustration of  the system, where the green horizontal straight arrows and the blue curved ones represent the stochastic force $\FWN$ and the stochastic torque $\MWN$ in Equations~\eqref{eq:stochastic}, respectively. \textbf{(B)} Mechanism of injection-dissipation of energy in the introduced model. Dissipative collisions produce a cooling effect on the translational and rotational temperatures ($\Tt$ and $\Tr$), together with a transfer between the translational and rotational energies (equipartition effect). Additionally, the external white-noise force and torque inject energy (heating effect).}
    \label{fig:illustration}
\end{figure}

In the case $\alpha\neq 1$ or $|\beta|\neq1$, that is, when kinetic energy is dissipated upon collisions, the undriven system will evolve up to a completely frozen state. In order to avoid that quench, we will force the particles to externally receive energy via a homogeneous stochastic force $\FWN$ and a homogeneous stochastic torque $\MWN$ that inject translational and rotational kinetic energy, respectively, with the properties of a white noise. That is,
\begin{subequations}
\label{eq:stochastic}
\begin{align}
    \langle \FWN_i(t) \rangle =& 0, \qquad \langle \FWN_i(t)\FWN_j(t^\prime) \rangle = \mathsf{I}m^2\chit^2\delta_{ij}\delta(t-t^\prime), \\
    \langle \MWN_i(t) \rangle =& 0, \qquad \langle \MWN_i(t)\MWN_j(t^\prime) \rangle = mI\chir^2\delta_{ij}\delta(t-t^\prime),
\end{align}
\end{subequations}
where $\mathsf{I}$ is the $2\times 2$ identity matrix, $i$ and $j$ are particle indices, and $\chit^2$ and $\chir^2$ are the intensities of the noises applied to the translational and rotational degrees of freedom, respectively. The combination of the stochastic force and torque is characterized by the pair of parameters $(\chit^2,\chir^2)$ and defines the ST, as described in section \ref{sec:1}. In section \ref{sec:2.1} we will introduce a more manageable pair of equivalent parameters. An illustration of the system is represented in Figure~\textbf{\ref{fig:illustration}}.

To dynamically describe the system, we will work under the assumptions of the homogeneous Boltzmann--Fokker--Planck equation (BFPE). That is, we consider a homogeneous and isotropic  gas in a dilute regime, such that the evolution due to the collisional process is determined by just binary collisions, assuming \emph{Stosszahlansatz} (or molecular chaos). The BFPE for this collisional model together, with the ST, is written as follows
\begin{equation}\label{eq:BFPE}
    \frac{\partial}{\partial t}f(\vv,\omega;t)-\frac{\chit^2}{2}\left(\frac{\partial}{\partial \vv}\right)^2f(\vv,\omega;t)-\frac{\chir^2}{2}\left(\frac{\partial}{\partial \omega}\right)^2f(\vv,\omega;t) = J[\vv,\omega|f],
\end{equation}
where $f$ is the one-particle VDF, $J[\vv,\omega|f]$ is the usual Boltzmann collision operator for hard disks defined as \cite{MS19,MS19b,S18}
\begin{equation}
    J[\vv_1,\omega_1|f] = n\sigma \int \dif\vv_2\int_{-\infty}^\infty\dif\omega_2\int_{+}\dif\widehat{\boldsymbol{\sigma}}|\vv_{12}\cdot\widehat{\boldsymbol{\sigma}}|\left[\frac{f(\vv_1^{\prime\prime},\omega_1^{\prime\prime};t)f(\vv_2^{\prime\prime},\omega_2^{\prime\prime};t)}{\een|\eet|}-f(\vv_1,\omega_1;t)f(\vv_2,\omega_2;t) \right],
\end{equation}
where $n$ is the number density, the subscript $+$ in the integral over $\widehat{\boldsymbol{\sigma}}$ means the constraint $\vv_{12}\cdot\widehat{\boldsymbol{\sigma}}>0$, and double primed quantities are precollisional velocities, which are given by \cite{MS19,MS19b,S18}
\begin{equation}
    \vv_{1/2}^{\prime\prime} = \vv_{1/2}\mp \left\{\frac{1+\een^{-1}}{2}(\vv_{12}\cdot\widehat{\boldsymbol{\sigma}})\widehat{\boldsymbol{\sigma}}+\frac{\kappa}{1+\kappa}\frac{1+\eet^{-1}}{2}\left[\vv_{12}\cdot\widehat{\boldsymbol{\sigma}}_\perp-\frac{\sigma}{2}(\omega_1+\omega_2) \right]\widehat{\boldsymbol{\sigma}}_\perp\right\}.
\end{equation}

\subsection{Dynamics}\label{sec:2.1}

The time evolution of the system is fully described by the BFPE, Equation~\eqref{eq:BFPE}, which allows one to determine the dynamics of macroscopic quantities. The most important and basic quantities to study the dynamics of the system will be the translational and rotational \emph{granular} temperatures defined at a certain time $t$ as
\begin{equation}\label{eq:partial_temp}
    \Tt(t) = \frac{1}{2}m\langle v^2\rangle, \qquad \Tr(t) =I\langle \omega^2\rangle,
\end{equation}
where the notation $\langle\medspace\cdot\medspace \rangle$ means the average over the instantaneous VDF,
\begin{equation}
    \langle X(\vv,\omega;t) \rangle = n^{-1}\int \dif\vv\int_{-\infty}^\infty\dif\omega X(\vv,\omega;t)f(\vv,\omega;t).
\end{equation}
One should notice that the \emph{partial} noise intensities $\chit^2$ and $\chir^2$ affect directly the evolution equations of $\Tt(t)$ and $\Tr(t)$, respectively. On the other hand, those quantities are coupled due to the transfer of energy during collisions (see Figure~\textbf{\ref{fig:illustration}}\textbf{B}). The description of the dynamics from $\Tt(t)$ and $\Tr(t)$ is equivalent to consider the mean granular temperature, $T(t)$, and the rotational-to-translational temperature, $\theta(t)$, defined as follows
\begin{equation}\label{eq:Ttheta}
    T(t)=\frac{2}{3}\Tt(t)+\frac{1}{3}\Tr(t), \qquad \theta(t)=\frac{\Tr(t)}{\Tt(t)}.
\end{equation}
In the definition of $T(t)$ we have taken into account that there are two translational and one rotational degrees of freedom.
This new pair of variables will be useful to study ME, which will be related to the evolution of the mean granular temperature $T(t)$. In addition, this change of dynamical quantities in Equation~\eqref{eq:Ttheta} induces a change of parameters describing the ST. Thus, we introduce the \emph{total} noise intensity, $\chi^2$, and the \emph{rotational-to-total} noise intensity ratio, $\varepsilon$, as
\begin{equation}\label{eq:noise_paramsB}
    \chi^2= \chit^2+\frac{I}{2m}\chir^2, \qquad \varepsilon = \frac{I}{2m}\frac{\chir^2}{\chi^2}.
\end{equation}
Notice that $\chit^2=(1-\varepsilon)\chi^2$. Therefore, $0\leq \varepsilon\leq 1$,  $\varepsilon=0$ and $\varepsilon=1$ corresponding to the purely translational and the purely rotational thermostat, respectively. The total noise intensity  $\chi^2$ is unbounded from above and, by dimensional analysis, can be equivalently characterized by  a \emph{noise temperature}
\begin{equation}\label{eq:Tn}
    \Tn \equiv m\left(\frac{\chi^2}{\sqrt{\pi}n\sigma}\right)^{2/3}.
\end{equation}
Therefore, from now on the ST will be characterized by the pair $(\Tn,\varepsilon)$. In terms of the new parameters, the BFPE, Equation~\eqref{eq:BFPE},  reads
\begin{equation}\label{eq:BFPE_v2}
    \frac{\partial}{\partial t}f(\vv,\omega;t)-\frac{\nun\Tn }{4m}\left[(1-\varepsilon)\left(\frac{\partial}{\partial \vv}\right)^2+\varepsilon\frac{2m}{I}\left(\frac{\partial}{\partial \omega}\right)^2\right]f(\vv,\omega;t) = J[\vv,\omega|f],
\end{equation}
\begin{equation}
\nun=2n\sigma \sqrt{\frac{\pi \Tn}{m}}
\end{equation}
being  a reference \emph{noise-induced} collisional frequency.

Inserting  the definitions of partial granular temperatures, Equation~\eqref{eq:partial_temp}, into Equation~\eqref{eq:BFPE_v2}, one obtains
\begin{align}
\label{eq:evol_Tt_Tr}
    \partial_t \Tt = -\xit \Tt+\frac{1-\varepsilon}{2} {\nun}\Tn, \quad
    \partial_t \Tr = -\xir \Tr+\varepsilon \nun \Tn,
\end{align}
where
\begin{align}
    \xit = -\frac{m}{2n\Tt}\int\dif \vv\int_{-\infty}^\infty \dif \omega \medspace v^2 J[\vv,\omega|f], \quad \xir = -\frac{I}{n\Tr} \int\dif \vv\int_{-\infty}^\infty \dif \omega \medspace \omega^2 J[\vv,\omega|f]
\end{align}
are the translational and rotational energy production rates \cite{S18,MS19}.

In terms of the quantities defined in Equation~\eqref{eq:Ttheta}, Equations~\eqref{eq:evol_Tt_Tr} become
\begin{align}\label{eqs:Tth_ev}
    \partial_t \widetilde{T} = -\zeta \widetilde{T}+\frac{1}{3}\nun, \quad \partial_t \theta= \theta(\xit-\xir)-  \nun\frac{2+\theta}{6\widetilde{T}}\left[\theta-\varepsilon(2+\theta) \right],
\end{align}
where $\widetilde{T}\equiv T/\Tn$ and
\begin{equation}
    \zeta \equiv \frac{2}{3} \frac{\xit\Tt}{T}+\frac{1}{3} \frac{\xir\Tr}{T} = \frac{2\xit+\xir\theta }{2+\theta}
\end{equation}
is the \emph{cooling rate}.

According to Equation~\eqref{eqs:Tth_ev}, the steady-state quantities $\widetilde{T}^\st=T^\st/\Tn$ and $\theta^\st=\Tr^\st/\Tt^\st$ satisfy the conditions
\begin{equation}
\label{eq:ss}
\zeta^\st \widetilde{T}^\st=\frac{1}{3}\nun,\quad \varepsilon \xit^\st=\frac{1-\varepsilon}{2}\xir^\st\theta^\st,
\end{equation}
which imply a balance between collisional cooling and external heating.
The steady-state temperature can be used to define a reduced temperature $T^*\equiv \widetilde{T}/\widetilde{T}^\st=T/T^{\st}$ and a reduced time $t^*=\frac{1}{2}\nu^\st t$, where
\begin{equation}
\nu^\st=2n\sigma \sqrt{\frac{\pi \Tt^\st}{m}}=\nun\sqrt{\frac{3\widetilde{T}^\st}{2+\theta^\st}}
\end{equation}
is the steady-state collision frequency. More in general, the time-dependent collision frequency is
\begin{equation}
\nu(t)=2n\sigma \sqrt{\frac{\pi \Tt(t)}{m}}=\nun\sqrt{\frac{3\widetilde{T}(t)}{2+\theta(t)}}=\nu^\st G(T^*(t),\theta(t)),\quad G(T^*,\theta)\equiv \sqrt{T^*\frac{2+\theta^\st}{2+\theta}}.
\end{equation}
The above collision frequency can be used to nondimensionalize the energy production rates as
\begin{equation}
\mu_{20}\equiv\frac{\xit}{\nu}, \quad \mu_{02}\equiv\frac{\xir}{2\nu},
\end{equation}
where
\begin{equation}\label{eq:coll_mom}
    \mu_{k\ell} = -\int\dif\cc\int_{-\infty}^\infty\dif w\medspace c^k w^\ell \mathcal{J}[\cc,w|f]
\end{equation}
are the reduced collisional moments. In Equation~\eqref{eq:coll_mom}, $\cc=\vv/\sqrt{2\Tt/m}$ and $w=\omega/\sqrt{2\Tr/I}$ are the reduced translational and angular velocities, respectively, and
$\mathcal{J}\equiv (2\Tt/m)\sqrt{2\Tr/I}/(n\nu) J$ is the reduced collision operator.
Thus, the steady-state conditions \eqref{eq:ss} become
\begin{align}
\label{eq:ss2}
    2\left(\mu_{20}^{\st}+\mu_{02}^{\st}\theta^{\st}\right) = \left(\frac{2+\theta^\st}{3 {\widetilde{T}}^{\st}}\right)^{3/2}\equiv \gamma^{\st}, \qquad
    \varepsilon \mu_{20}^{\st}=(1-\varepsilon)\mu_{02}^{\st}\theta^{\st}.
\end{align}

Using these dimensionless definitions, Equations~\eqref{eqs:Tth_ev} yield
\begin{subequations}
\label{eq:ev_eqs_1&2}
\begin{align}\label{eq:ev_eqs_1}
    \frac{1}{2}\partial_{t^{*}}\ln T^{*} &= -2{G(T^*,\theta)}\frac{\mu_{20}+\mu_{02}\theta}{2+\theta}+\frac{2}{T^*}\frac{\mu_{20}^\st+\mu_{02}^\st\theta^\st}{2+\theta^\st},\\ \label{eq:ev_eqs_2}
    \frac{1}{2}\partial_{t^{*}}\ln \theta &=
G(T^*,\theta)\left(\mu_{20}-2\mu_{02}\right)-\frac{1}{T^*}\frac{2+\theta}{2+\theta^\st}\frac{\theta^\st}{\theta}\frac{\theta-\varepsilon(2+\theta)}{\theta^\st-\varepsilon(2+\theta^\st)}\left(\mu_{20}^\st-2\mu_{02}^\st\right),
\end{align}
\end{subequations}
where use has been made of Equations~\eqref{eq:ss} and \eqref{eq:ss2}.

According to the definition of collisional moments, Equation~\eqref{eq:coll_mom}, they depend on the whole VDF. This implies that Equations~\eqref{eq:ev_eqs_1&2}  do not make a closed set of equations. The same applies to the steady-state solution, Equation~\eqref{eq:ss2}. This shortcoming, however, can be circumvented if an \emph{approximate} closure is applied. This is the subject of section \ref{sec:2.2}.

\subsection{Maxwellian approximation}
\label{sec:2.2}
In order to get explicit results from Equations~\eqref{eq:ss2} and \eqref{eq:ev_eqs_1&2} by using the simplest possible closure, we resort to the two-temperature MA
\begin{equation}
\label{eq:Max}
f(\mathbf{v},\omega)\to n \frac{m}{2\pi\Tt}\sqrt{\frac{I}{2\pi\Tr}}\exp\left(-\frac{mv^2}{2\Tt}-\frac{Iv^2}{2\Tr}\right).
\end{equation}
This approximation does a very good job in the three-dimensional case with $\varepsilon=0$ \cite{VS15} and it is reasonably expected to perform also well in the case of disks with $\varepsilon\neq 0$.

\begin{figure}
	\centering
	\includegraphics[width=0.45\textwidth]{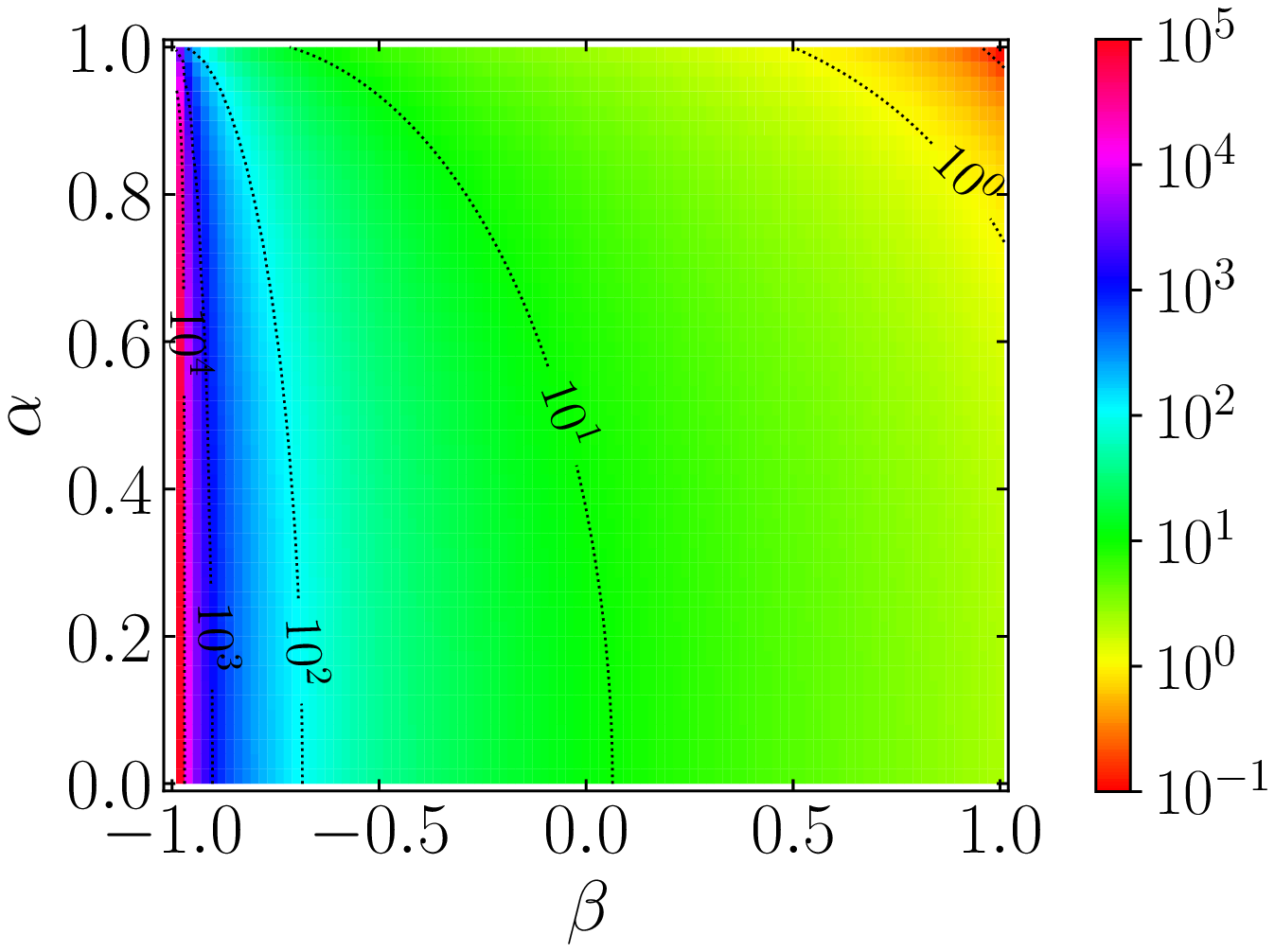}
	\caption{Difference $\theta^\st(\varepsilon=1)-\theta^\st(\varepsilon=0)$ [see Equations~\eqref{eq:theta_st_0&1}] in the plane $(\alpha,\beta)$.}
	\label{fig:theta_st_diff}
\end{figure}

Within this approximation, the relevant collisional moments can be evaluated with the result \cite{MS19,MS19b,S18}
\begin{subequations}
\label{eq:mu}
\begin{align}
    \mu_{20} =& \frac{1-\een^2}{2}+\frac{\mathcal{K}}{2}\left[\frac{\kappa(1-\eet)}{2}\left(1+\frac{\theta}{\kappa} \right)+1-\theta \right], \\
    \mu_{02} =& \frac{\mathcal{K}}{2}\left[\frac{1-\eet}{2\kappa}\left(1+\frac{\kappa}{\theta} \right)+1-\theta^{-1} \right],\quad \mathcal{K}\equiv \kappa\frac{1+\eet}{(1+\kappa)^2}.
\end{align}
\end{subequations}
Solving Equations~\eqref{eq:ss2}, we get the steady-state expressions
\begin{align}\label{eq:steady_values}
    \theta^\st = \kappa\left[\frac{2}{\mathcal{K}}\frac{(1-\een^2)\varepsilon+\mathcal{K}(1+\kappa)}{(1-\eet)[1-\varepsilon(1+\kappa)]+2\kappa}-1\right], \quad {\widetilde{T}}^{\st} = \frac{2+\theta^{\mathrm{st}}}{3(\gamma^\st)^{2/3}},
\end{align}
with
\begin{equation}
\label{eq:gamma}
    \gamma^\st = 1-\een^2+\mathcal{K}\frac{1+\kappa}{2\kappa}(1-\eet)\left(\kappa+\theta^{\st}\right).
\end{equation}
In particular,
\begin{subequations}
\label{eq:theta_st_0&1}
\begin{align}
\label{eq:theta_st_0}
  \varepsilon=0\Rightarrow\theta^\st=&\frac{1+\beta}{2+\kappa^{-1}(1-\beta)},\\
  \label{eq:theta_st_1}
  \varepsilon=1\Rightarrow\theta^\st=&2\frac{1+\mathcal{K}^{-1}(1-\alpha^2)+\frac{\kappa}{2}(1-\beta)}{1+\beta}.
\end{align}
\end{subequations}
Equation~\eqref{eq:theta_st_0} agrees with a previous result \cite{MS19}. Notice that, for the special value $\varepsilon=0$, $\theta^\st$ is independent of the coefficient of normal restitution $\alpha$ both for disks and spheres \cite{VS15,S18,MS19,MS19b,TLLVPS19}. However, this property  is broken down when energy is injected into the rotational degree of freedom ($\varepsilon\neq 0$).

{}From Equation~\eqref{eq:steady_values} one can observe that $\theta^\st$ is independent of $\Tn$ and, at given $\alpha$ and $\beta$, it is a monotonically increasing function of $\varepsilon$.
This is physically expected since, by growing $\varepsilon$, we are increasing the relative amount of rotational energy injected with respect to the total energy; therefore, it is presumed that the stationary value of $\Tr$ rises with respect to $\Tt$ at fixed $\Tn$. Thus, the most disparate values of $\theta^\st$ correspond to $\varepsilon=1$ and $0$, their difference being plotted in Figure~\textbf{\ref{fig:theta_st_diff}} as a function of  $\alpha$ and $\beta$.

It is interesting to note that, in the MA, Equation~\eqref{eq:Max}, one simply has
\begin{subequations}
\begin{equation}
  2(\mu_{20}+\mu_{02}\theta)=\gamma^\st+\mathcal{K}(1-\eet)\frac{1+\kappa}{2\kappa}(\theta-\theta^\st),
  \end{equation}
\begin{equation}
   \mu_{20}-2\mu_{02}=\left(1-\varepsilon\frac{2+\theta^\st}{\theta^\st}\right)\frac{\gamma^\st}{2}-\mathcal{K}\frac{1+\beta}{4}\left(1+\frac{2}{\theta\theta^\st}\right)(\theta-\theta^\st).
\end{equation}
\end{subequations}
As a consequence, Equations~\eqref{eq:ev_eqs_1&2} can be recast as,
\begin{subequations}
\label{eq:Tstar-theta-ev}
\begin{align}
\label{eq:Tstar}
    \frac{1}{2}\partial_{t^{*}}\ln T^{*} = & \Phi(T^{*},\theta)\equiv -\gamma^\st\left[\frac{G(T^*,\theta)}{2+\theta}-\frac{1/T^*}{2+\theta^\st}\right]-G(T^*,\theta)\mathcal{K}(1-\eet)\frac{1+\kappa}{2\kappa}\frac{\theta-\theta^\st}{2+\theta},\\
    \frac{1}{2}\partial_{t^{*}}\ln \theta =&
\left(1-\varepsilon\frac{2+\theta^\st}{\theta^\st}\right)\frac{\gamma^\st}{2}\left[G(T^*,\theta)-\frac{1}{T^*}\frac{2+\theta}{2+\theta^\st}\frac{\theta^\st}{\theta}
\frac{\theta-\varepsilon(2+\theta)}{\theta^\st-\varepsilon(2+\theta^\st)}\right]\nonumber\\
\label{eq:theta-ev}
&-G(T^*,\theta)\mathcal{K}\frac{1+\beta}{4}\left(1+\frac{2}{\theta\theta^\st}\right)(\theta-\theta^\st).
\end{align}
\end{subequations}
Equations~\eqref{eq:Tstar-theta-ev} make a closed set of two ordinary differential equations that can be (numerically) solved with arbitrary initial conditions $T^*(0)\equiv T_0^*$ and $\theta(0)\equiv \theta_0$.
Although the theoretical results have been derived for arbitrary values of the reduced moment of inertia $\kappa$, henceforth all the graphs are obtained for the conventional case of a uniform mass distribution of the disks, i.e., $\kappa=\frac{1}{2}$.

\begin{figure}
    \centering
    \begin{minipage}[t]{.3\linewidth}
    	\textbf{A}\\
    \includegraphics[width = \linewidth]{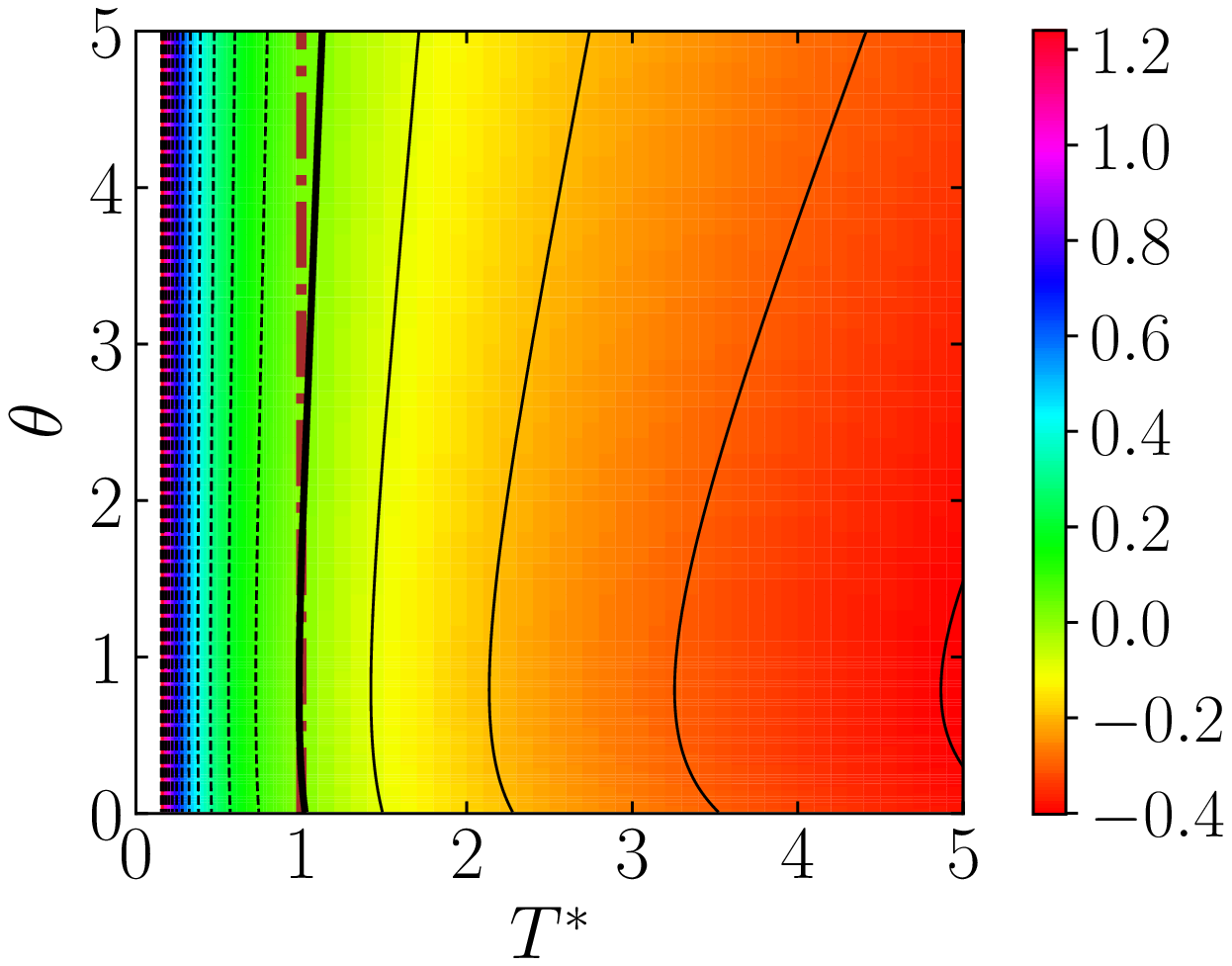}
    \end{minipage}
    \begin{minipage}[t]{.3\linewidth}
    	\textbf{B}\\
    \includegraphics[width = \linewidth]{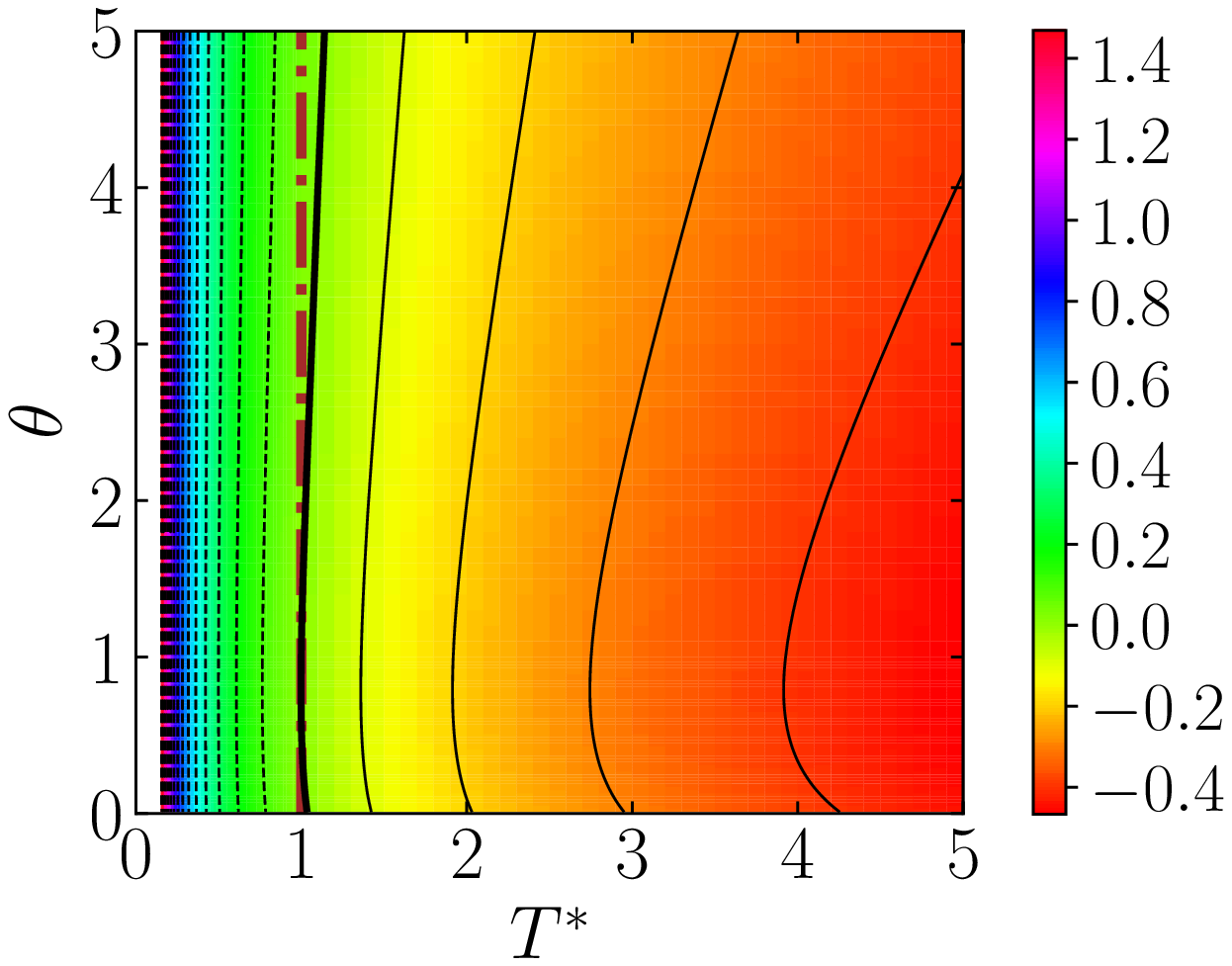}
    \end{minipage}
    \begin{minipage}[t]{.3\linewidth}
    	\textbf{C}\\
    \includegraphics[width = \linewidth]{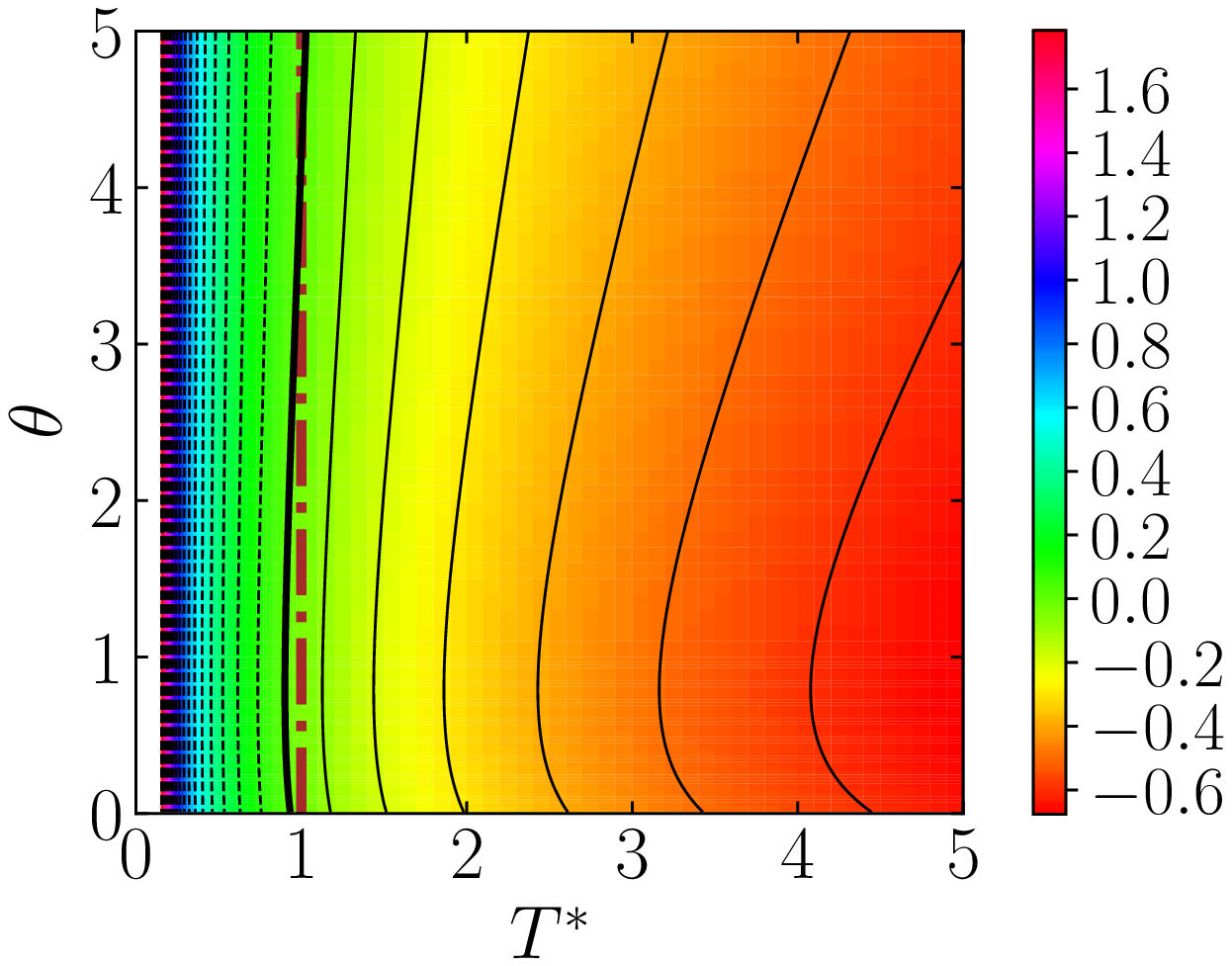}
    \end{minipage}\\
    \begin{minipage}[t]{.3\linewidth}
    	\textbf{D}\\
    \includegraphics[width = \linewidth]{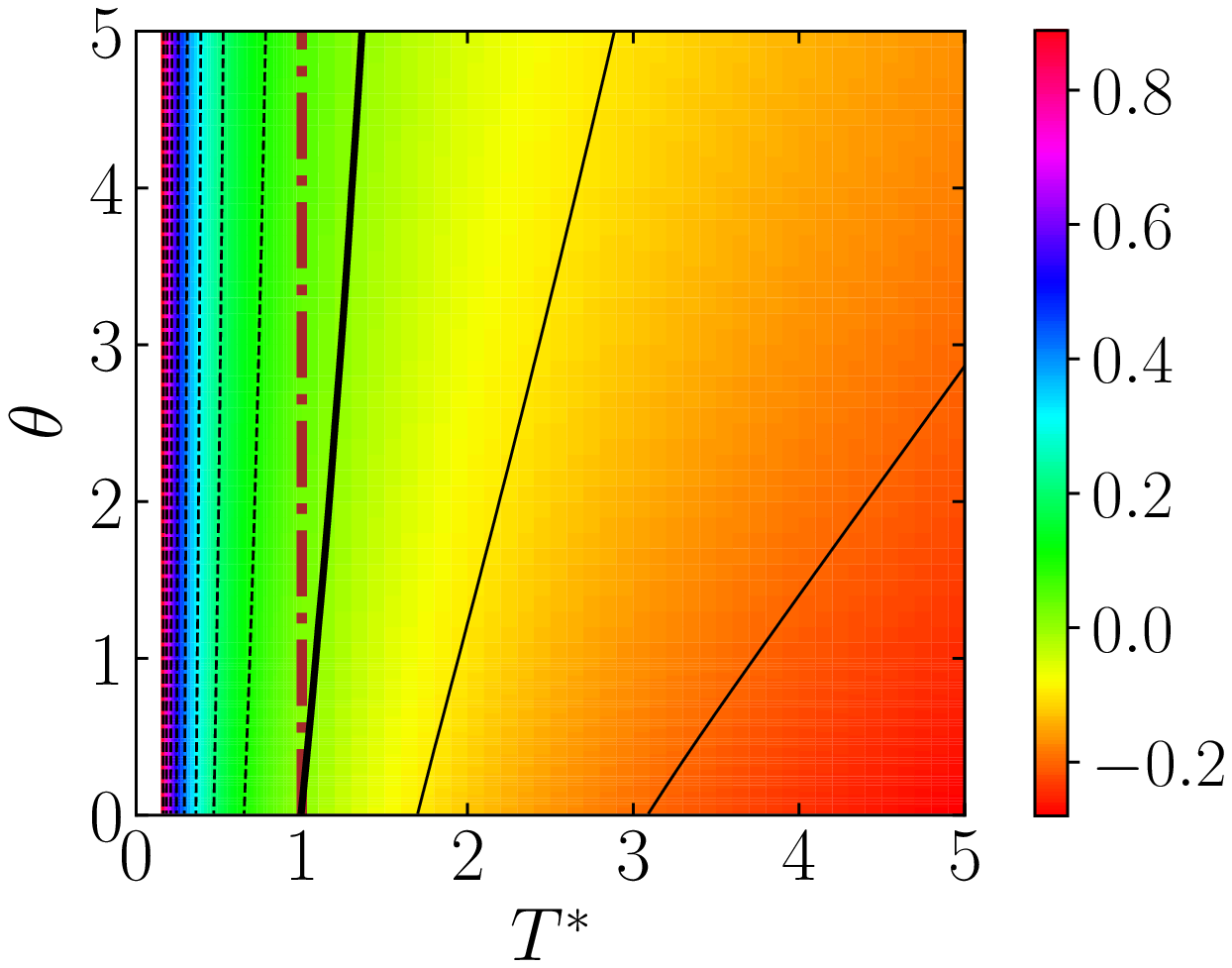}
    \end{minipage}
    \begin{minipage}[t]{.3\linewidth}
    	\textbf{E}\\
    \includegraphics[width = \linewidth]{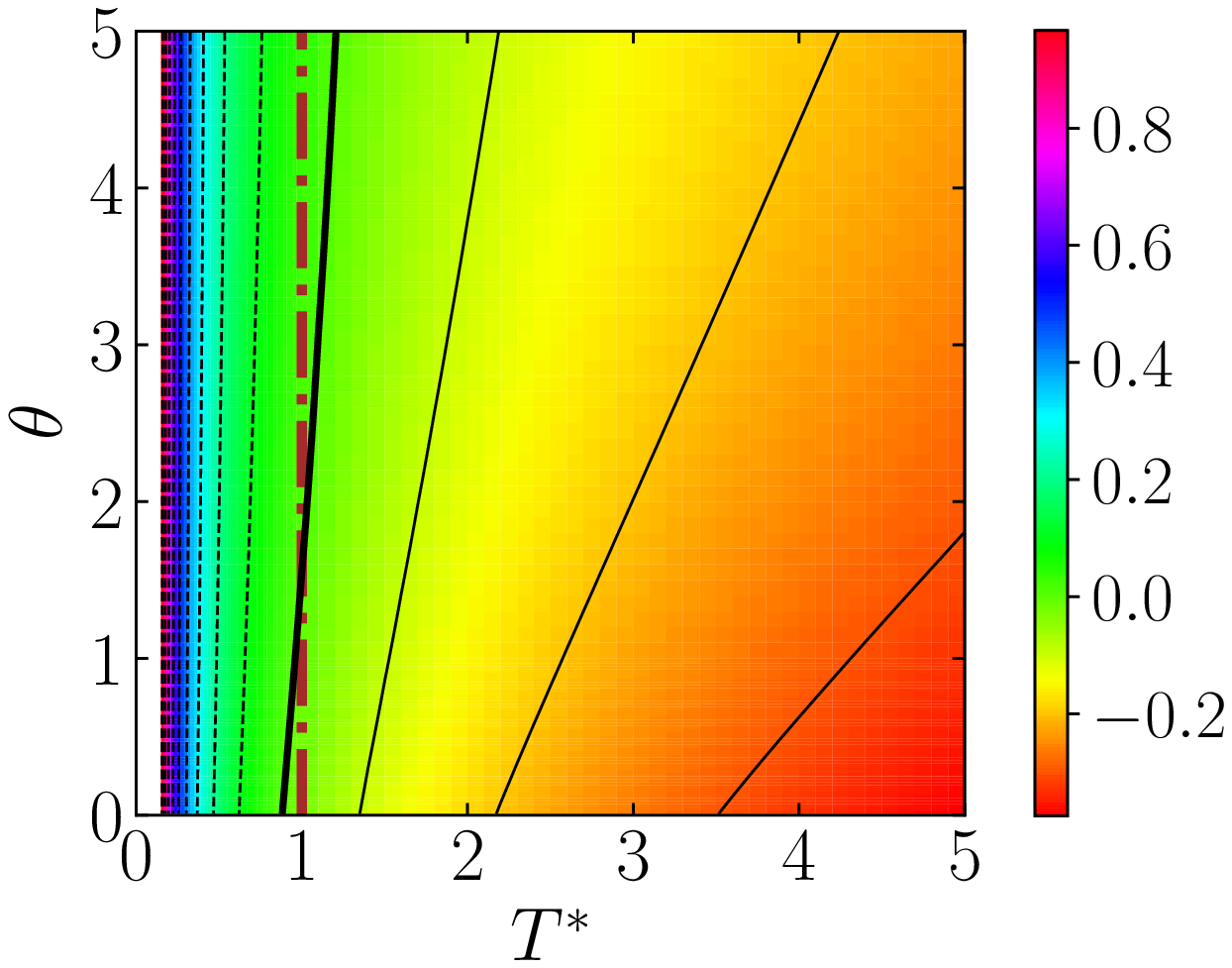}
    \end{minipage}
    \begin{minipage}[t]{.3\linewidth}
    	\textbf{F}\\
    \includegraphics[width = \linewidth]{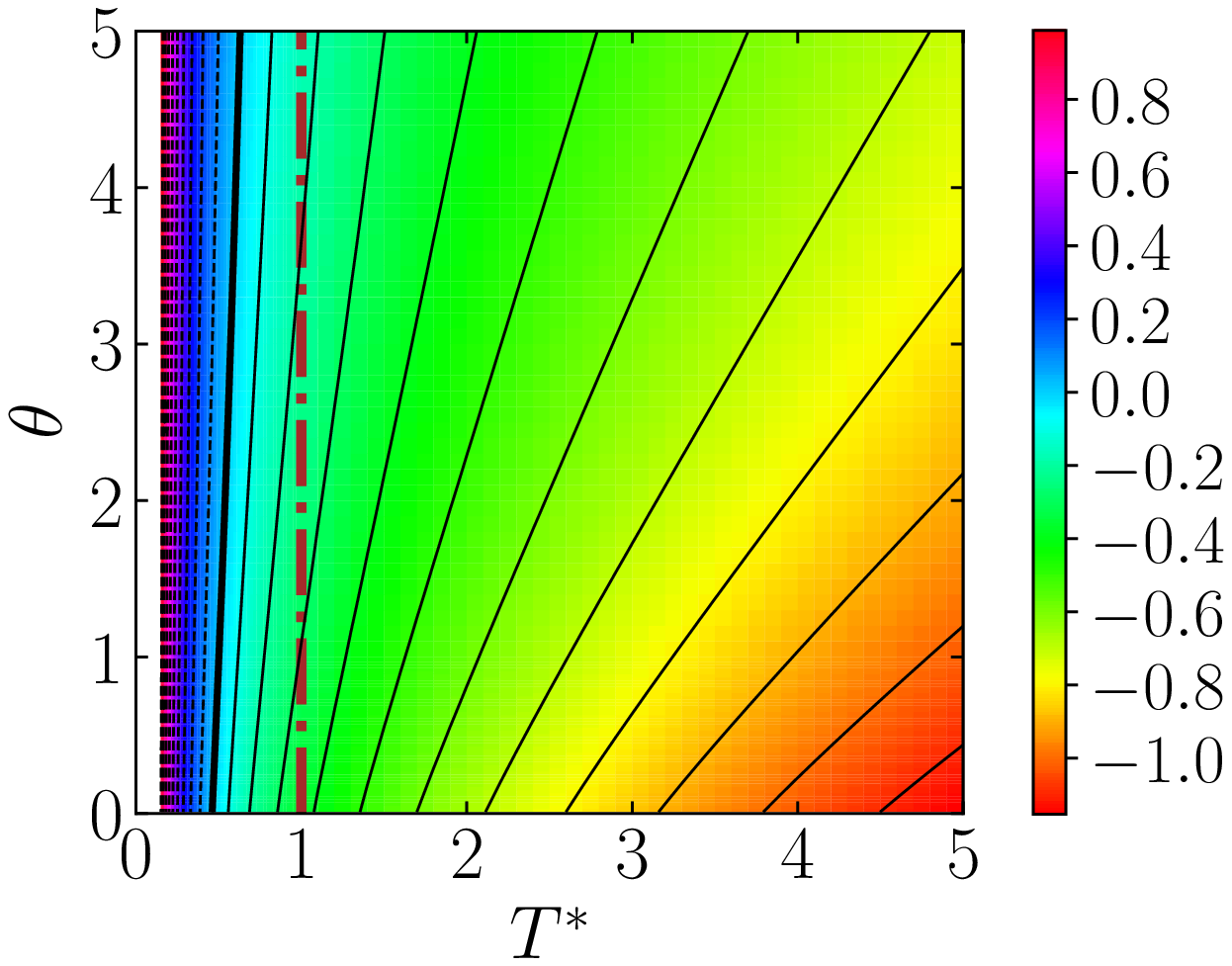}
    \end{minipage}
    \caption{Density plots of $\Phi(T^*,\theta)$ for $\alpha=0.9$ and \textbf{(A)} $\beta=0$, $\varepsilon=0$; \textbf{(B)} $\beta=0$, $\varepsilon=0.5$; \textbf{(C)} $\beta=0$, $\varepsilon=1$; \textbf{(D)} $\beta=-0.7$, $\varepsilon=0$; \textbf{(E)} $\beta=-0.7$, $\varepsilon=0.5$; and \textbf{(F)} $\beta=-0.7$, $\varepsilon=1$.
    The contour lines (solid for $\Phi<0$, dashed for $\Phi>0$) are separated by an amount $\Delta\Phi=0.1$. The thick solid line is the locus
    $\Phi(T^*,\theta)=0$. It intercepts the (brown dash-dotted) vertical line $T^*=1$  at $\theta=\theta^\st$.
}
    \label{fig:slopes_1}
\end{figure}

\begin{figure}
    \centering
    
    \begin{minipage}[t]{.3\linewidth}
    	\textbf{A}\\
    \includegraphics[width = \linewidth]{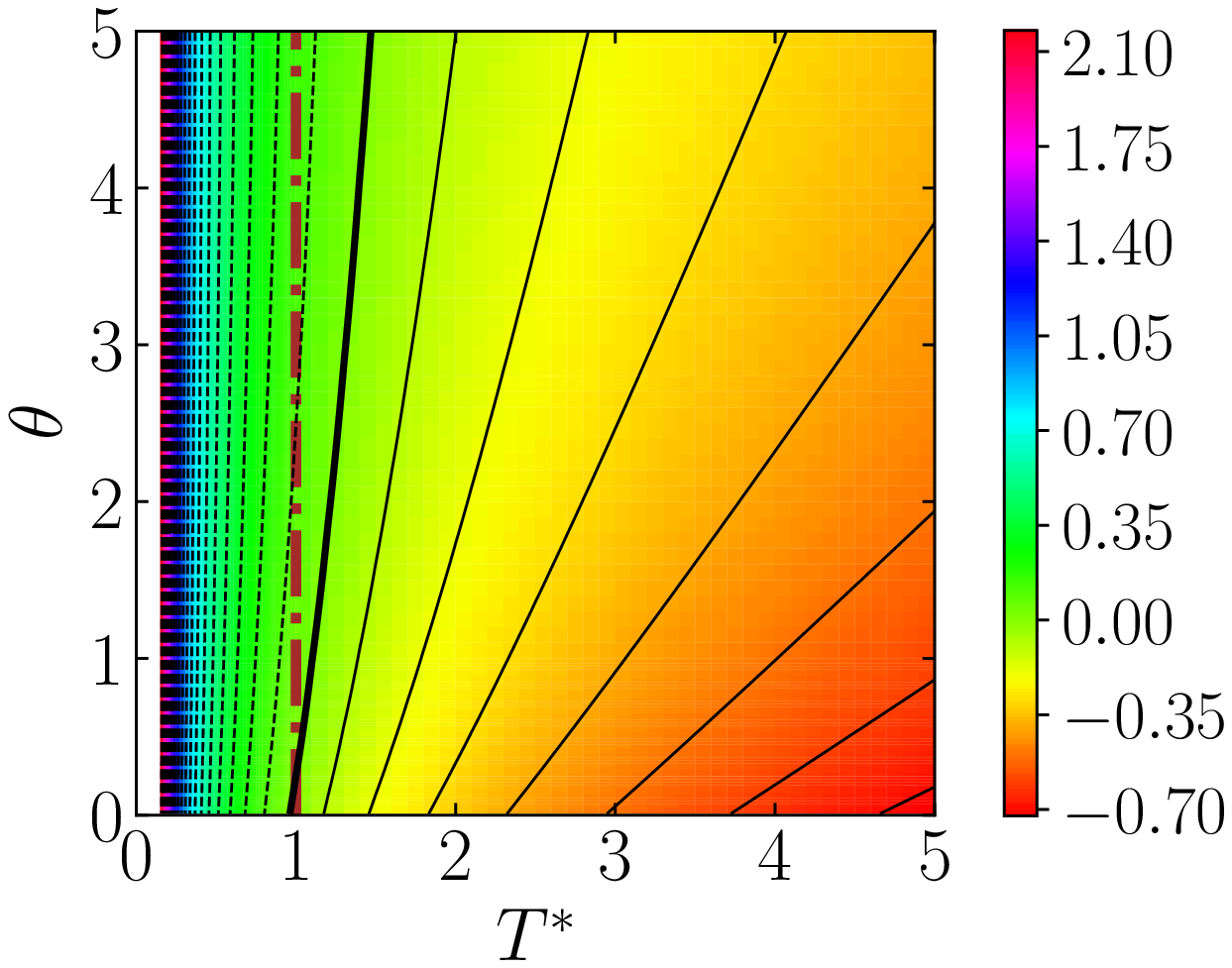}
    \end{minipage}
    \begin{minipage}[t]{.3\linewidth}
    	\textbf{B}\\
    \includegraphics[width = \linewidth]{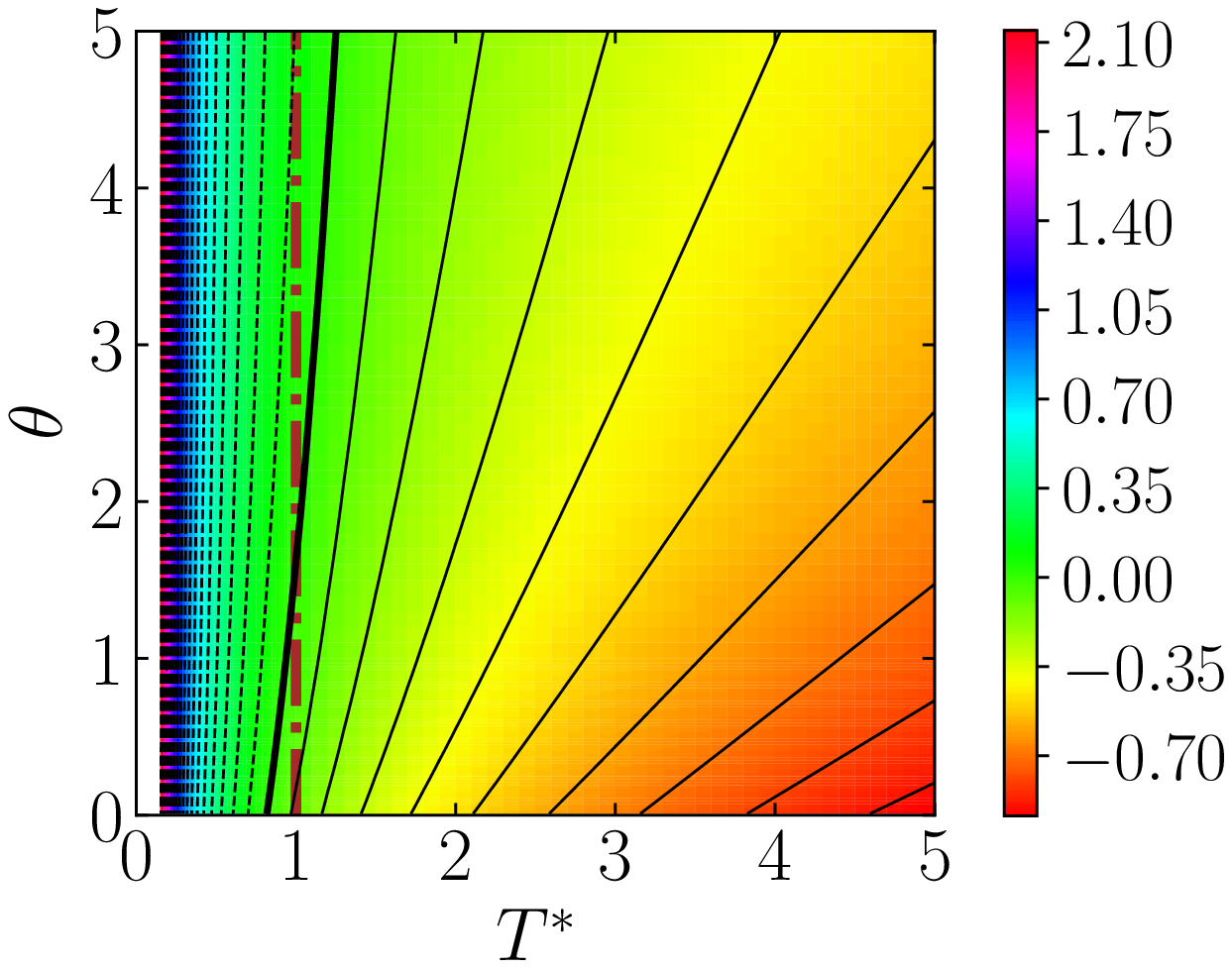}
    \end{minipage}
    \begin{minipage}[t]{.3\linewidth}
    	\textbf{C}\\
    \includegraphics[width = \linewidth]{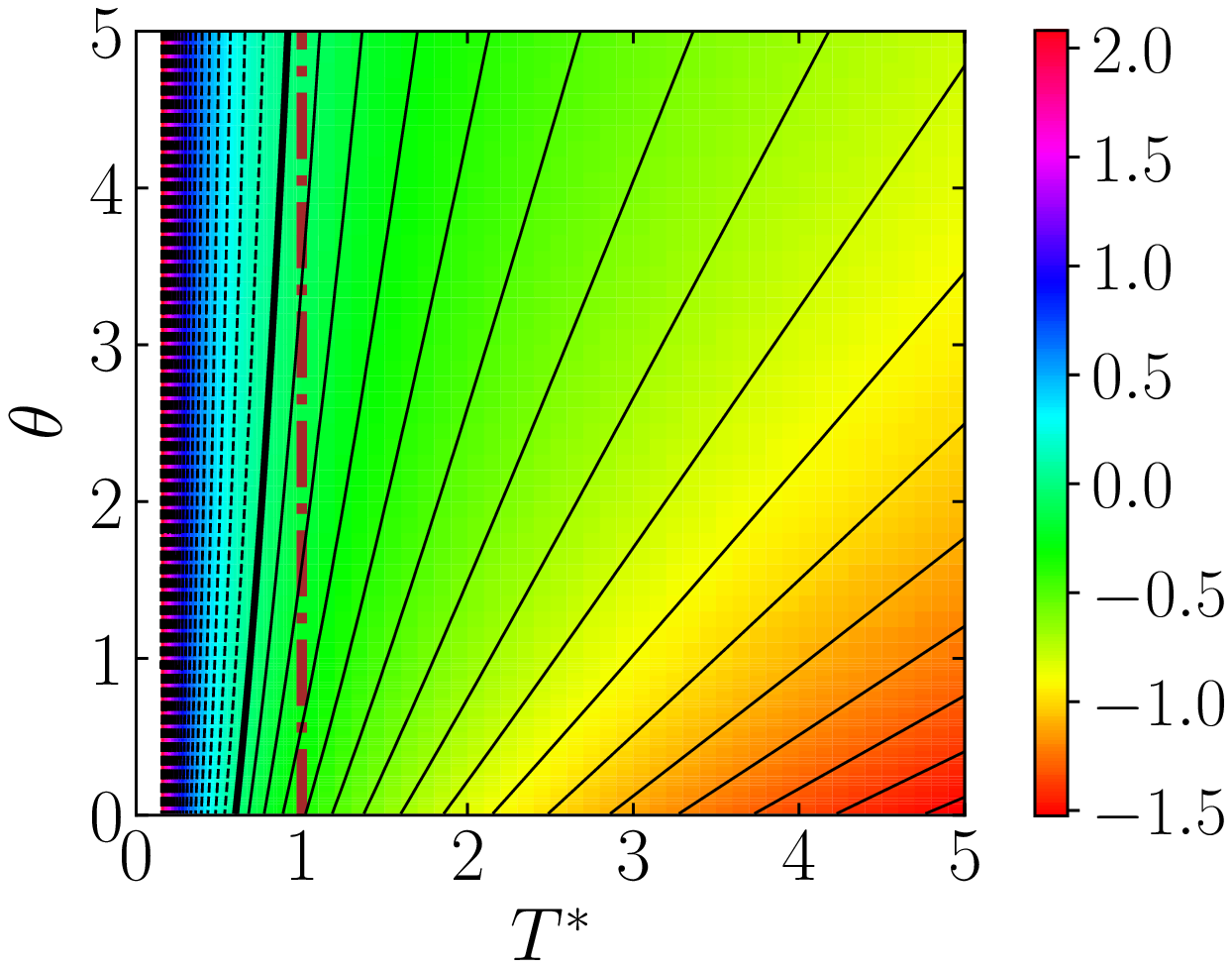}
    \end{minipage}\\
    \begin{minipage}[t]{.3\linewidth}
    	\textbf{D}\\
    \includegraphics[width = \linewidth]{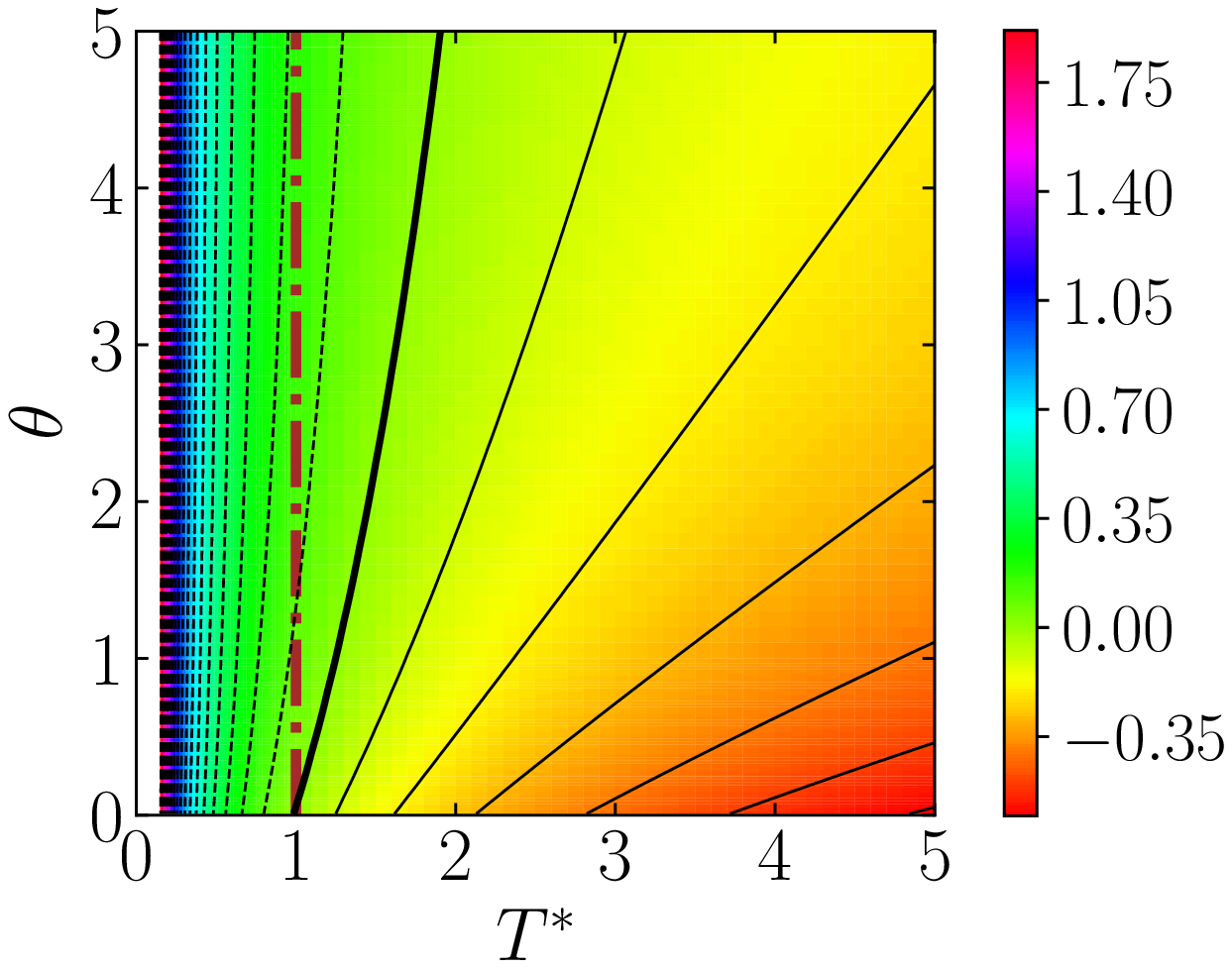}
    \end{minipage}
    \begin{minipage}[t]{.3\linewidth}
    	\textbf{E}\\
    \includegraphics[width = \linewidth]{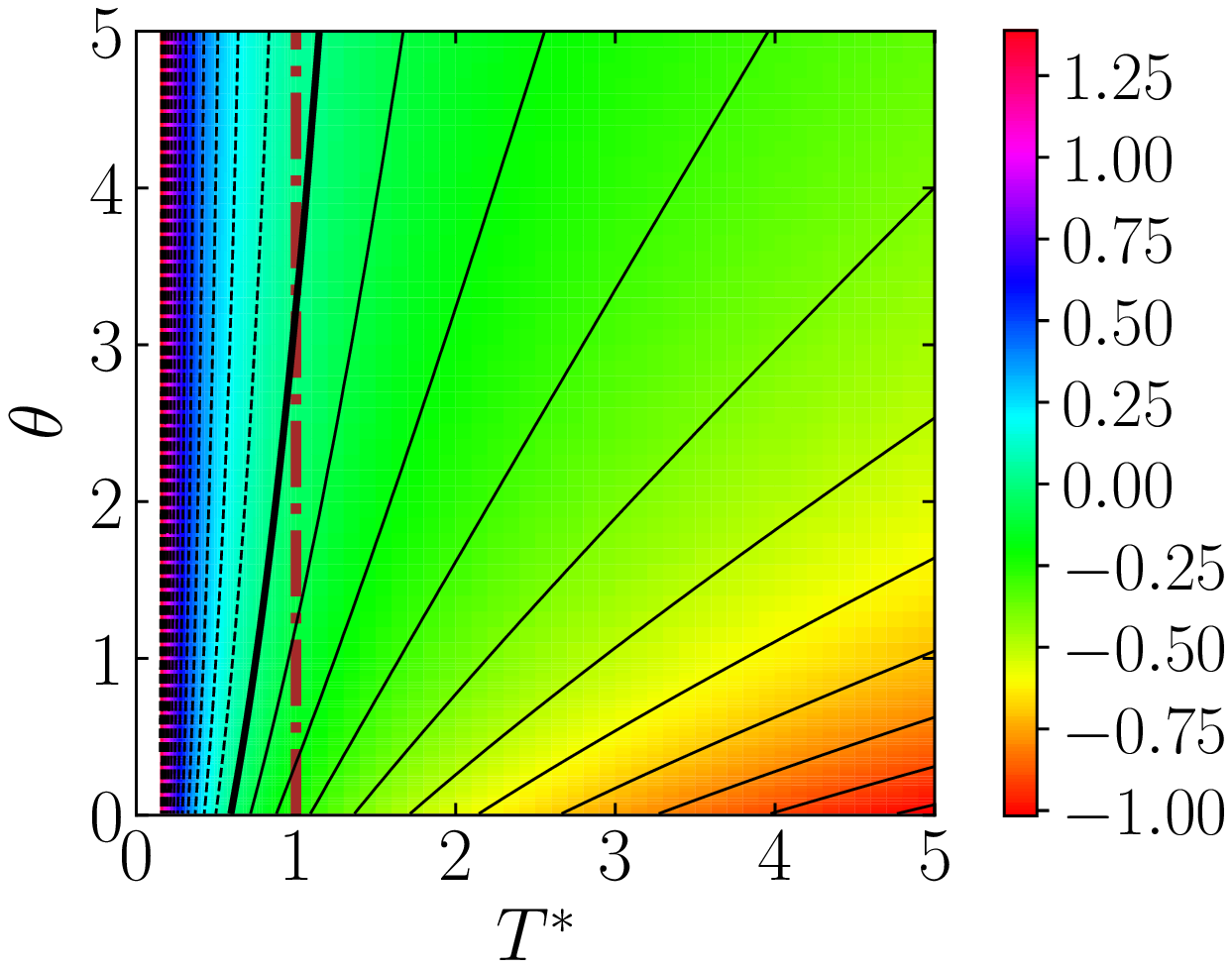}
\end{minipage}
    \begin{minipage}[t]{.3\linewidth}
    	\textbf{F}\\
    \includegraphics[width = \linewidth]{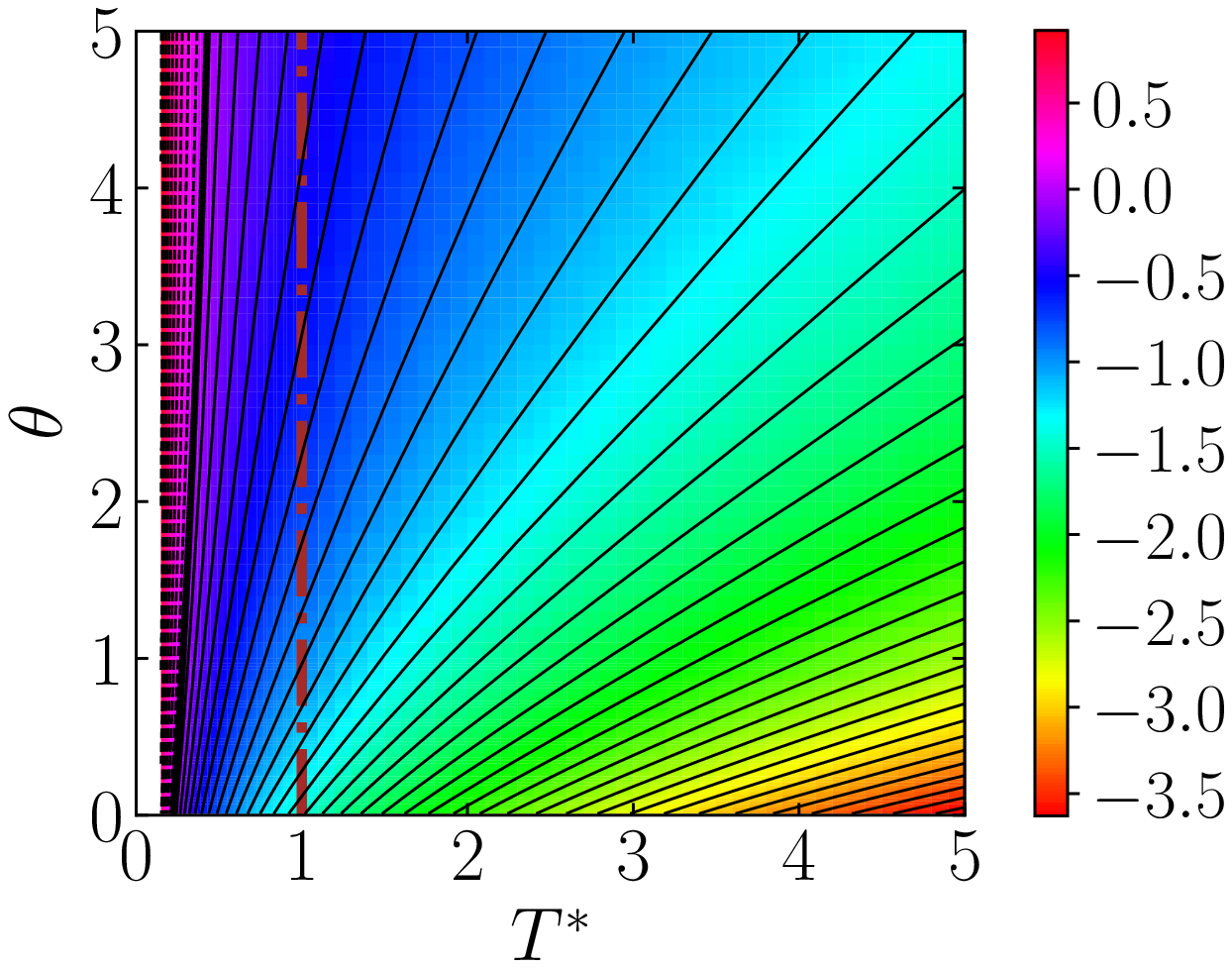}
    \end{minipage}
    \caption{Same as Figure~\ref{fig:slopes_1} but for $\alpha=0.7$.}
    \label{fig:slopes_2}
\end{figure}

Figures~\textbf{\ref{fig:slopes_1}} and \textbf{\ref{fig:slopes_2}} show density plots of $\Phi(T^*,\theta)$ for $\alpha=0.9$ and $0.7$, respectively. In each case, two values of $\beta$ ($0$ and $-0.7$) and three values of $\varepsilon$ ($0$, $0.5$, and $1$) are considered. We observe that, typically, $\Phi(T^*,\theta)$ increases with increasing $\theta$ at fixed $T^*$, while it decreases with increasing $T^*$ at fixed $\theta$.

\subsection{Comparison with simulation results}
\label{sec:2.3}

\begin{figure}
    \centering
    \begin{minipage}[t]{.36\linewidth}
    	\textbf{A}\\
    \includegraphics[width = \linewidth]
{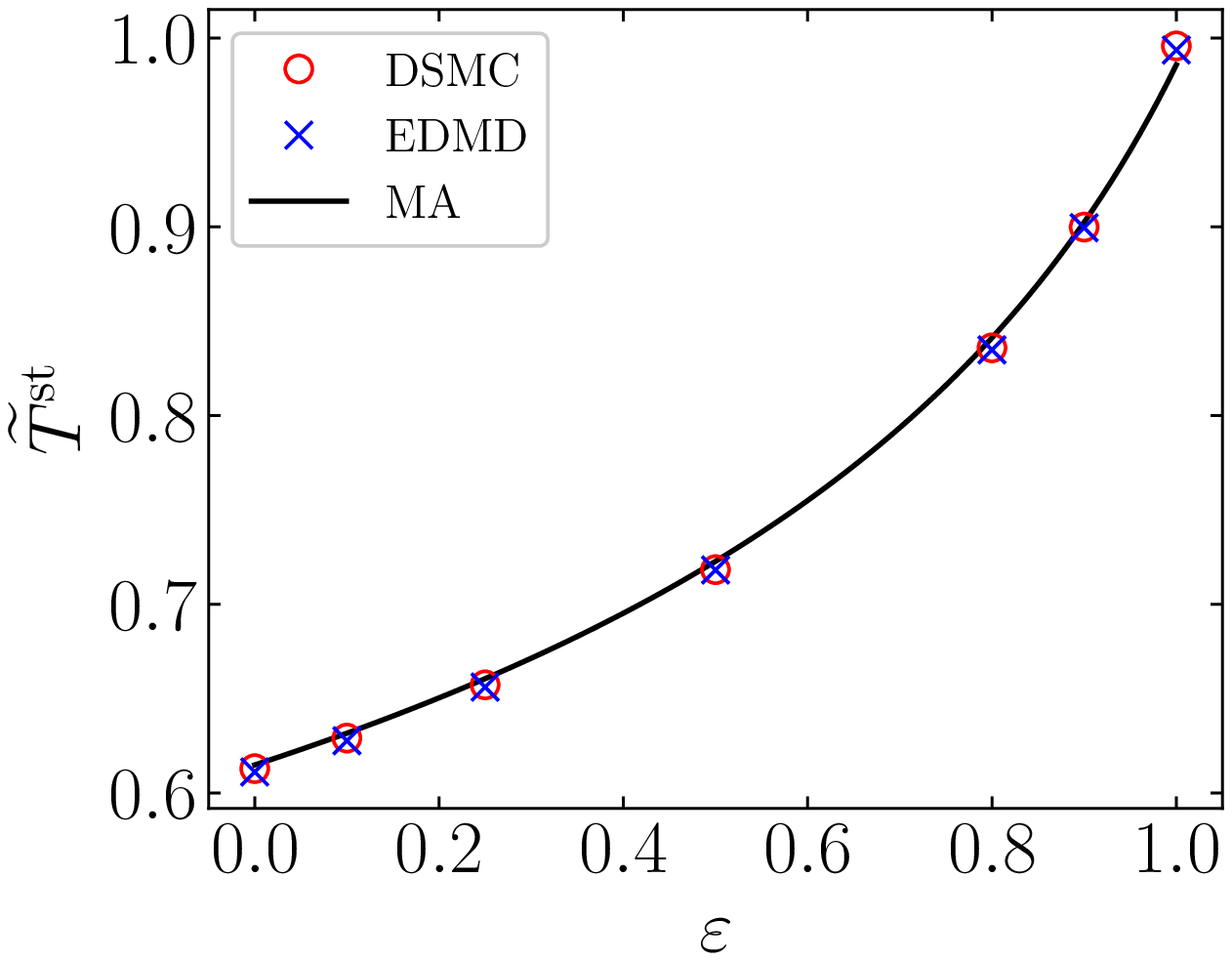}  
\end{minipage}\hspace{2cm} 
    \begin{minipage}[t]{.36\linewidth}
    	\textbf{B}\\
    \includegraphics[width = \linewidth]{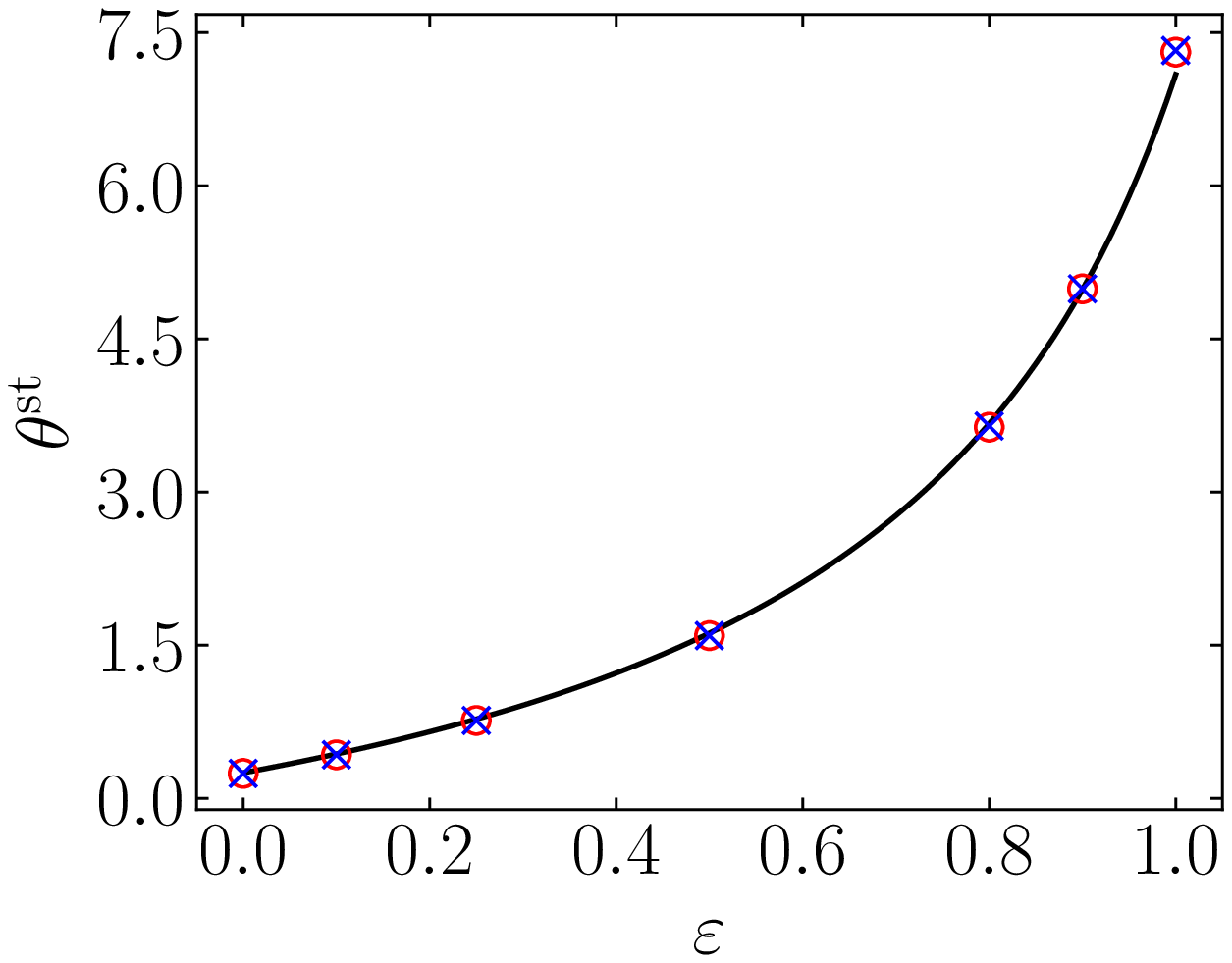}
      \end{minipage}\\
    \begin{minipage}[t]{.36\linewidth}
    	\textbf{C}\\
    \includegraphics[width = \linewidth]{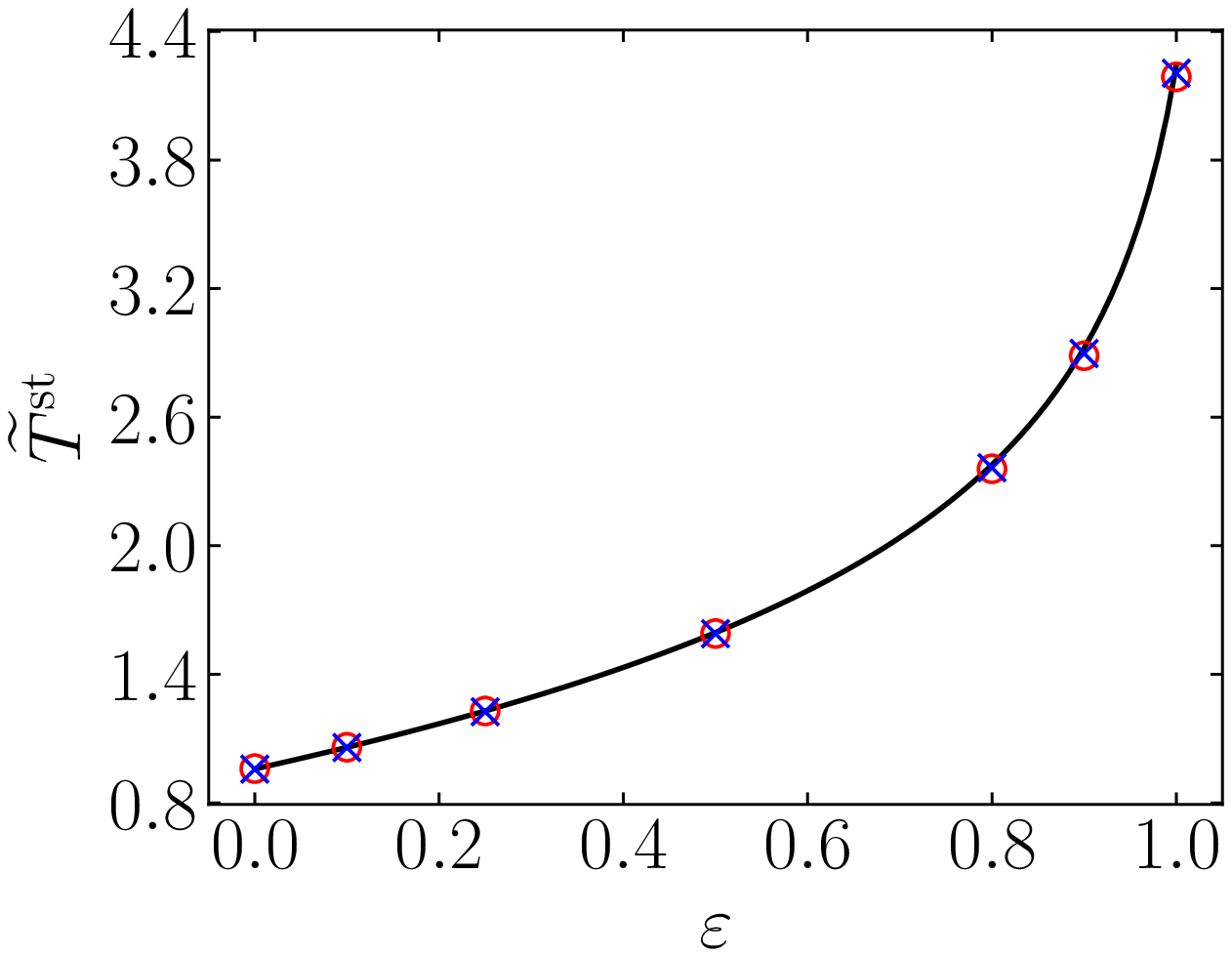}   
    \end{minipage}\hspace{2cm} 
    \begin{minipage}[t]{.36\linewidth}
    	\textbf{D}\\
    \includegraphics[width = \linewidth]{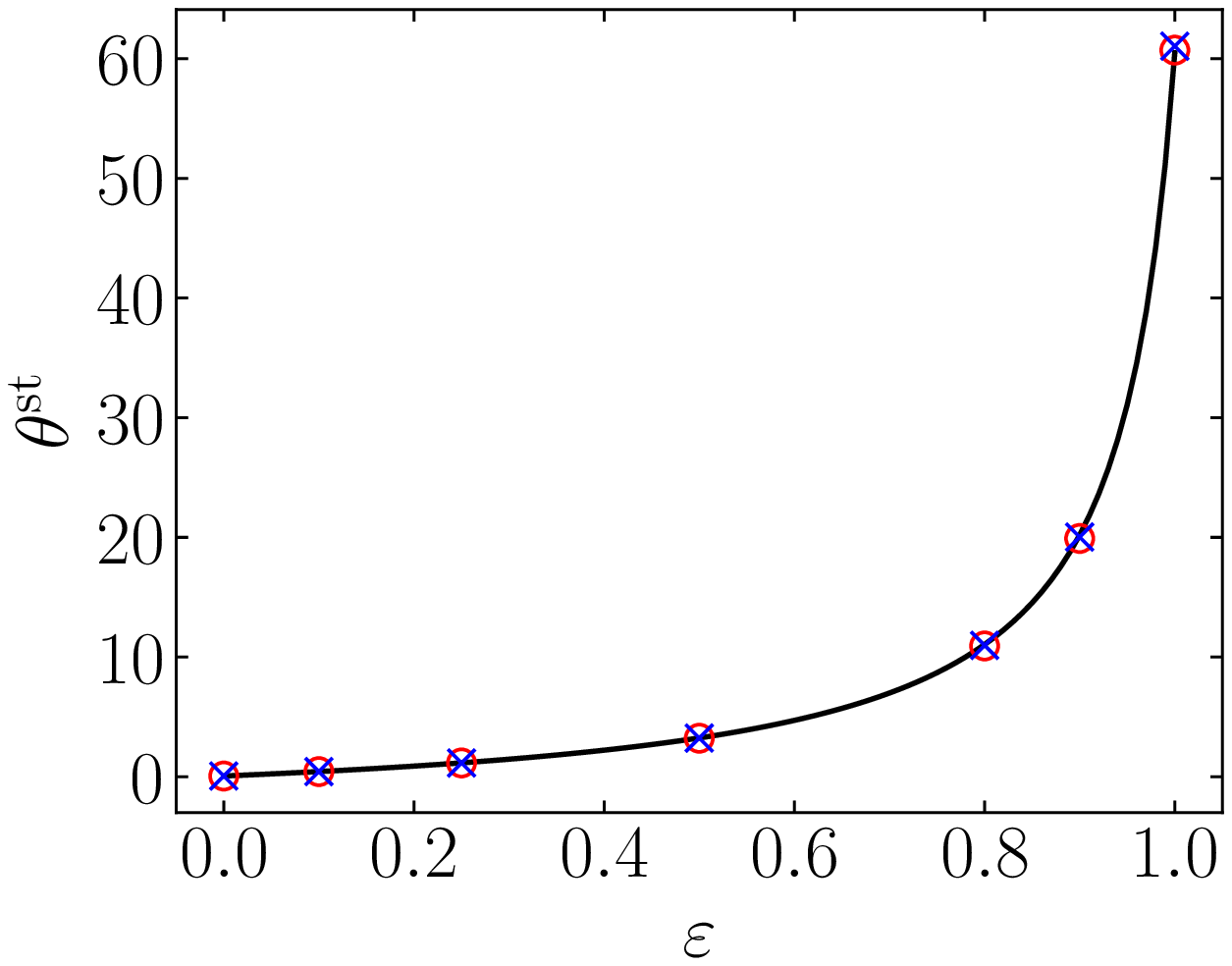}
     \end{minipage}\\
    \begin{minipage}[t]{.36\linewidth}
    	\textbf{E}\\
    \includegraphics[width = \linewidth]{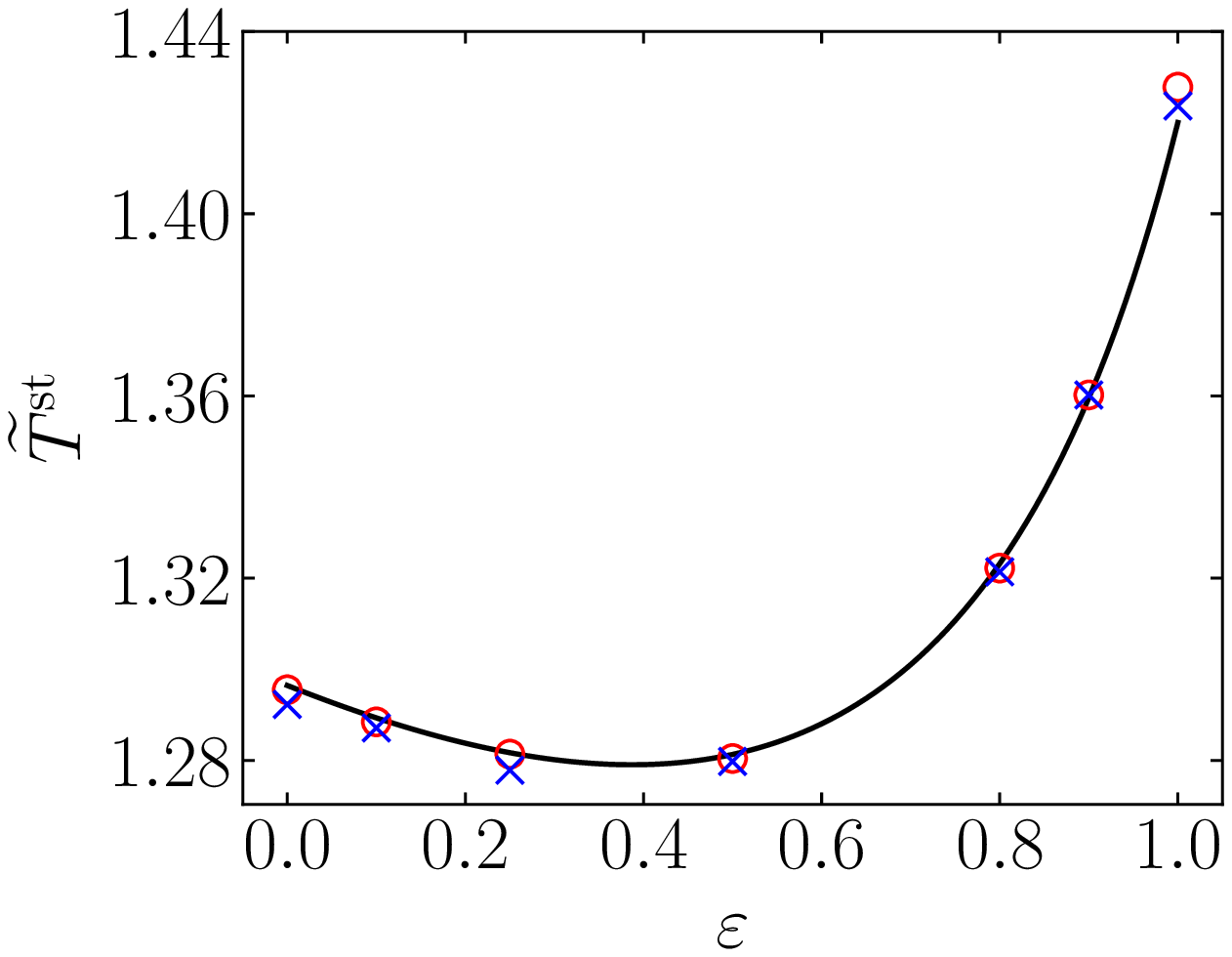}  
    \end{minipage} \hspace{2cm} 
    \begin{minipage}[t]{.36\linewidth}
    	\textbf{F}\\
    \includegraphics[width = \linewidth]{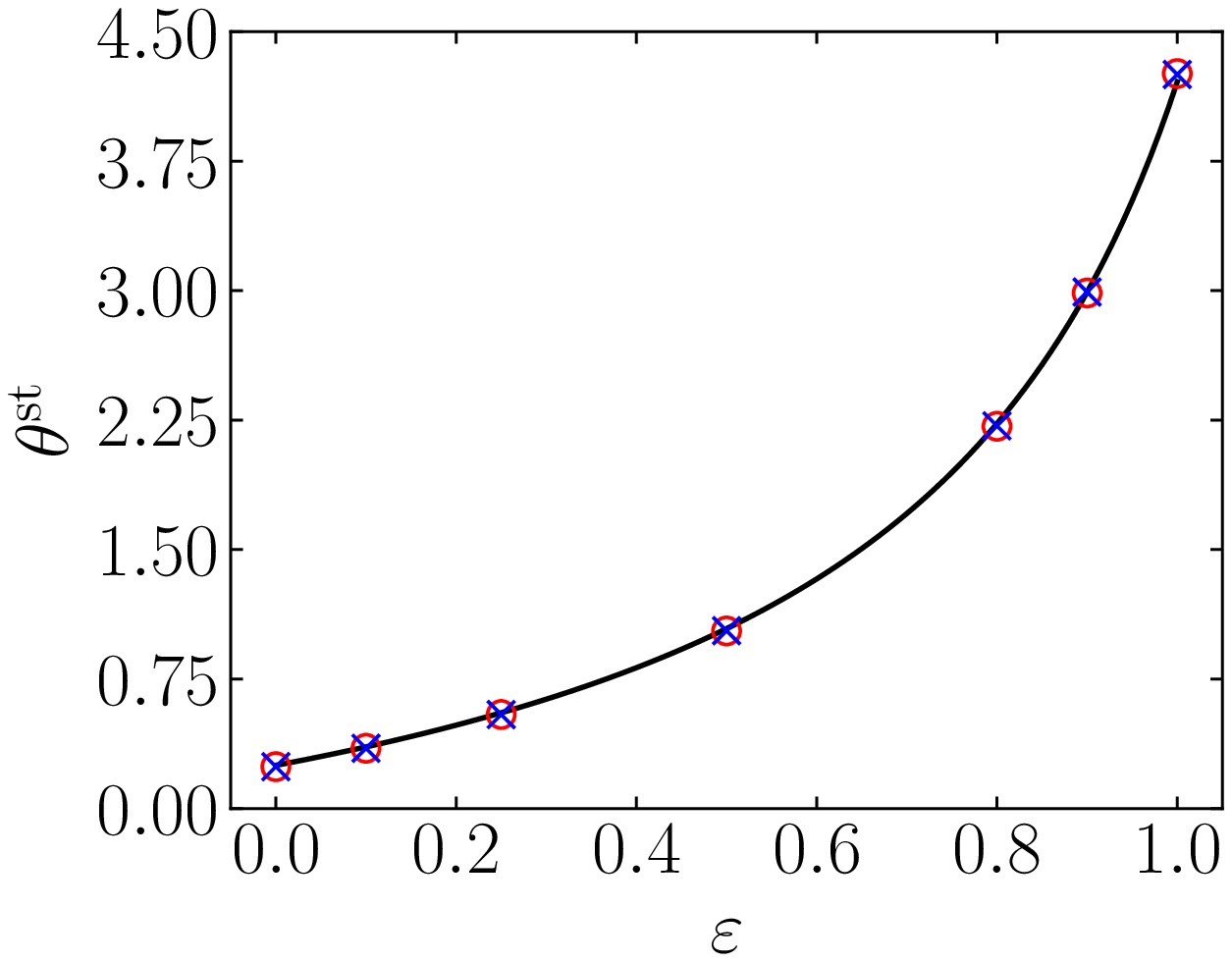}
        \end{minipage}\\
    \begin{minipage}[t]{.36\linewidth}
    	\textbf{G}\\
    \includegraphics[width = \linewidth]{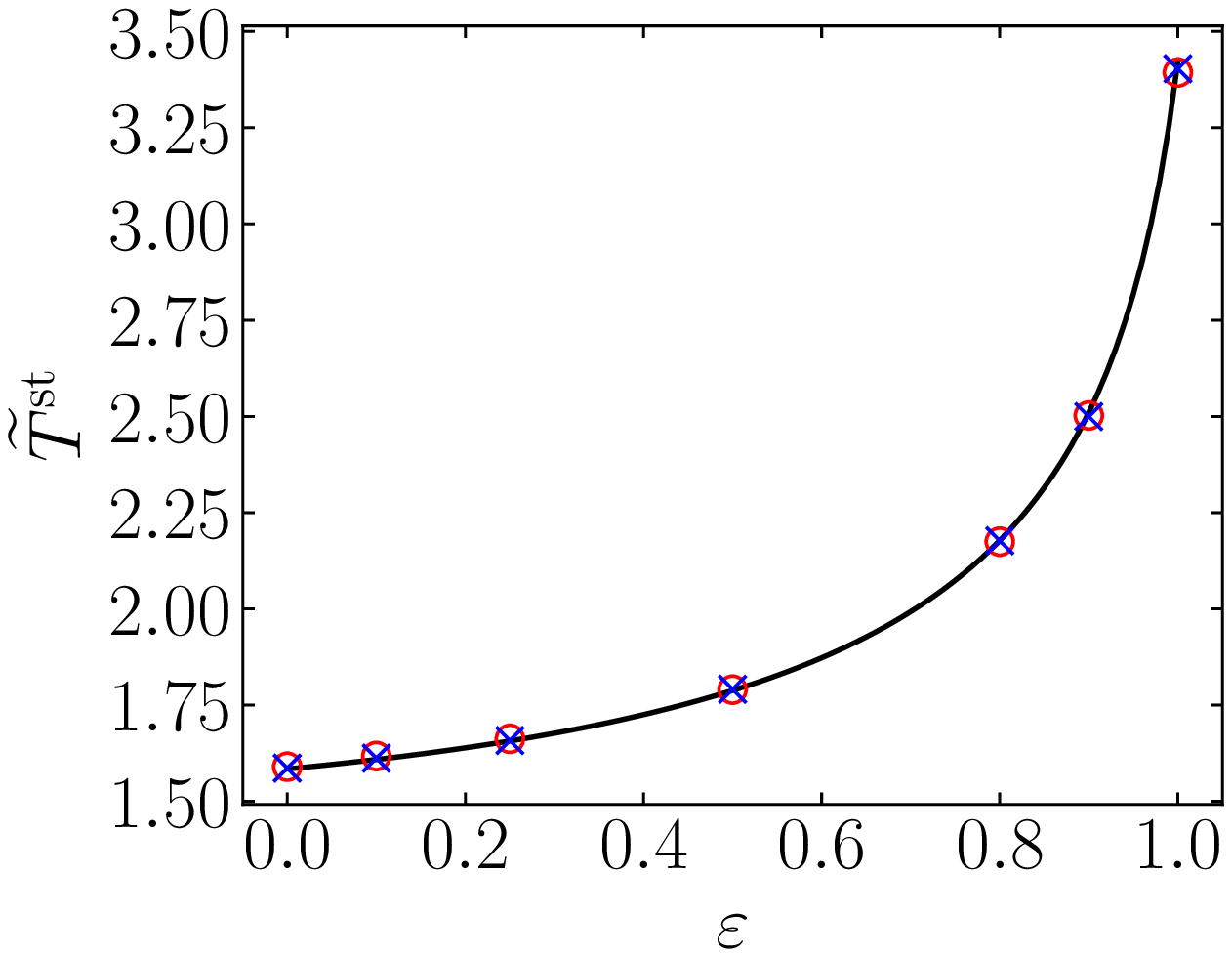} 
    \end{minipage}\hspace{2cm}  
    \begin{minipage}[t]{.36\linewidth}
    	\textbf{H}\\
    \includegraphics[width = \linewidth]{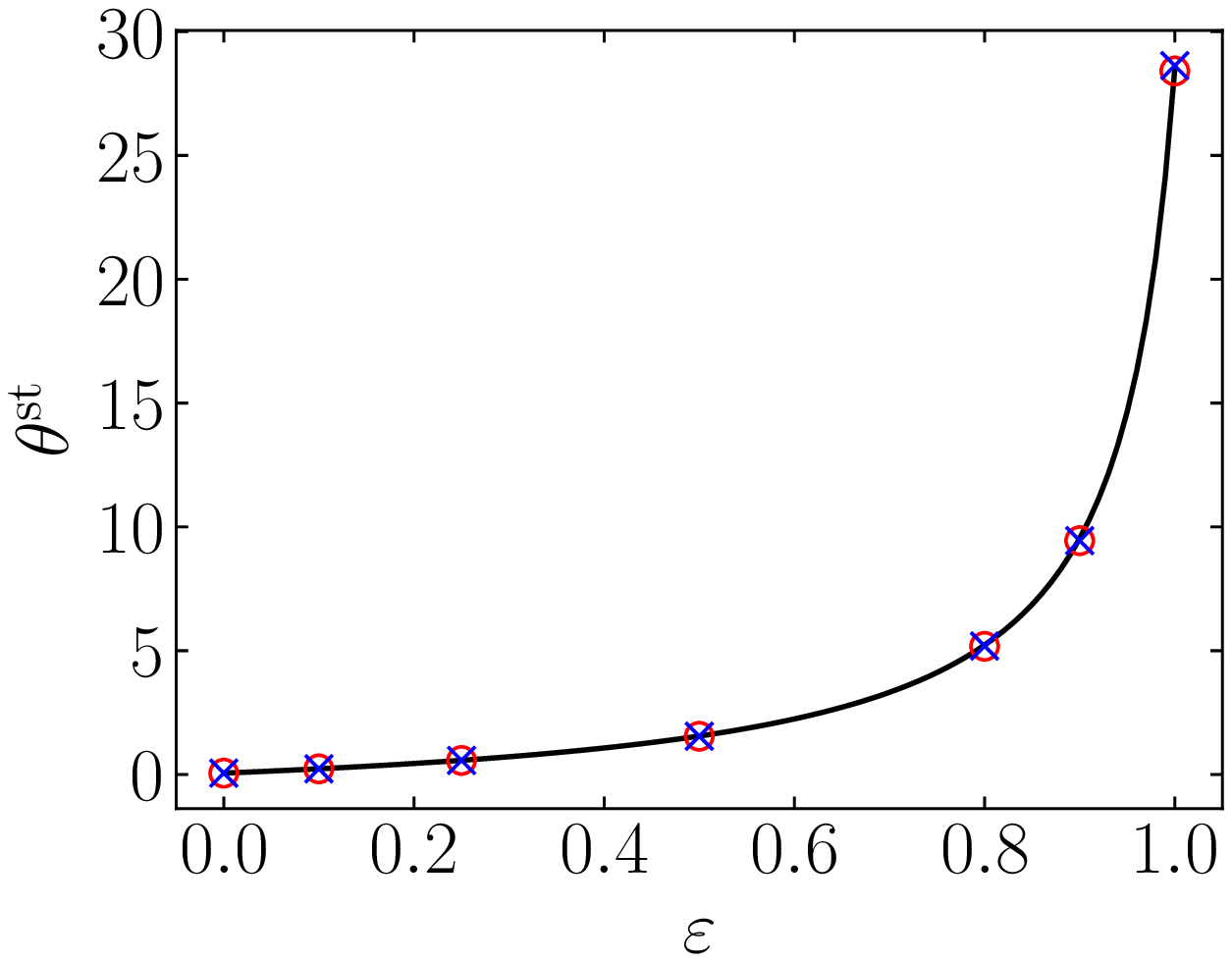}
    \end{minipage}
    \caption{Steady-state values ${\widetilde{T}}^\st$ and $\theta^\st$ as a function of $\varepsilon$. The values of the coefficients of restitution are \textbf{(A)} and \textbf{(B)}: $(\alpha,\beta)=(0.7,0)$; \textbf{(C)} and \textbf{(D)}: $(\alpha,\beta)=(0.7,-0.7)$; \textbf{(E)} and \textbf{(F)}: $(\alpha,\beta)=(0.9,0)$; and \textbf{(G)} and \textbf{(H)}: $(\alpha,\beta)=(0.9,-0.7)$.  Thick black lines correspond to the theoretical prediction in Equation~\eqref{eq:steady_values}, and symbols refer to DSMC ($\circ$) and EDMD ($\times$) simulation results for $\varepsilon=0,\,0.1,\,0.25,\,0.5,\,0.8,\,0.9$, and $1$. }
    \label{fig:steady_state}
\end{figure}

\begin{figure}
    \centering
    \begin{minipage}[t]{.36\linewidth}
    	\textbf{A}\\
    \includegraphics[width = \linewidth]{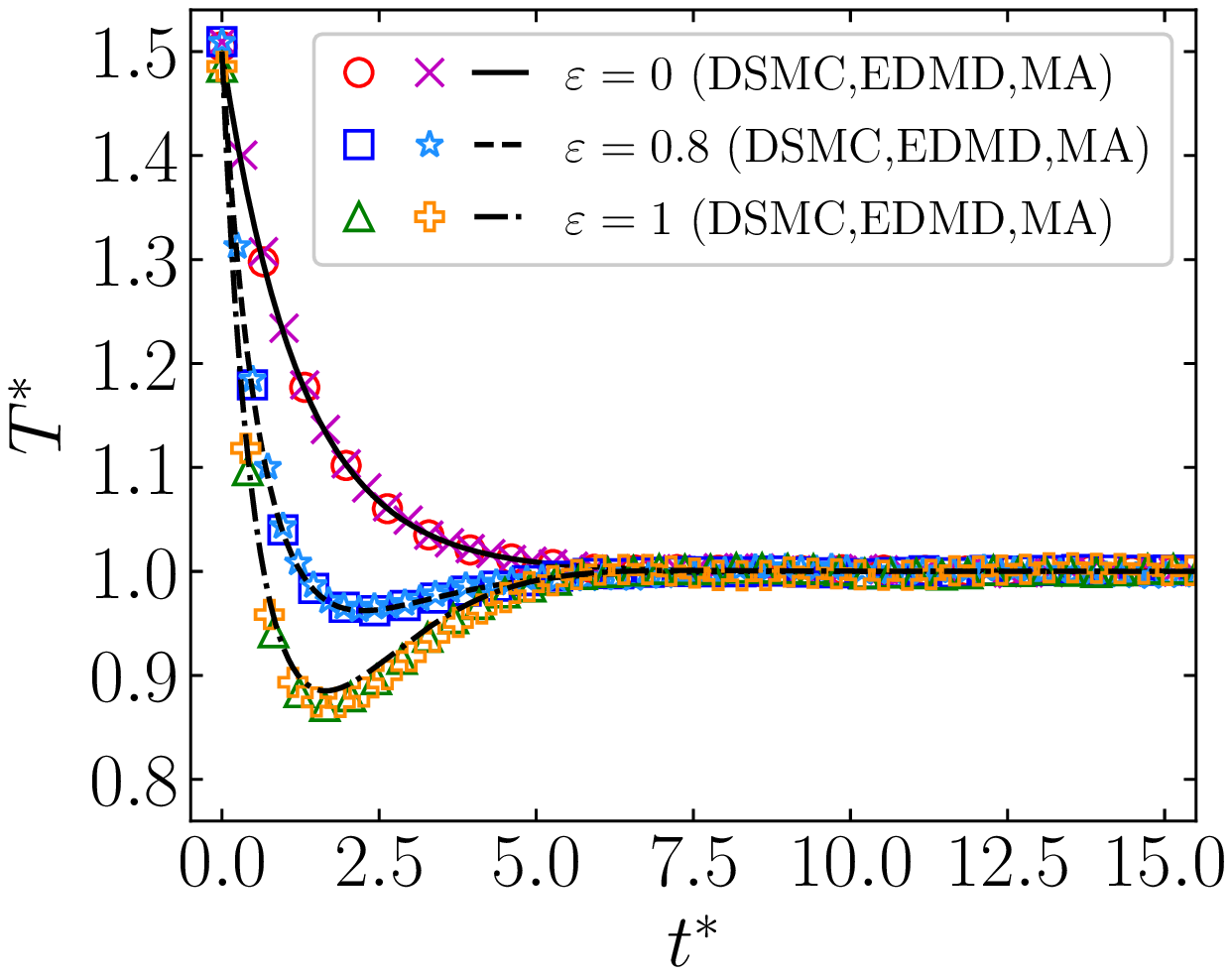}  		\end{minipage}\hspace{2cm}  
    \begin{minipage}[t]{.36\linewidth}
    	\textbf{B}\\
    \includegraphics[width = \linewidth]{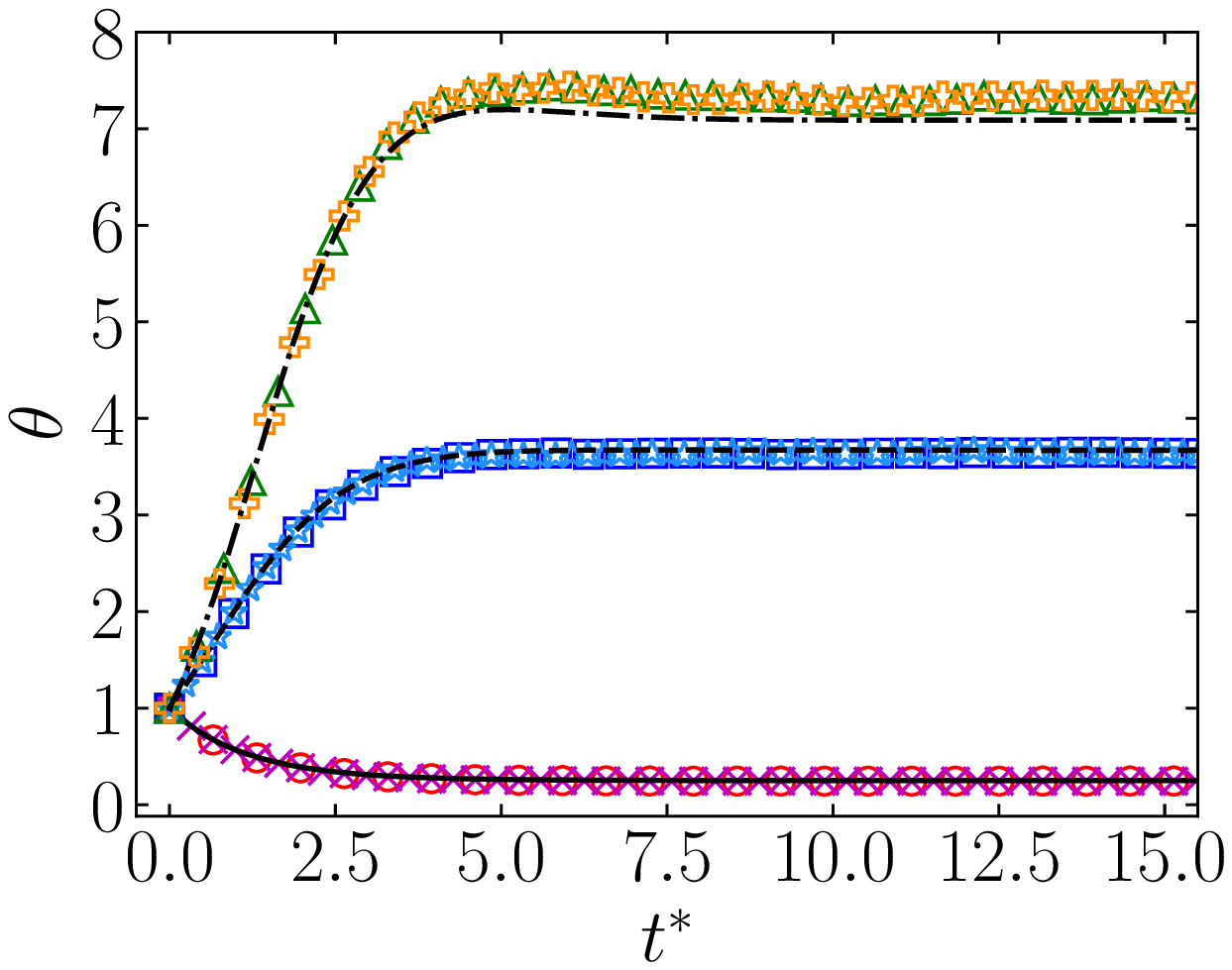}
    \end{minipage}\\
    \begin{minipage}[t]{.36\linewidth}
    	\textbf{C}\\
    \includegraphics[width = \linewidth]{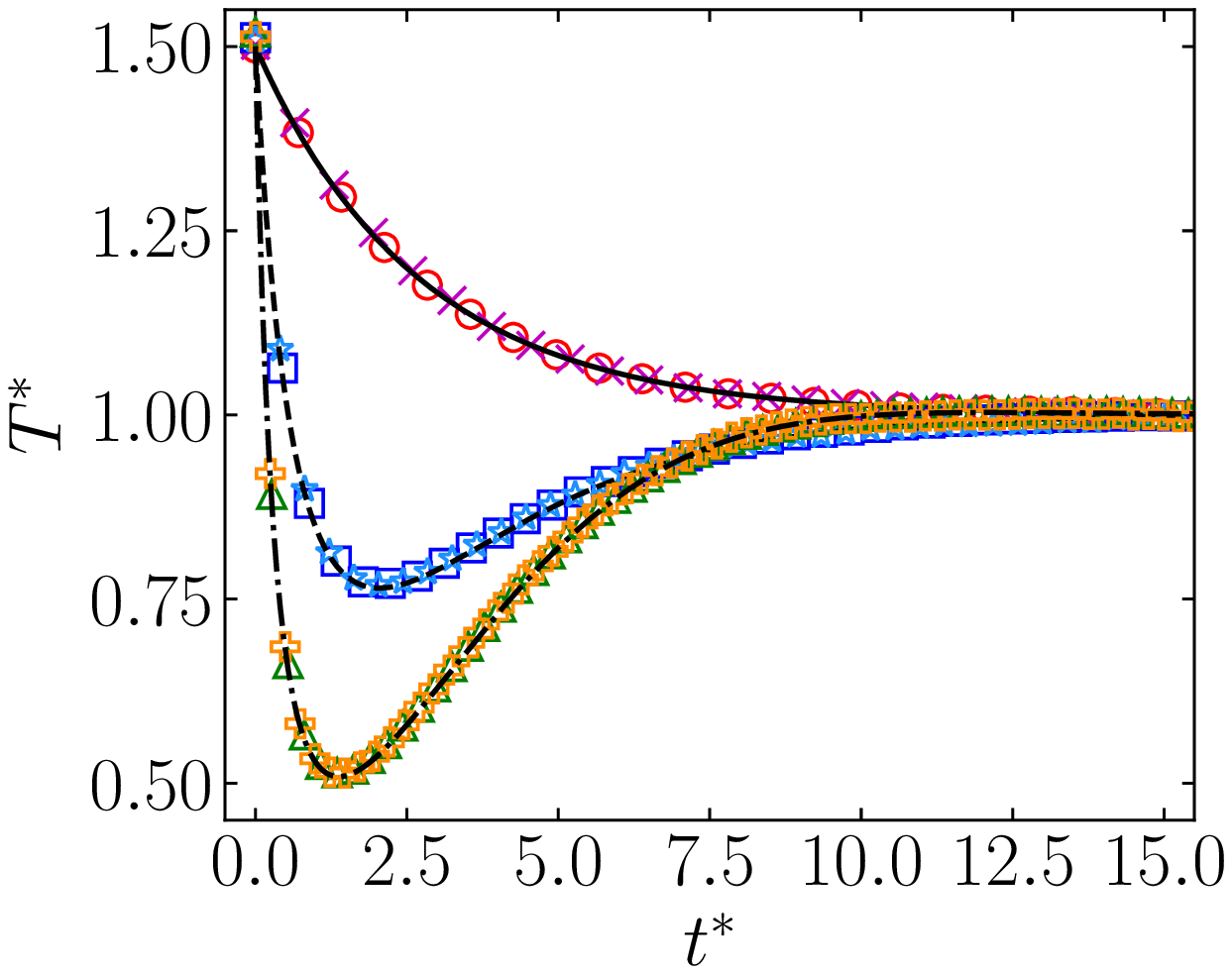} 
    \end{minipage}\hspace{2cm}  
    \begin{minipage}[t]{.36\linewidth}
    	\textbf{D}\\
    	\includegraphics[width = \linewidth]{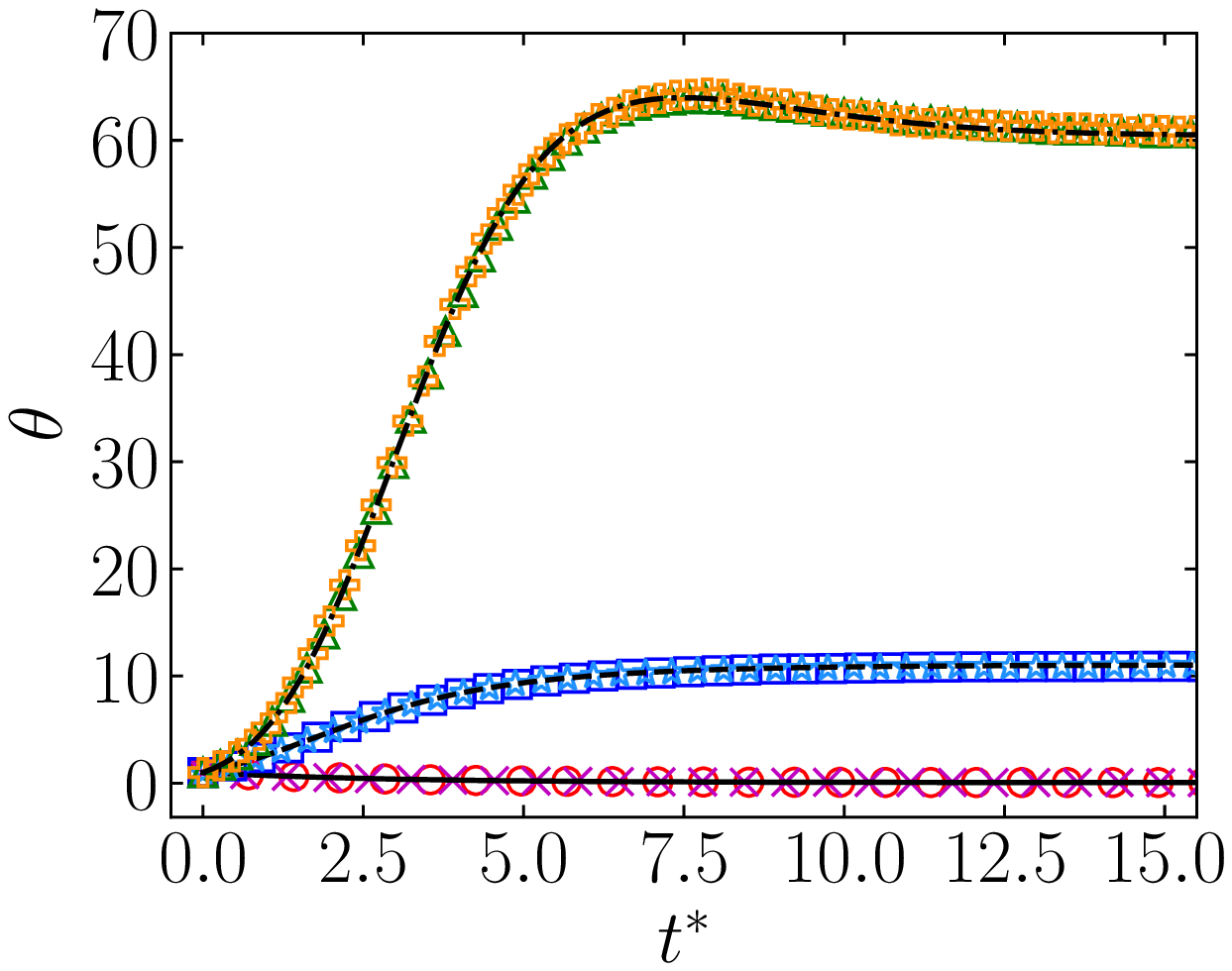}
        \end{minipage}\\  
    	\begin{minipage}[t]{.36\linewidth}
    	\textbf{E}\\
    	\includegraphics[width = \linewidth]{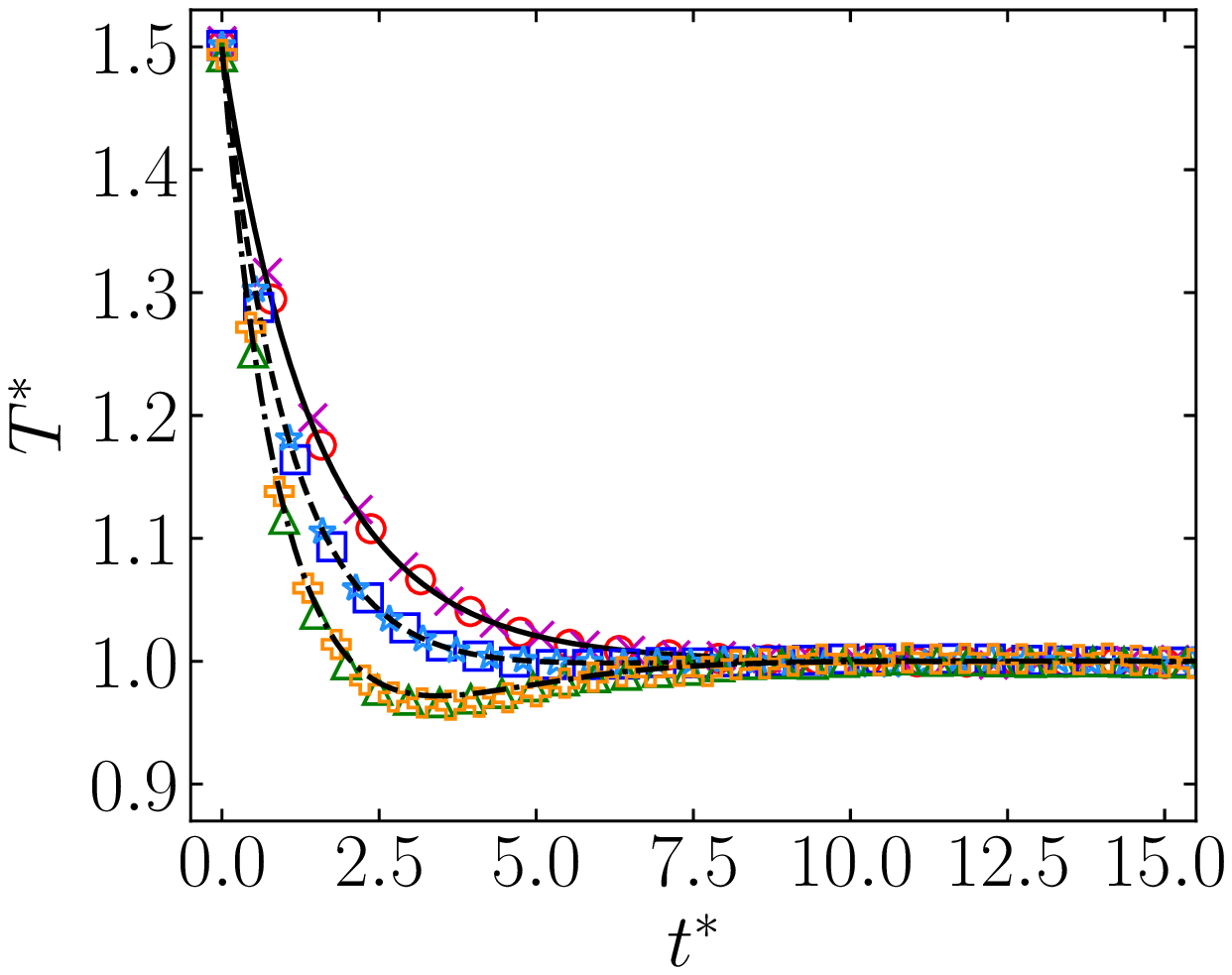}
    	\end{minipage}\hspace{2cm}  
    \begin{minipage}[t]{.36\linewidth}
    	\textbf{F}\\
    	\includegraphics[width = \linewidth]{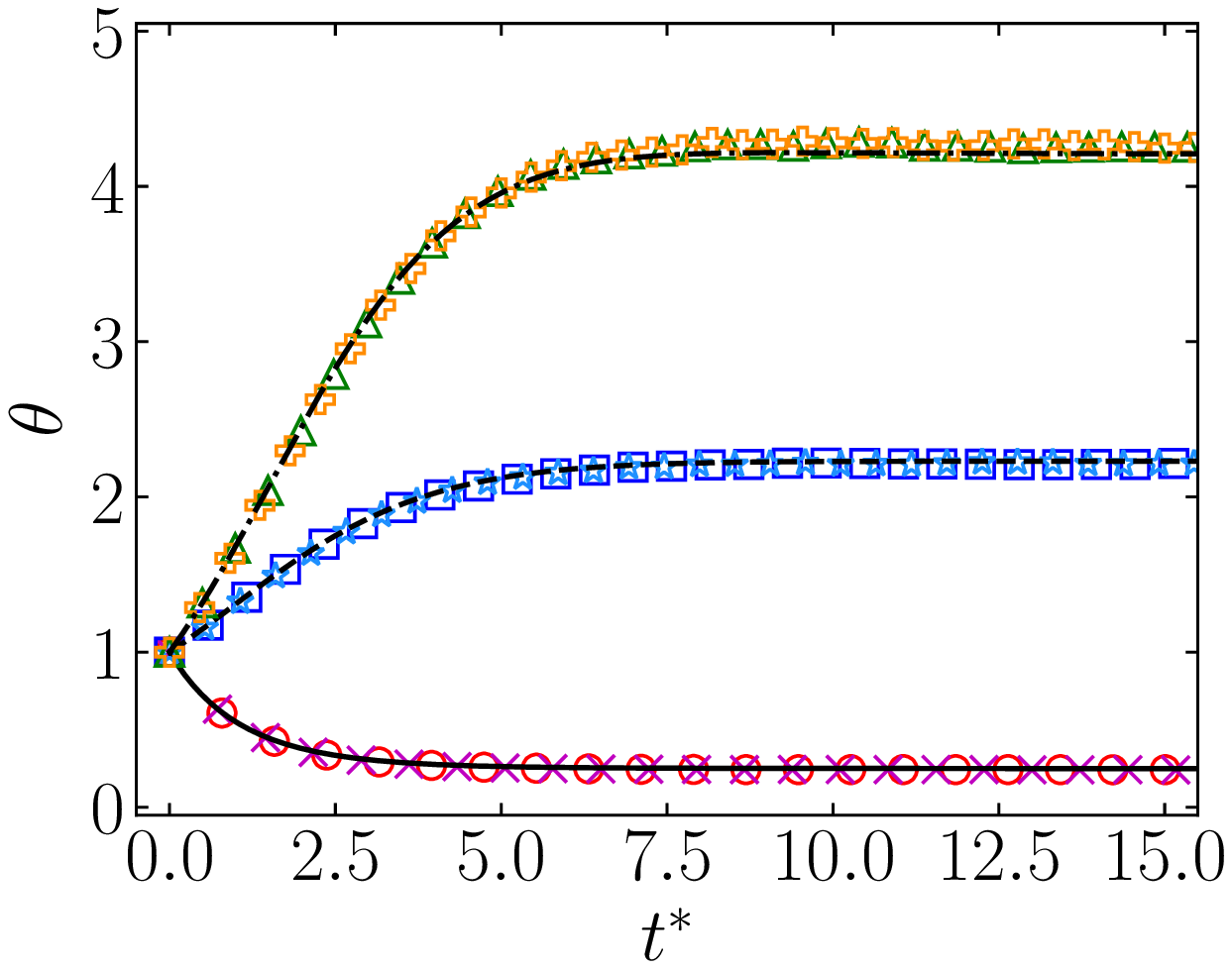}
        \end{minipage}\\  
    \begin{minipage}[t]{.36\linewidth}
    	\textbf{G}\\
    	\includegraphics[width = \linewidth]{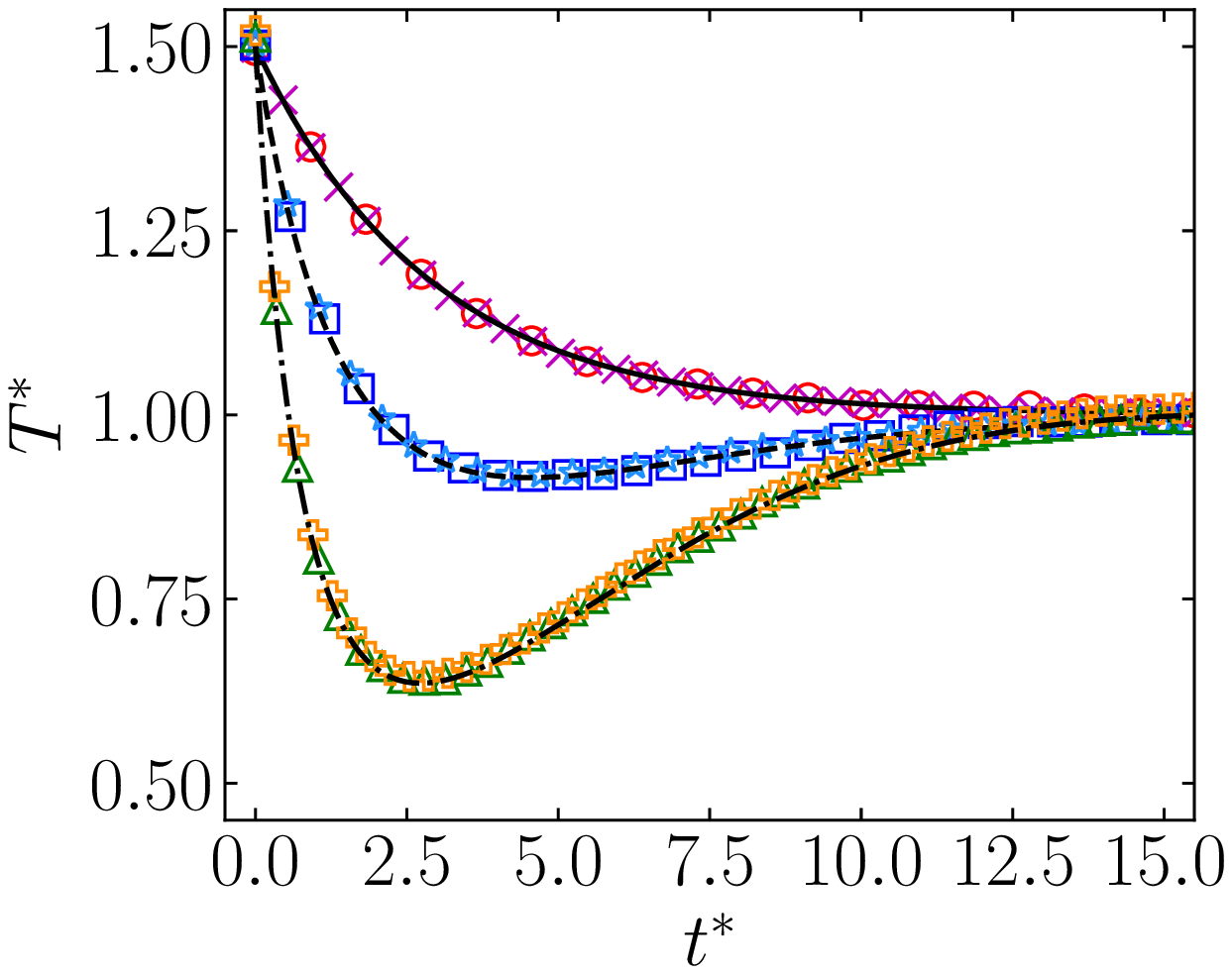}  
    	\end{minipage}\hspace{2cm}  
    	\begin{minipage}[t]{.36\linewidth}
    	\textbf{H}\\
    	\includegraphics[width = \linewidth]{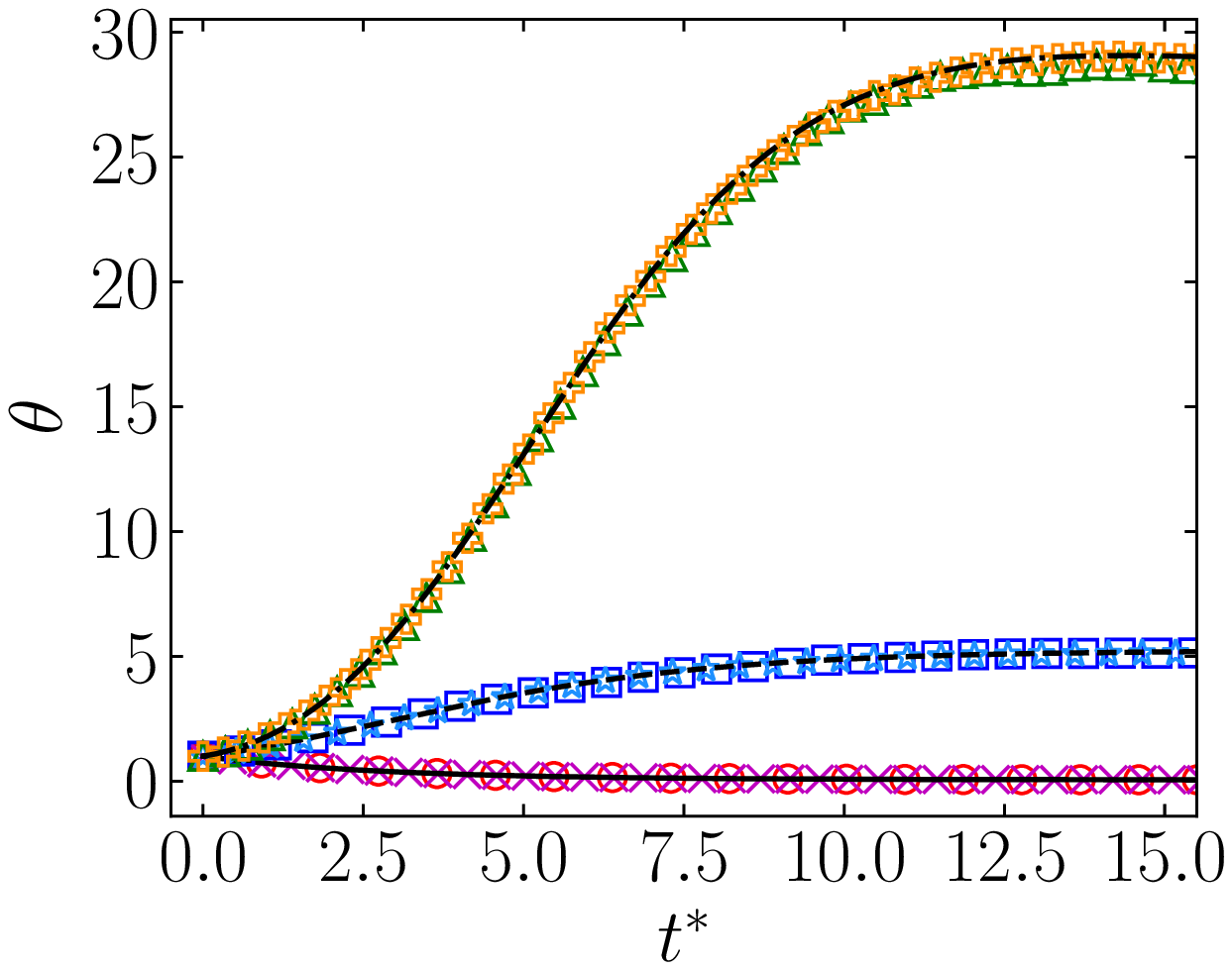}
		\end{minipage}        
        \caption{Time evolution of $T^*$ and $\theta$ for $\varepsilon=0$, $0.8$, and $1$. The initial conditions are $T_0^*=1.5$ and $\theta_0=1$ in all the cases. The values of the coefficients of restitution are \textbf{(A)} and \textbf{(B)}: $(\alpha,\beta)=(0.7,0)$; \textbf{(C)} and \textbf{(D)}: $(\alpha,\beta)=(0.7,-0.7)$; \textbf{(E)} and \textbf{(F)}: $(\alpha,\beta)=(0.9,0)$; and \textbf{(G)} and \textbf{(H)}: $(\alpha,\beta)=(0.9,-0.7)$.  Lines correspond to the theoretical prediction from  Equations~\eqref{eq:Tstar-theta-ev}, and symbols refer to DSMC  and EDMD  simulation results. }
    \label{fig:evol}
\end{figure}

In order to check the validity of the MA, we have compared our theoretical predictions against DSMC and EDMD simulation results both for transient and steady-state values.

The DSMC algorithm used is based on the one presented in, e.g., Refs.~\cite{B13,MS00}, and adapted for the model presented in this work. For our DSMC simulations we have dealt with $N=10^4$ particles and chosen a time step $\Delta t=4\times 10^{-5}/\nun$. In addition, the way of implementing the stochastic force and torque in the EDMD code is based on the \emph{approximate Green function} algorithm \cite{S12}, as applied to the ST. We have chosen $N=3.6\times 10^{3}$ particles, a density $n\sigma^2= 5\times 10^{-4}$ (implying a box length of $L/\sigma \approx 1897.37$), and a time step $\Delta t \approx 4\times 10^{-4}/\nun$. No instabilities were observed.

In Figure~\textbf{\ref{fig:steady_state}}, results for the steady-state values ${\widetilde{T}}^\st$ and $\theta^\st$ as functions of $\varepsilon$  and different values of $\alpha$ and $\beta$ are presented. Simulation results for DSMC and EDMD come from averages over $100$ replicas and over $50$ data points, once the steady state is ensured to be reached. A very good agreement between DSMC and EDMD with expressions in Equation~\eqref{eq:steady_values} is observed.
Whereas $\theta^\st$ is an increasing function of $\varepsilon$, this is not the case, in general, with ${\widetilde{T}}^\st$, as can be observed in Figure~\textbf{\ref{fig:steady_state}}\textbf{E}.  

As a test of the transient stage, we present in Figure~\textbf{\ref{fig:evol}} the evolution of $T^*$ and $\theta$ (starting from a Gaussian-generated VDF with $T_0^*=1.5$ and $\theta_0=1$) for the same choices of $\alpha$ and $\beta$ as in Figure~\textbf{\ref{fig:steady_state}} and for the representative values $\varepsilon=0$ (translational noise only), $0.8$ (both translational and rotational noise), and $\varepsilon=1$ (rotational noise only). We observe again an excellent agreement of the MA, Equations~\eqref{eq:Tstar-theta-ev}, with simulation results.

\subsection{Temperature overshoot}
\label{sec:2.4}

\begin{figure}
    \centering
    \includegraphics[width=0.5\textwidth]{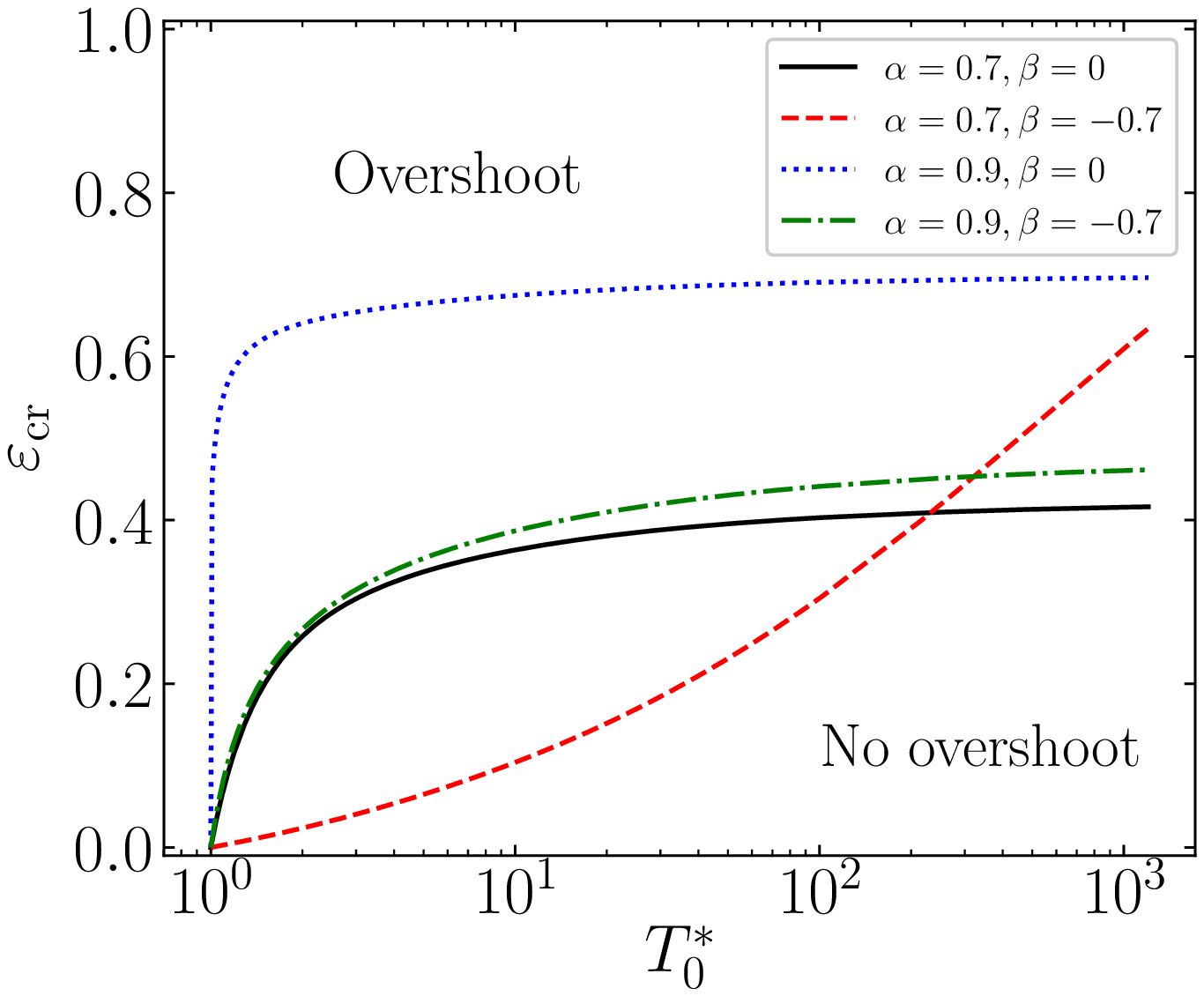}
    \caption{Phase diagram for the emergence of overshoot. The numerical critical value $\varepsilon_\ct(T_0^*,\theta_0)$ is presented as a function of $T_0^*$ for $\theta_0$ given by Equation~\eqref{eq:theta_st_0} and four different pairs of the coefficients of restitution: $(\alpha,\beta)=(0.7,0)$, $(0.7,-0.7)$, $(0.9,0)$, and $(0.9,-0.7)$.}
    \label{fig:PD_overshoot}
\end{figure}

As illustrated in Figure~\textbf{\ref{fig:evol}}, the evolution of $T^*(t^*)$ for certain initial states might experiment an overshoot $T^*(t_O^*)=1$ at a finite time $t^*_O$, followed by a minimum, and then relax to the steady state from below.
This overshoot effect becomes more pronounced as $\Phi(T_0^*,\theta_0)$ takes more negative values, i.e., as $\varepsilon$ increases and/or $\theta_0$ decreases.

In general, at a given initial condition $(T_0^*,\theta_0)$, there exists a critical value $\varepsilon_\ct(T_0^*,\theta_0)$, such that the $T^*(t^*)$  exhibits overshoot if $\varepsilon>\varepsilon_\ct(T_0^*,\theta_0)$. The determination of $\varepsilon_\ct(T_0^*,\theta_0)$ within the MA requires  the numerical solution of Equations~\eqref{eq:Tstar-theta-ev}. We have analyzed those numerical solutions up to $t^*=15$ since the overshoot typically takes place in the first stage of the evolution. Thus, the numerical value $\varepsilon=\varepsilon_\ct$ corresponds to $t_O^*=15$.

Figure~\textbf{\ref{fig:PD_overshoot}} shows $\varepsilon_\ct(T_0^*,\theta_0)$ as a function of $T_0^*$ at a specific value of $\theta_0$, namely the one given by Equation~\eqref{eq:theta_st_0}, and for the same four pairs of coefficients of restitution as in Figures~\textbf{\ref{fig:slopes_1}}--\textbf{\ref{fig:evol}}. In each case, the curve $\varepsilon_\ct(T_0^*,\theta_0)$ splits the plane $\varepsilon$ vs $T^*_0$ into two regions: the region above the curve, where the overshoot effect is present, and the one below the curve, where temperature relaxes to the steady-state value from above.
Note that the shape of the curve $\varepsilon_\ct$ in Figure~\textbf{\ref{fig:PD_overshoot}} associated with the pair $(\alpha,\beta)=(0.7,-0.7)$ differs in curvature from the curves associated with the other three pairs.

\section{Mpemba effect}
\label{sec:3}

As already said in section \ref{sec:1}, ME refers to the counterintuitive phenomenon according to which an initially hotter sample of a given fluid relaxes earlier to the steady state than an initially colder one. In a recent paper \cite{MSP22}, we distinguished  between the thermal ME---where the relaxation process is  described by the temperature of the system (second moment of the VDF)--- and the entropic ME---where the deviation from the final steady state is monitored  by the Kullback--Leibler divergence (thus involving  the full VDF). Whereas this distinction is interesting and the relationship between the thermal ME and the entropic ME is not always biunivocal \cite{MSP22}, we focus this paper on the thermal version due to its simpler characterization and its relationship with the original results  \cite{MO69}. Morover, only cooling processes will be considered throughout this work.

Let us assume two samples---denoted by A and B---of the same gas, subject to the same noise temperature $\Tn_\rf$ and the same splitting parameter $\varepsilon_\rf$, so that the final steady-state values $\widetilde{T}^\st_\rf$ and $\theta^\st_\rf$ will also be the same. Both samples differ in the initial conditions $(T_{0A}^*,\theta_{0A})$ and $(T_{0B}^*,\theta_{0B})$, respectively, where  $T_{0A}^*>T_{0B}^*>1$, that is, A refers to the initially hotter sample and we are considering a cooling experiment.

\subsection{Standard Mpemba effect}
\label{sec:3.1}

Let us first consider the standard form of thermal ME  \cite{LVPS17,TLLVPS19,SP20,GKG21,MSP22}, where both $T_A^*(t^*)$ and $T_B^*(t^*)$  cross over at a certain crossing time $t_c^*$ and then relax from above, i.e., $T_A^*(t_c^*)=T_B^*(t_c^*)$ and $T_B^*(t^*)>T_A^*(t^*)>1$ for $t^*>t_c^*$. The emergence of ME can be subdued to the appearance of a crossing time (or an odd number of them) in the absence of any overshoot effect in the thermal evolution \cite{MSP22}. We will refer to this situation as the standard ME (SME).
According to the discussion in section \ref{sec:2.4}, the SME implies that $\varepsilon_\rf<\min\{\varepsilon_\ct(T_{0A}^*,\theta_{0A}),\varepsilon_\ct(T_{0B}^*,\theta_{0B})\}$.  The case when a temperature overshoot  takes place will be discussed in section \ref{sec:3.2}.

It can be reasonably expected that a necessary condition for the occurrence of the SME is that the initial slope is smaller in sample A than in sample B, i.e.,
\begin{equation}\label{eq:nec_cond}
\Phi(T_{0A}^*,\theta_{0A})<\Phi(T_{0B}^*,\theta_{0B}).
\end{equation}
Note here that the usual situation is that both slopes are negative, in which case $|\Phi(T_{0A}^*,\theta_{0A})|>|\Phi(T_{0B}^*,\theta_{0B})|$.
Of course, Equation~\eqref{eq:nec_cond} is not sufficient for the SME since the latter also depends on how close $T_{0A}^*$ and $T_{0B}^*$ are and how far both initial temperatures  are from unity.
From Figures~\textbf{\ref{fig:slopes_1}} and \textbf{\ref{fig:slopes_2}} we can conclude that, in general, the inequality \eqref{eq:nec_cond} is best satisfied if $\theta_{0A}\ll\theta_{0B}$.

\subsection{Overshoot Mpemba effect}
\label{sec:3.2}

The emergence of the temperature overshoot described in section \ref{sec:2.4} makes the crossover criterion   employed in the SME become meaningless. Imagine that such a crossover takes place with $T_A^*(t^*_c)=T_B^*(t_c^*)>1$, but then $T_B(t^*)$ relaxes from above while $T_A^*(t^*)$ overshoots the steady-state value, $T_A^*(t_O^*)=1$. It is then possible that $T_A^*(t^*)$ relaxes (from below) later than $T_B^*(t^*)$. In that case, the initially hotter sample (A) would reach the steady state later than the initially colder sample (B), thus contradicting the existence of a ME, despite the crossover.

Reciprocally, imagine that $T^*_A(t^*)$ and $T^*_B(t^*)$ never cross each other but $T^*_B(t^*)$ overshoots the steady-state value and then relaxes from below. It is now possible that the initially hotter sample (A) reaches the steady state earlier than the initially colder sample (B), thus qualifying as a ME, despite the absence of any crossover. We will refer to this phenomenon  as {overshoot} ME (OME) \cite{MSP22}. From the discussion in section \ref{sec:2.4} we conclude that the OME requires $\varepsilon_\rf>\varepsilon_\ct(T_{0B}^*,\theta_{0B})$.

To characterize the existence of OME without a thermal crossover,  we adopt the quantity \cite{MSP22}
\begin{equation}
\label{eq:KLD}
    \mathfrak{D}(T^*(t)) \equiv T^*(t)-1-\ln T^*(t).
\end{equation}
This quantity is (except for a factor) the Kullback--Leibler divergence of the Maxwellian VDF given by Equation~\eqref{eq:Max} (with $\Tr/\Tt\to\theta^\st$) with respect to the steady-state Maxwellian. Note that $\mathfrak{D}(T^*)$ is a positive-definite convex function of $T^*$. Therefore, we can define the OME by the crossover of $\mathfrak{D}(T_A^*)$ and $\mathfrak{D}(T_B^*)$ with, however,  $T_A^*<1$ and $T_A^*<T_B^*$.

In order to look for OME, the necessary condition for SME [given by Equation~\eqref{eq:nec_cond}] must be reversed. That is,
\begin{equation}\label{eq:nec_cond_OME}
    \Phi(T^*_{0A},\theta_{0A})>\Phi(T^*_{0B},\theta_{0B}).
\end{equation}
Establishing the most favorable conditions for OME  is not as simple as just declaring the opposite of the SME condition.
Firstly, we want for the colder sample to overshoot as much as possible the steady state, so that the relaxation from below is retarded maximally. This reasoning is translated into the condition of highly negative initial slope $\Phi(T_{0B}^*,\theta_{0B})$, which implies  small $\theta_{0B}$ (see Figures~\textbf{\ref{fig:slopes_1}} and \textbf{\ref{fig:slopes_2}}).
On the other hand, two competing phenomena exist for the hotter sample: we want to either avoid any overshoot or  force it to be as weak as possible, but we also want the relaxation to be faster than in the colder sample. Therefore, one needs $\theta_{0A}>\theta_{0B}$ but, for very large values of $\theta_{0A}$, one might not find OME due to a slower relaxation of the hotter sample.

\subsection{Initial preparation protocols}
\label{sec:3.3}
In order to study the absence or existence of ME, one needs to specify the initial conditions $(T_{0A}^*,\theta_{0A})$ and $(T_{0B}^*,\theta_{0B})$ of both samples. In previous studies \cite{SP20,MSP22,TLLVPS19} the values of $\theta_{0A}$ and $\theta_{0B}$ were freely chosen, without a specific reference to a previous protocol to initially prepare the samples.

Most of the interest of the present work resides in the proposal of protocols to generate the initial states of the samples involved in a ME experiment. The protocols are based on the proposed ST, and the initial states will be generated by assuming \emph{prior} thermostat values $(\Tn_i,\varepsilon_i)$, $i=A,B$, and allowing  both samples to reach their respective steady states $(T_i^\st,\theta_i^\st)$ before switching to the common \emph{posterior} thermostat ($\Tn_\rf,\varepsilon_\rf)$ at $t=0$. The values of the prior thermostats will be chosen to optimize the necessary conditions \eqref{eq:nec_cond} and \eqref{eq:nec_cond_OME} for  SME and OME, respectively.

According to Equations~\eqref{eq:steady_values} and \eqref{eq:gamma},  the ratios between the prior and posterior noise temperatures  for desired  values of $T_{0i}^*=T_i^\st/T_\rf^\st$, $i=A,B$, where $T^\st_i=\Tn_i\widetilde{T}^\st(\varepsilon_i)$ and  $T^\st_\rf=\Tn_\rf\widetilde{T}^\st(\varepsilon_\rf)$, are
\begin{equation}
\label{eq:prior/posterior}
\frac{\Tn_i}{\Tn_\rf}=T_{0i}^*\frac{2+\theta^\st_\rf}{2+\theta^\st_i}\left(\frac{\gamma_i^\st}{\gamma_\rf^\st}\right)^{2/3},\quad i=A,B.
\end{equation}

\subsubsection{Protocol for the standard Mpemba effect}
\label{sec:3.3.1}

\begin{figure}
    \centering  
    \begin{minipage}[t]{.4\linewidth}
    	\textbf{A}\\
    	\includegraphics[width = \linewidth]{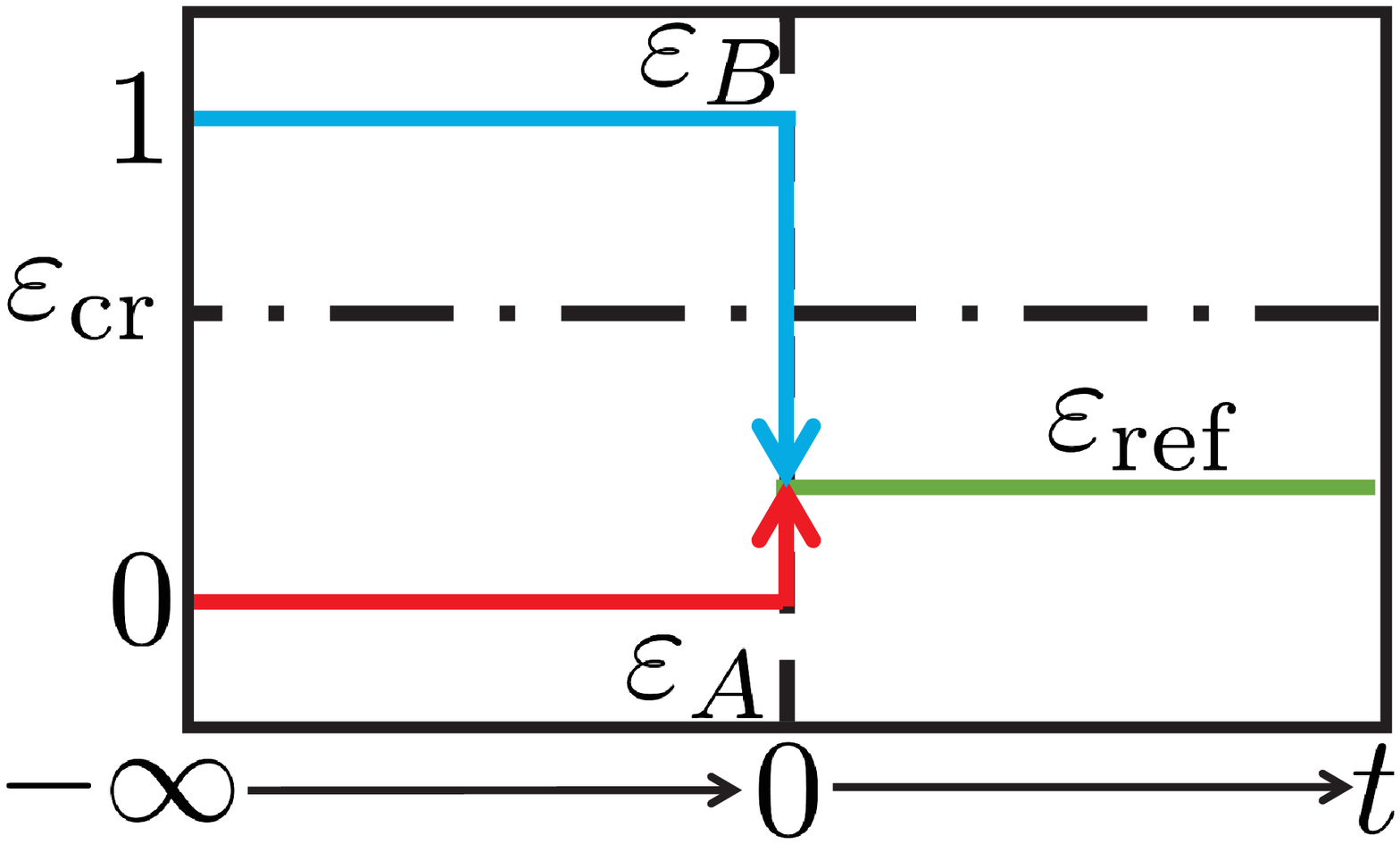}
    \end{minipage}\hspace{2cm}  
    \begin{minipage}[t]{.4\linewidth}
    	\textbf{B}\\
    	\includegraphics[width = \linewidth]{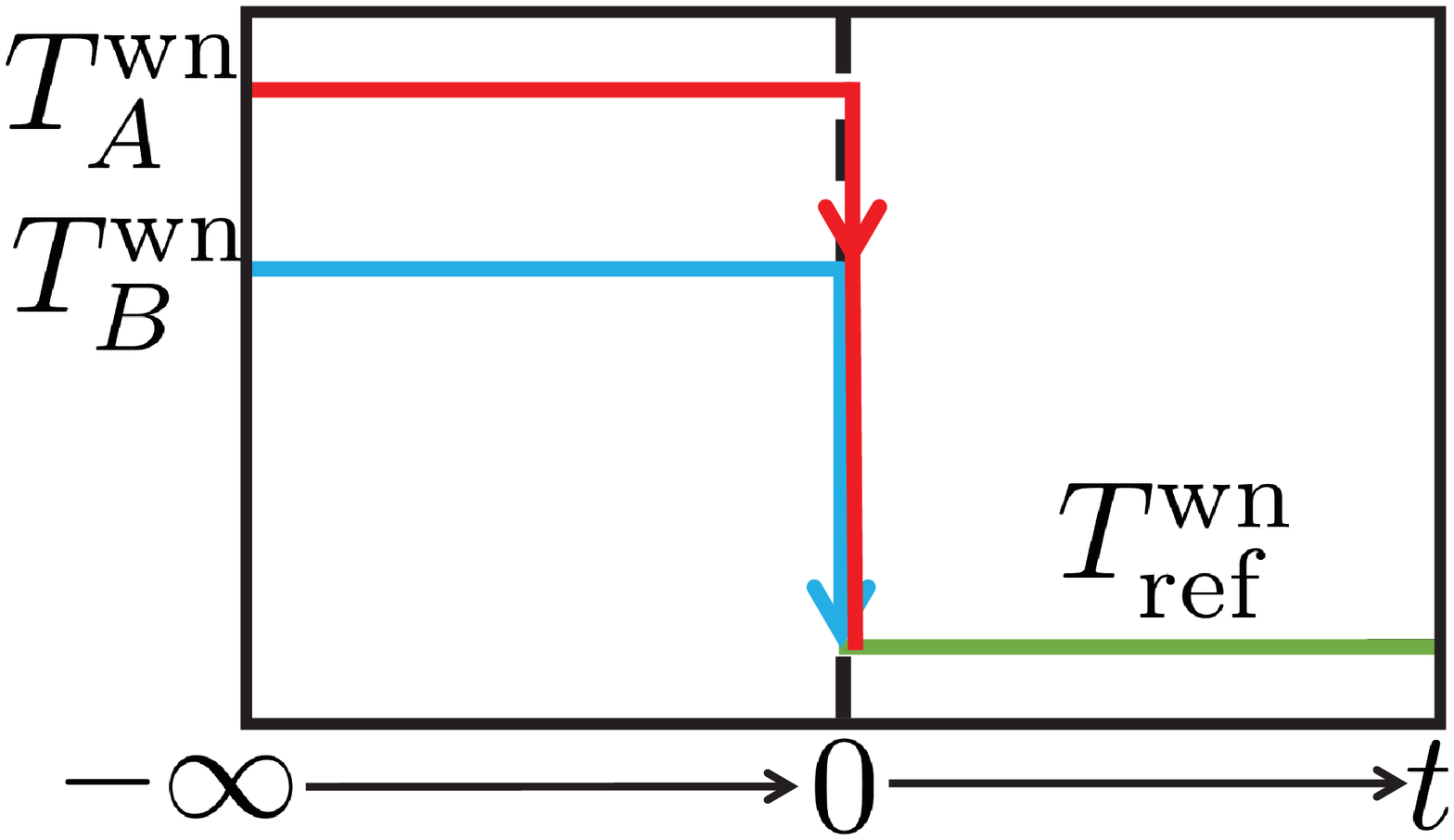}
    \end{minipage}
    \caption{Scheme of the protocol for SME. \textbf{(A)} Choice of the splitting parameters $\varepsilon$. \textbf{(B)} Choice of the noise temperatures $\Tn$. }
    \label{fig:scheme_1}
\end{figure}

In this case, we want to have $\theta_{0A}\ll\theta_{0B}$. According to Equation~\eqref{eq:steady_values}, and as observed in Figure~\textbf{\ref{fig:steady_state}}, $\theta^\st$ is an increasing function of $\varepsilon$ and independent of $\Tn$. Therefore, the most disparate values of $\theta_{0A}$ and $\theta_{0B}$  are obtained if the prior thermostats of samples A and B have $\varepsilon_A=0$ and $\varepsilon_B=1$, respectively. According to Figure~\textbf{\ref{fig:theta_st_diff}}, SME would be stronger and/or easier to find for lower values of $\beta$ at fixed $\alpha$ and for lower values of $\alpha$  at fixed $\beta$. In addition, in order to define a cooling process, we need to choose proper values of $\Tn_{A}/\Tn_{\rf}$ and $\Tn_{B}/\Tn_{\rf}$ [see Equation~\eqref{eq:prior/posterior}], such that
$T_{0A}^*=T_{A}^\st/T^\st_\rf>T_{0B}^*= T_{B}^\st/T^\st_\rf>1$. Finally, one must fix $\varepsilon_\rf<\min\{\varepsilon_\ct(T_{0A}^*,\theta_{0A}),\varepsilon_\ct(T_{0B}^*,\theta_{0B})\}$ to prevent any possible overshoot. This minimum of the critical rotational-to-total noise intensity parameter is expected to be $\varepsilon_\ct(T_{0A}^*,\theta_{0A})$ because it corresponds to a more negative initial slope.

Thus, the designed protocol for SME reads as follows (see Figure~\textbf{\ref{fig:scheme_1}} for an illustrative scheme):
\begin{enumerate}
    \item Start by fixing $\varepsilon_A = 0$ and $\varepsilon_B=1$, in order to ensure $\theta_{0A} < \theta_{0B}$.
    \item Choose $\Tn_A/\Tn_{\rf}$ and $\Tn_B/\Tn_{\rf}$,  such that $T_{0A}>T_{0B}>T_\rf^\st$.
    \item Let both samples evolve and reach the steady states corresponding to their respective prior thermostats. These steady states will play the role of  the initial conditions for our ME experiment.
    \item Switch the values of the thermostats of both samples to a common reference pair of values $(\Tn_\rf,\varepsilon_\rf)$, such that no overshoot is expected, that is, $\varepsilon_\rf<\min\{\varepsilon_\ct(T_{0A}^*,\theta_{0A}),\varepsilon_\ct(T_{0B}^*,\theta_{0B})\}$.  This thermostat switch fixes the origin of time, $t=0$.
    \item Finally, let both samples evolve and reach a common steady state.
\end{enumerate}


For given $(\alpha,\beta)$, the numerical solutions of Equations~\ref{eq:Tstar-theta-ev} for different values of $T_{0A}^*$ and $T_{0B}^*$---and with $\theta_{0A}$ and $\theta_{0B}$ given by Equations~\eqref{eq:theta_st_0} and \eqref{eq:theta_st_1}, respectively--- can be analyzed to determine whether SME is present or not. This provides the phase diagram  presented in Figure~\textbf{\ref{fig:PD_MP}A} for  $\varepsilon_\rf=0.1$ and some pairs of coefficients of restitution.

\subsubsection{Protocol for the overshoot Mpemba effect}
\label{sec:3.3.2}

\begin{figure}
    \centering
    \begin{minipage}[t]{.4\linewidth}
    	\textbf{A}\\
    	\includegraphics[width = \linewidth]{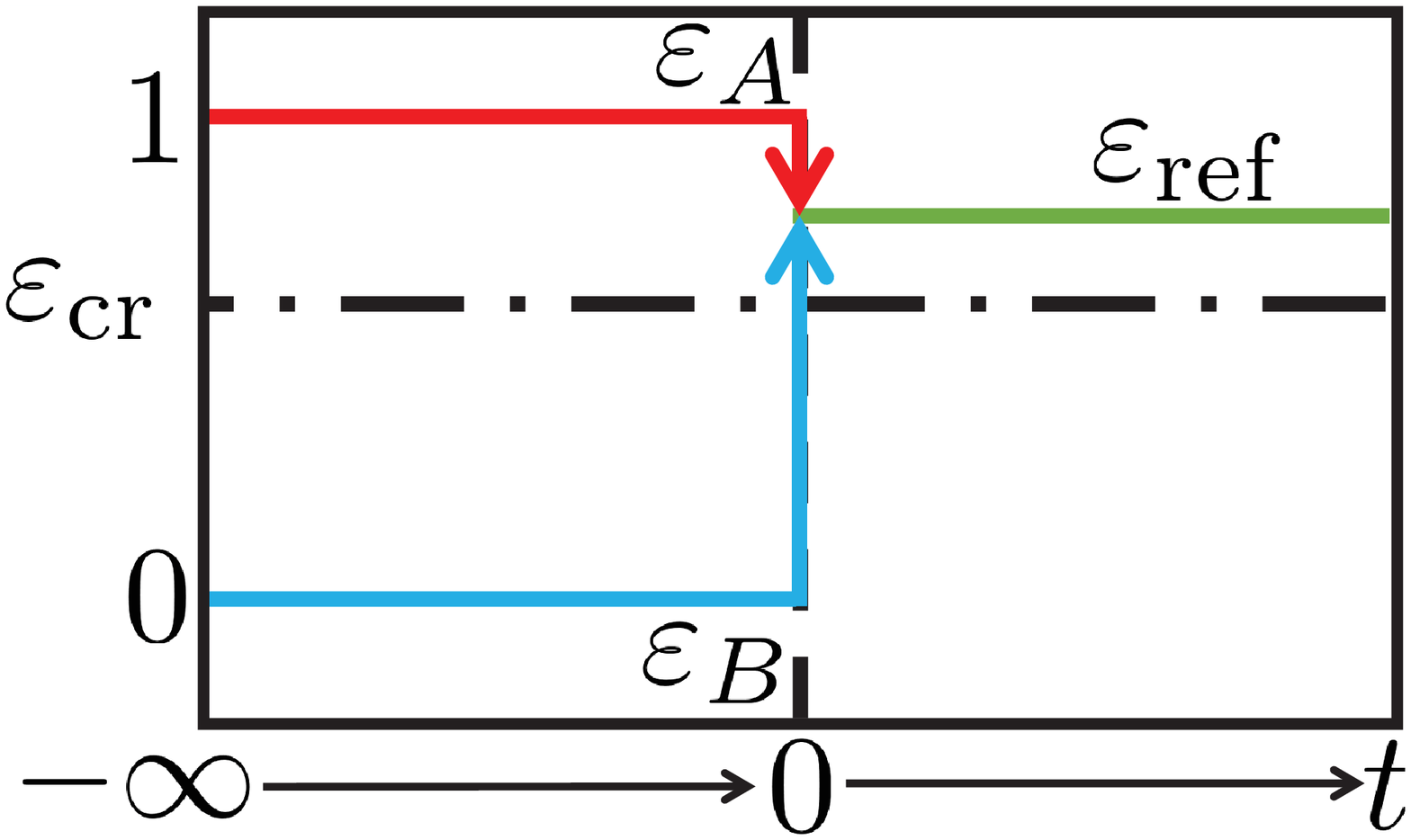}
    	\end{minipage}\hspace{2cm}  
    \begin{minipage}[t]{.4\linewidth}
    	\textbf{B}\\
    	\includegraphics[width = \linewidth]{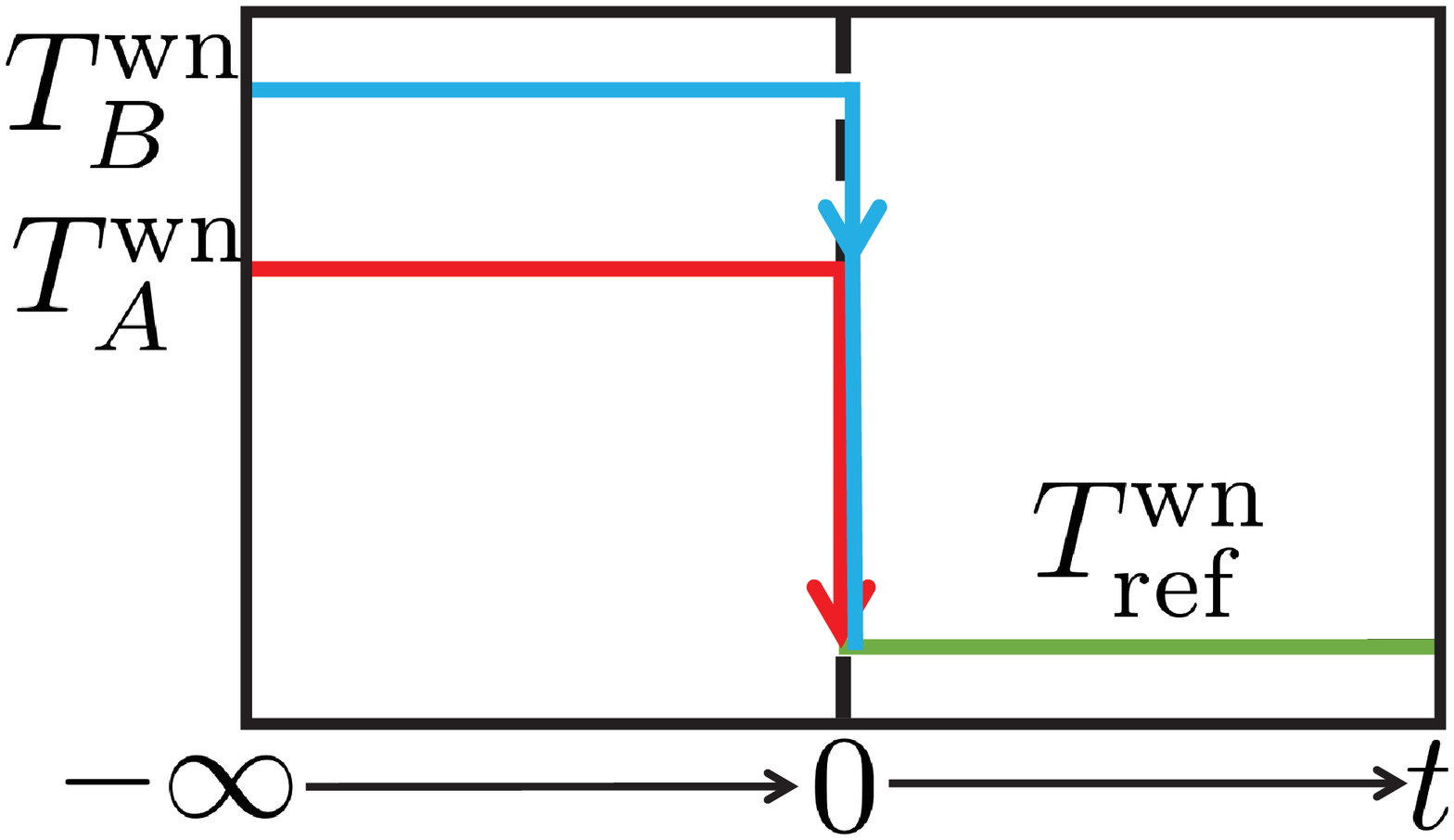}
    	\end{minipage}
    \caption{Same as in Figure~\ref{fig:scheme_1}, but for OME.}
    \label{fig:scheme_2}
\end{figure}

In the OME case, it is convenient to have $\theta_{0A}\gg\theta_{0B}$, so that  the adopted choices of $\varepsilon$ for the prior thermostats are the reverse of those of  SME, i.e.,  $\varepsilon_A=1$ and $\varepsilon_B=0$. Whereas in section~\ref{sec:3.2} we commented that the best situation for the OME is not always the opposite to that of the SME, the above choice helps us avoid or weaken a possible overshoot for the hotter sample. Again, in order to define a cooling process, we need to choose proper values of $\Tn_{A}/\Tn_{\rf}$ and $\Tn_{B}/\Tn_{\rf}$ [see Equation~\eqref{eq:prior/posterior}], and  such that
$T_{0A}>T_{0B}>T^\st_\rf$. Finally, $\varepsilon_\rf>\varepsilon_\ct(T_{0B}^*,\theta_{0B})$ to ensure overshoot of $T_B^*(t^*)$.

In analogy with the SME case, a protocol for observing OME is designed as follows (see Figure~\textbf{\ref{fig:scheme_2}}):

\begin{enumerate}
    \item Start by fixing $\varepsilon_A = 1$ and $\varepsilon_B=0$, in order to ensure $\theta_{0A} > \theta_{0B}$.
    \item Choose $\Tn_A/\Tn_{\rf}$ and $\Tn_B/\Tn_{\rf}$,  such that $T_{0A}>T_{0B}>T_\rf^\st$.
    \item Let both samples evolve and reach the steady states corresponding to their respective prior thermostats. These steady states will play the role of  the initial conditions for our ME experiment.
    \item Switch the values of the thermostats of both samples to a common reference pair of values $(\Tn_\rf,\varepsilon_\rf)$, such that  overshoot is ensured, that is, $\varepsilon_\rf>\varepsilon_\ct(T_{0B}^*,\theta_{0B})$.  This thermostat switch fixes the origin of time, $t=0$.
    \item Finally, let both samples evolve and reach a common steady state.
\end{enumerate}

\begin{figure}
    \centering  
    \begin{minipage}[t]{.45\linewidth}
    	\textbf{A}\\
    	\includegraphics[width = \linewidth]{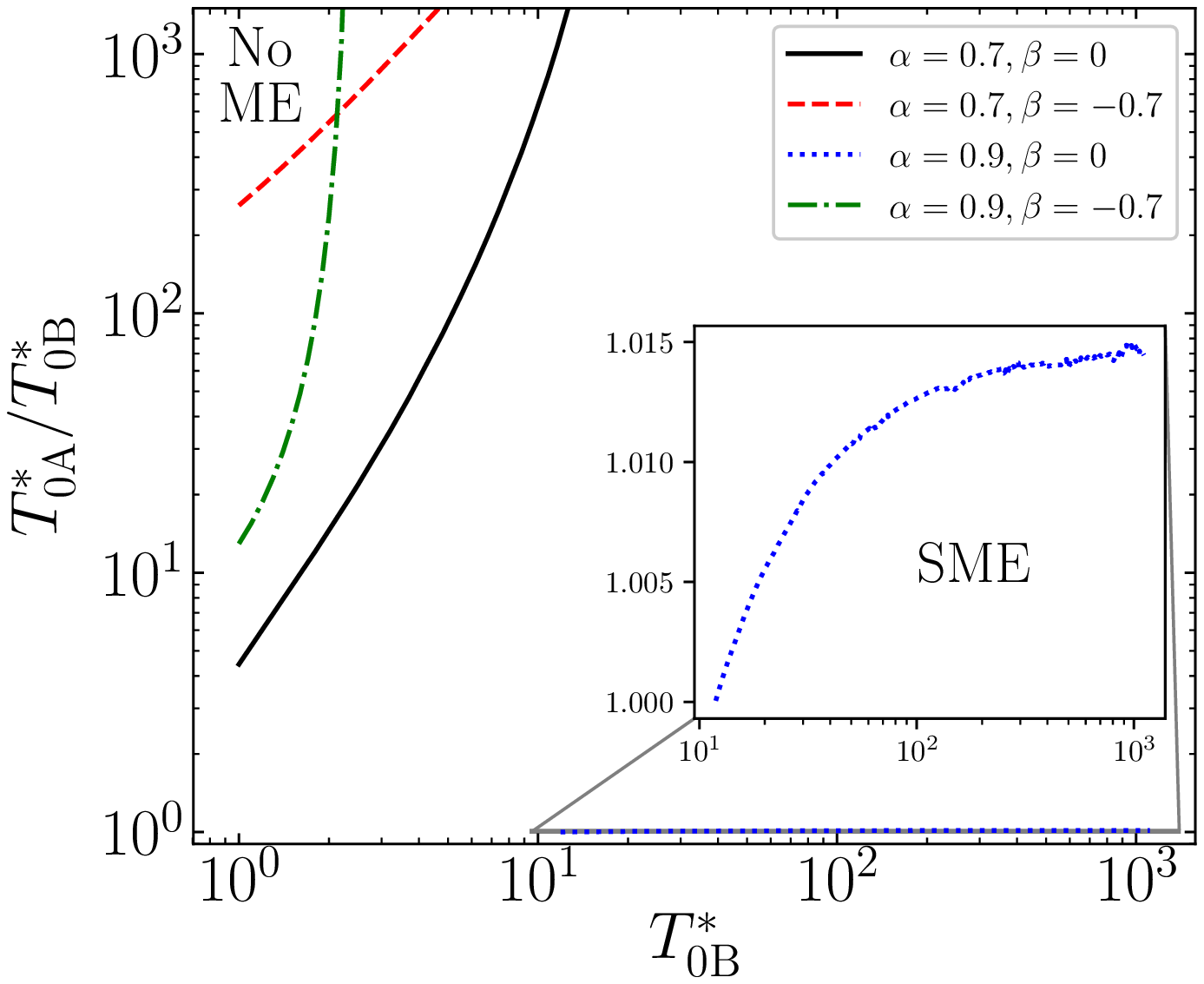}
    \end{minipage}\hspace{1cm}  
    \begin{minipage}[t]{.45\linewidth}
    	\textbf{B}\\
    	\includegraphics[width = \linewidth]{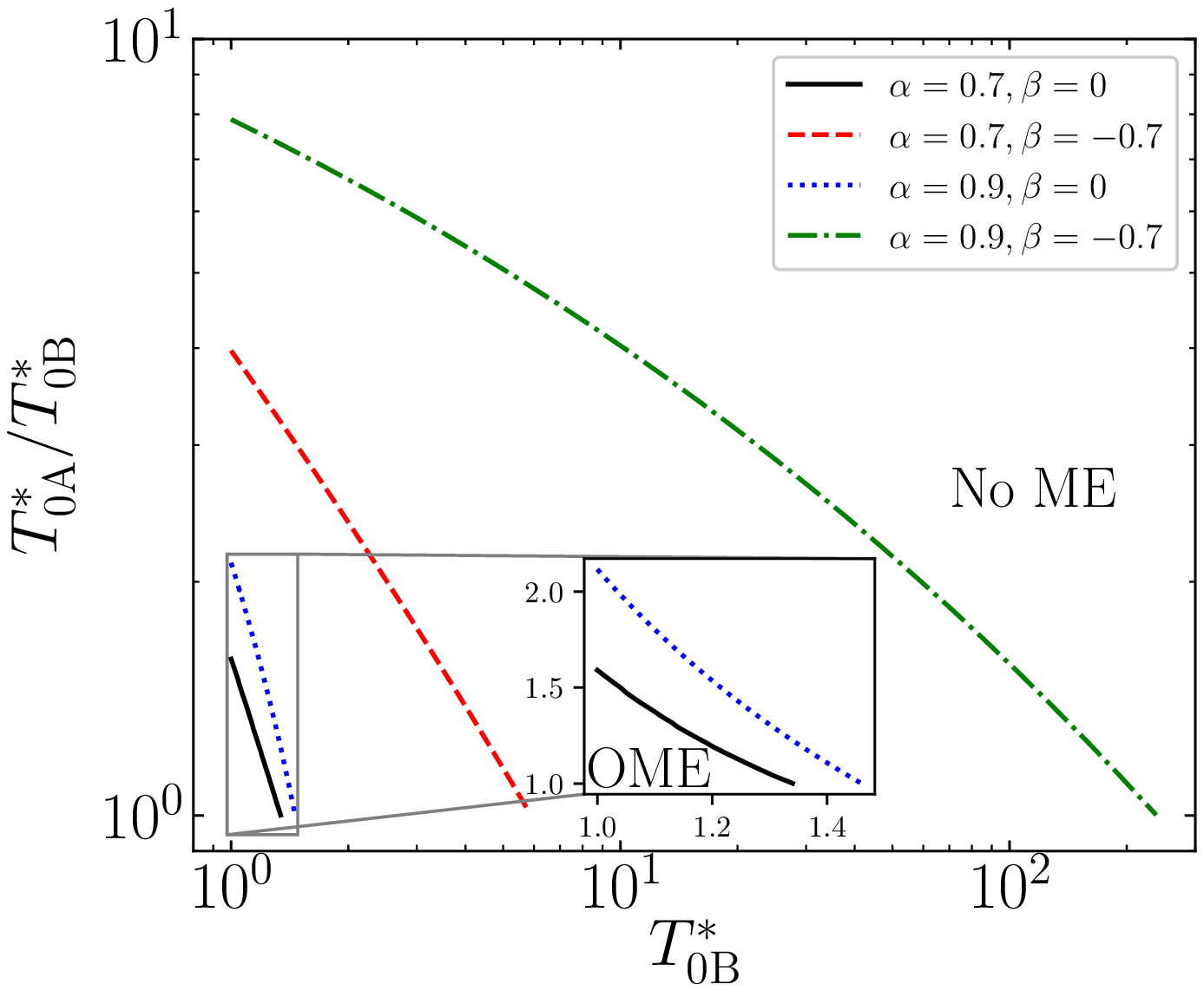}
    	\end{minipage}
    \caption{Phase diagrams in the plane $T^*_{0A}/T^*_{0B}$ vs $T^*_{0B}$ for the emergence of \textbf{(A)} SME (with $\varepsilon_\rf=0.1$) and \textbf{(B)} OME (with $\varepsilon_\rf=0.9$). Four different pairs of the coefficients of restitution are considered: ($\alpha,\beta)=(0.7,-0.7)$, $(0.7,0)$, $(0.9,-0.7)$, and $(0.9,0)$. The insets show magnified views of the indicated regions.}
    \label{fig:PD_MP}
\end{figure}

Figure~\textbf{\ref{fig:PD_MP}B} shows a phase diagram for the occurrence of OME (with $\varepsilon_\rf=0.9$) for the same pairs of coefficients of restitution as before.


\subsection{Comparison with simulation results}
\begin{figure}
        \centering 
    \begin{minipage}[t]{.355\linewidth}
    	\textbf{A}\\
    	\includegraphics[width = \linewidth]{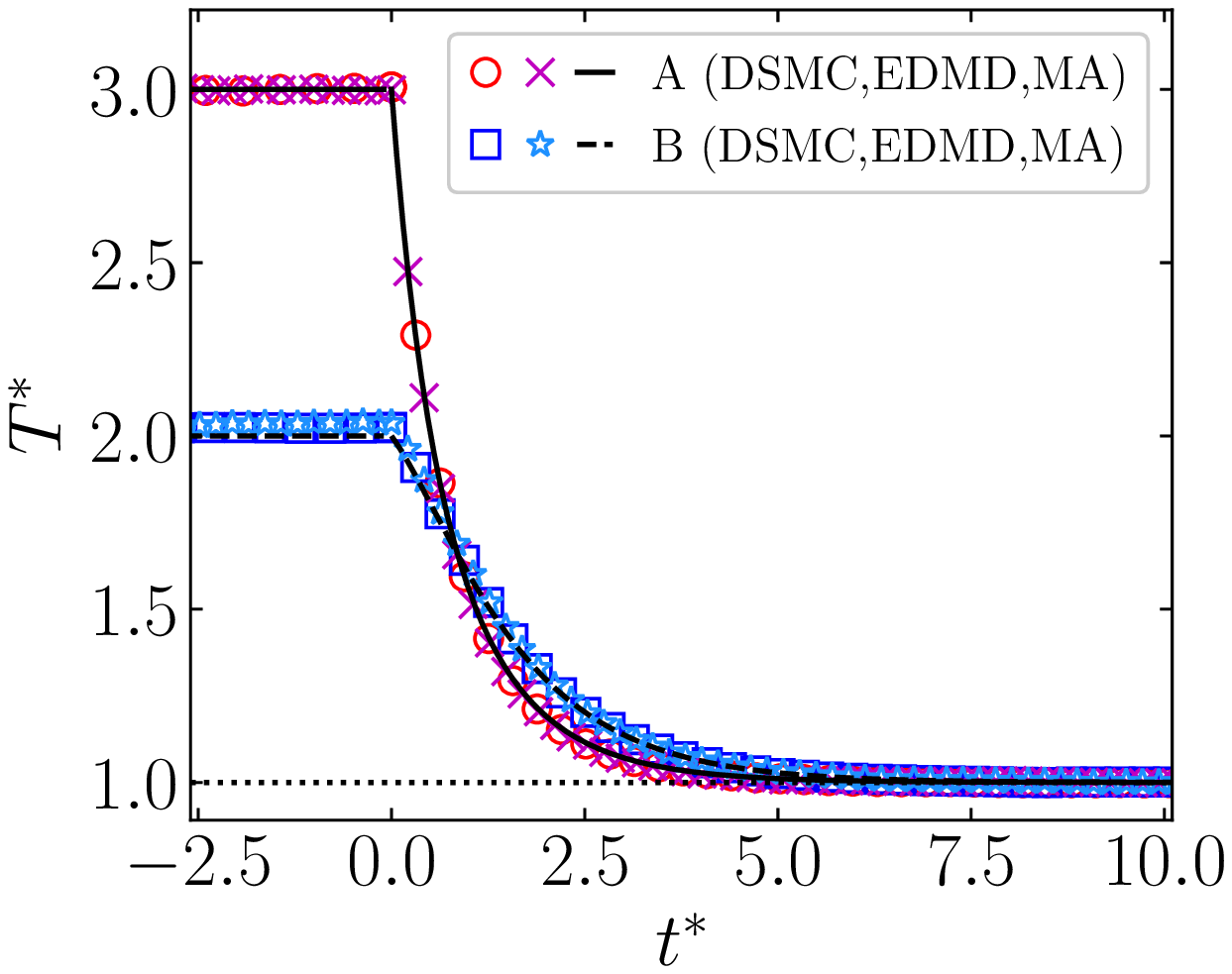}  
    \end{minipage}\hspace{2cm}  
    \begin{minipage}[t]{.355\linewidth}
    	\textbf{B}\\
    	\includegraphics[width = \linewidth]{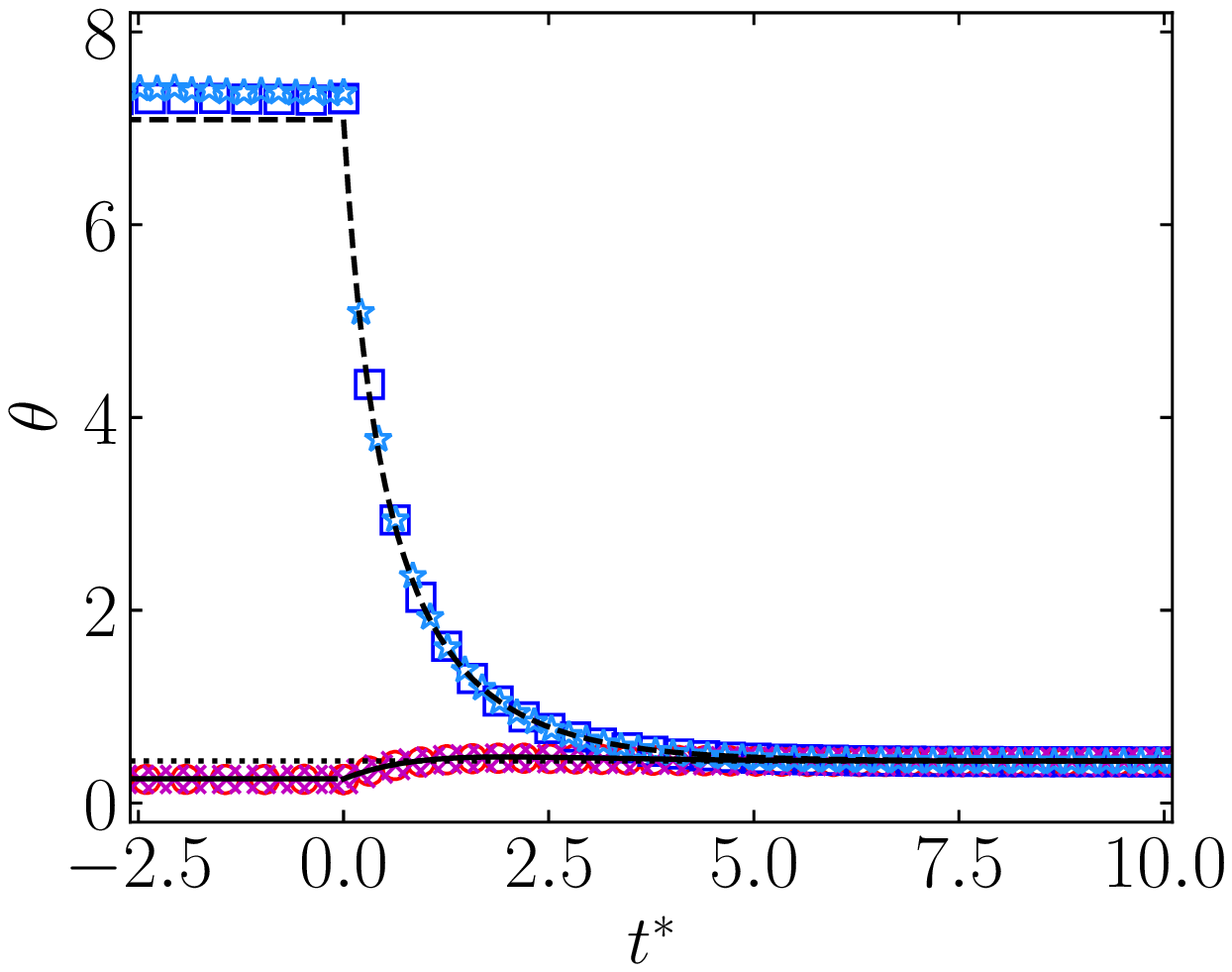}
    	\end{minipage}\\  
    \begin{minipage}[t]{.355\linewidth}
    	\textbf{C}\\
    	\includegraphics[width = \linewidth]{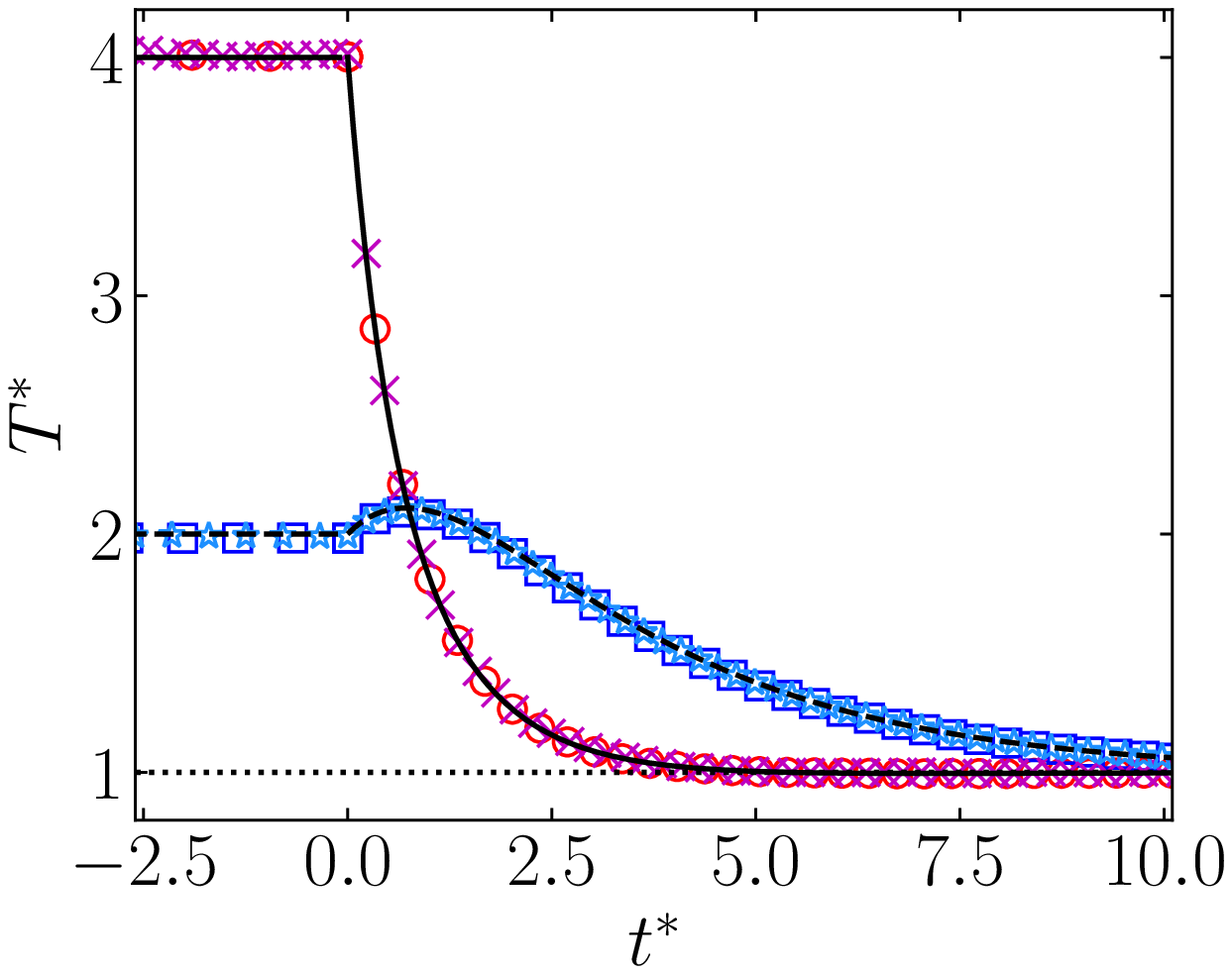}
    \end{minipage}\hspace{2cm}  
    \begin{minipage}[t]{.355\linewidth}
    	\textbf{D}\\
    	\includegraphics[width = \linewidth]{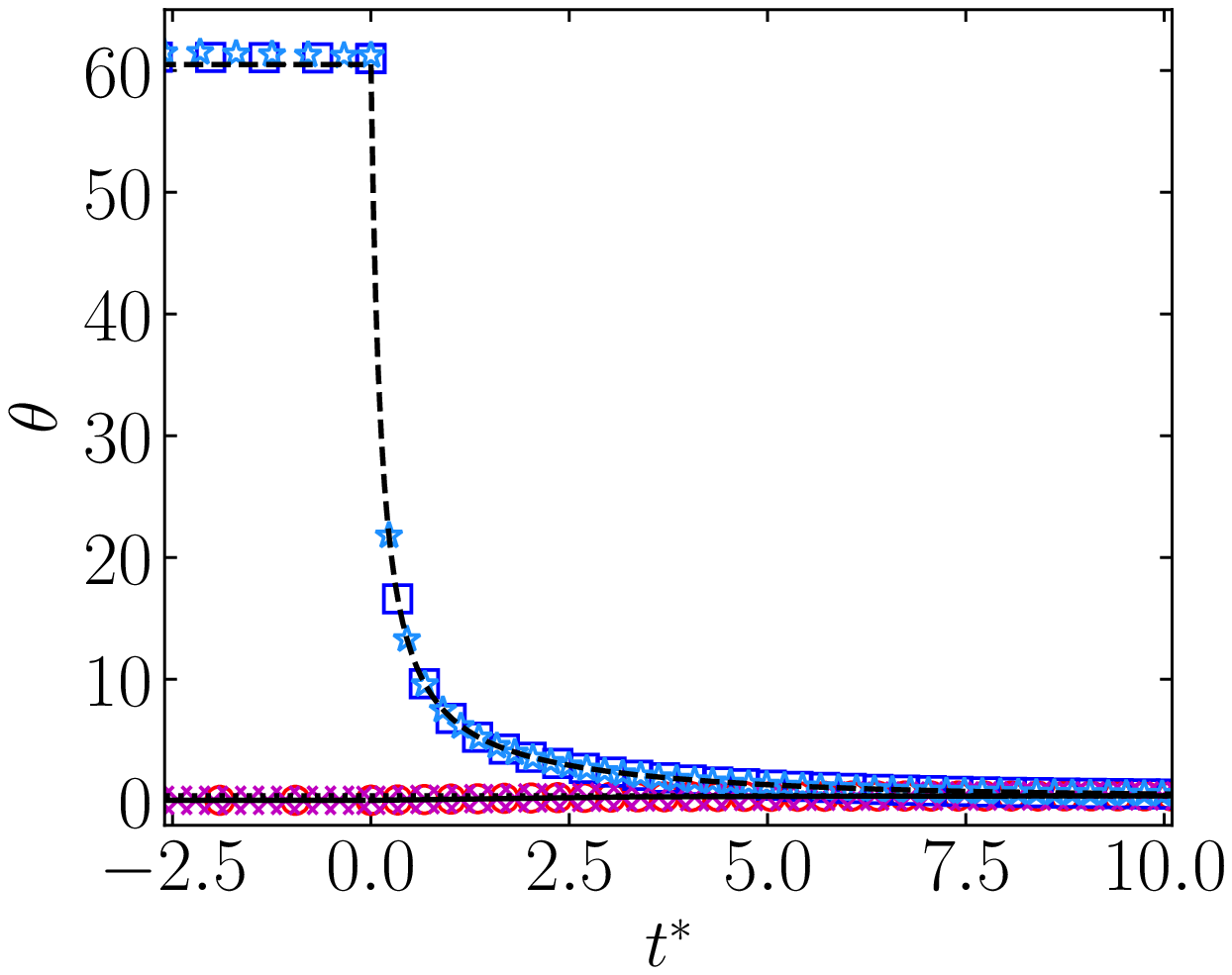}
    \end{minipage}\\ 
    \begin{minipage}[t]{.355\linewidth}
    	\textbf{E}\\
    	\includegraphics[width = \linewidth]{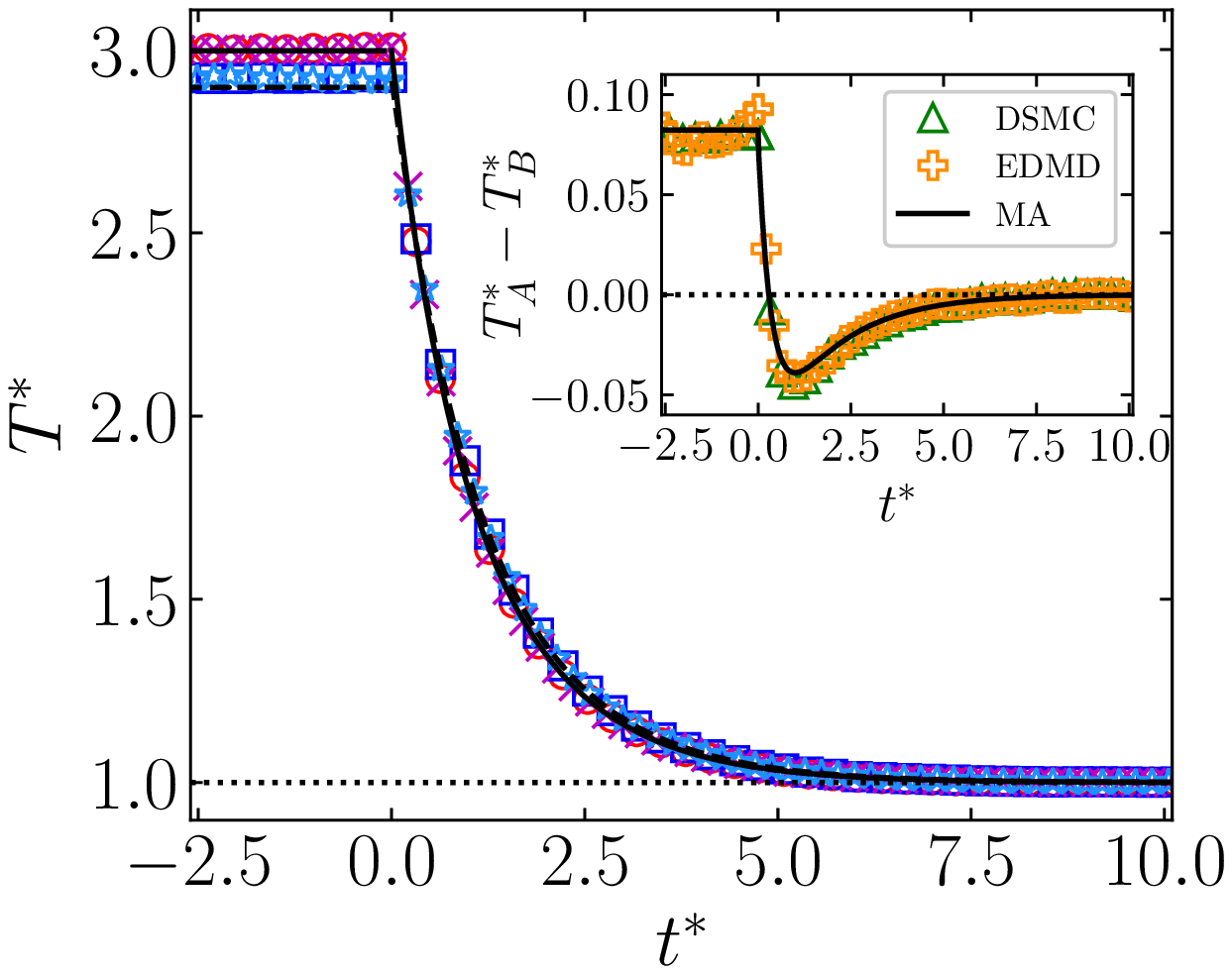}
    \end{minipage}\hspace{2cm}  
    \begin{minipage}[t]{.355\linewidth}
    	\textbf{F}\\
    	\includegraphics[width = \linewidth]{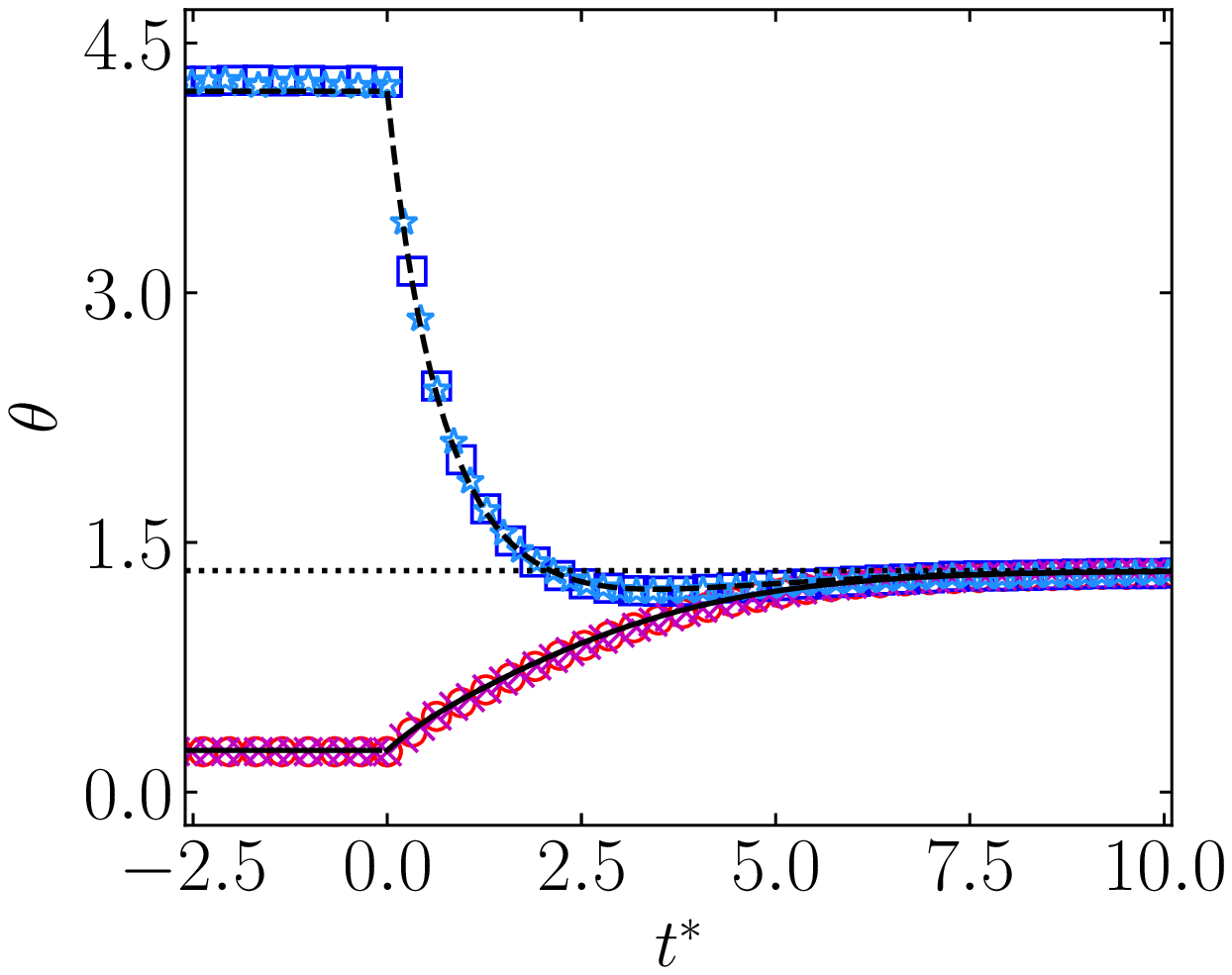}
    \end{minipage}\\ 
    \begin{minipage}[t]{.355\linewidth}
    	\textbf{G}\\
    	\includegraphics[width = \linewidth]{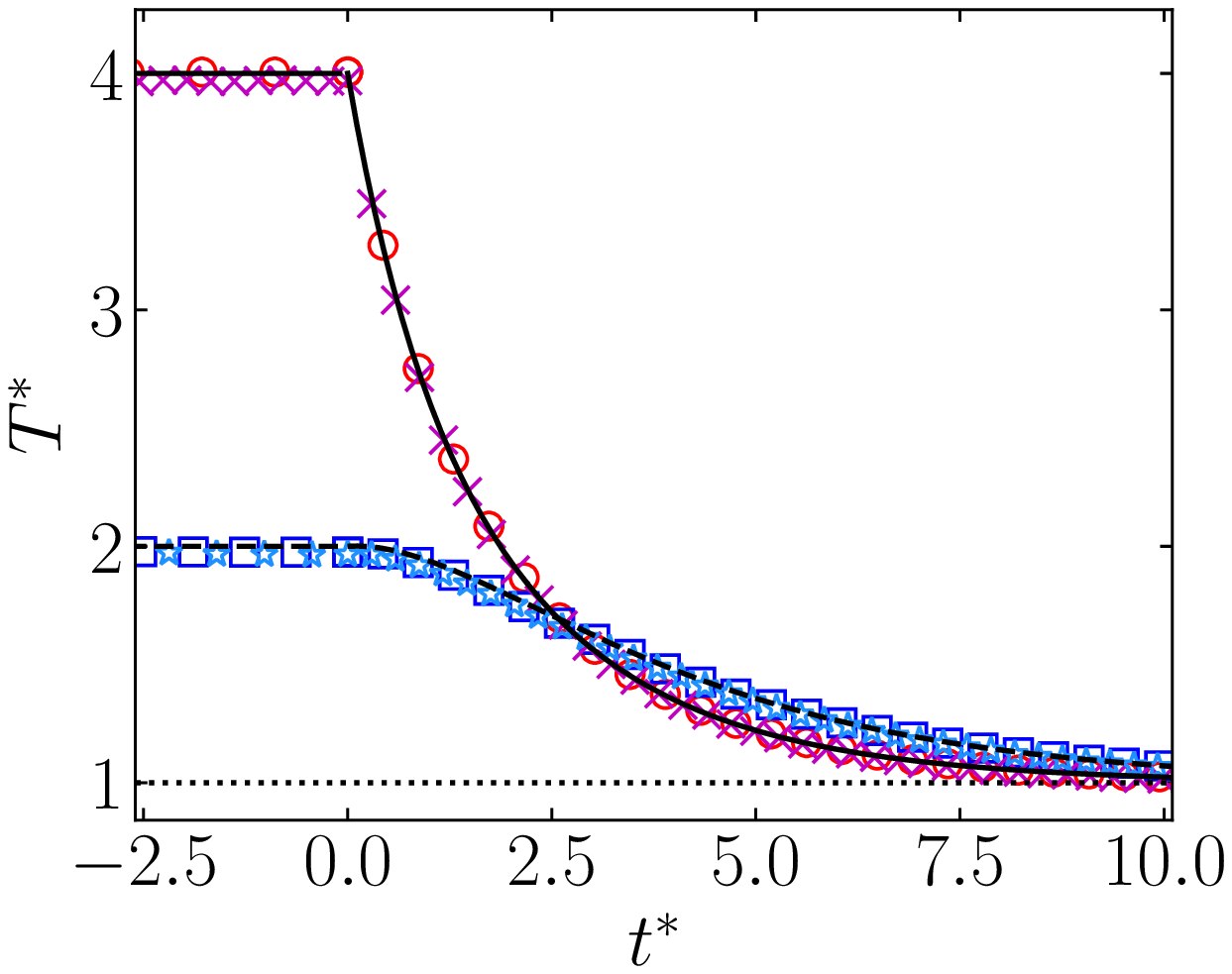}
    \end{minipage}\hspace{2cm}  
    \begin{minipage}[t]{.355\linewidth}
    	\textbf{H}\\
    	\includegraphics[width = \linewidth]{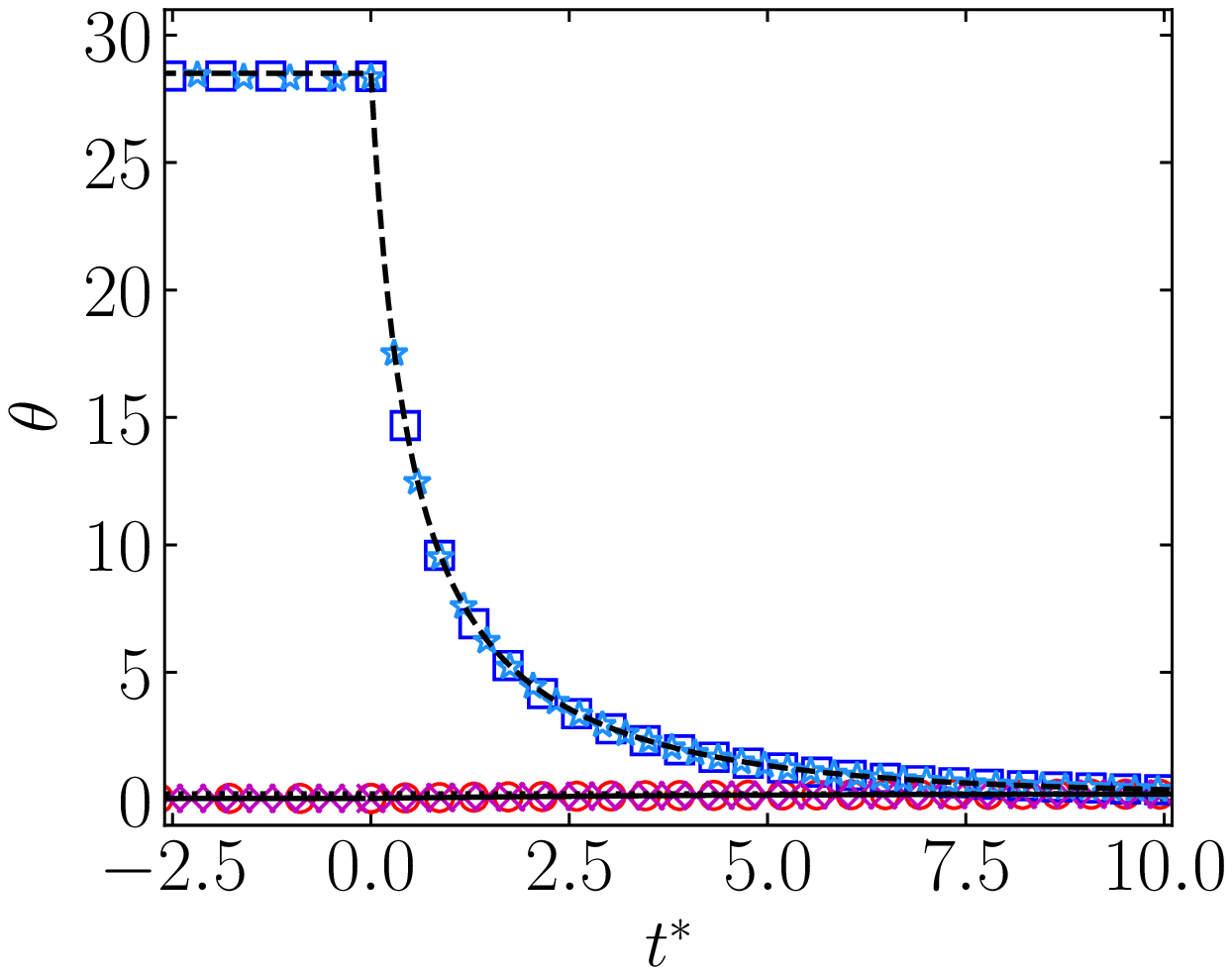}
    	\end{minipage}
        \caption{Time evolution of $T^*$ and $\theta$ for SME initialization protocol, as described in section~\ref{sec:3.3.1}. \textbf{(A)} and \textbf{(B)}: $(\alpha,\beta)=(0.7,0)$ and $\varepsilon_\rf=0.1$; \textbf{(C)} and \textbf{(D)}: $(\alpha,\beta)=(0.7,-0.7)$ and $\varepsilon_\rf=0.1$; \textbf{(E)} and \textbf{(F)}; $(\alpha,\beta)=(0.9,0)$ and $\varepsilon_\rf=0.6$; and \textbf{(G)} and \textbf{(H)}: $(\alpha,\beta)=(0.9,-0.7)$ and $\varepsilon_\rf=0.1$. Thick and dashed lines correspond to the theoretical prediction from  Equations~\eqref{eq:Tstar-theta-ev}, dotted lines represented the steady state value, and symbols refer to DSMC and EDMD simulation results. The inset in panel (\textbf{E}) shows the evolution of the temperature difference $T_A^*-T_B^*$.}
    \label{fig:SME}
\end{figure}

\begin{figure}[h!]
        \centering 
    \begin{minipage}[t]{.3\linewidth}
    	\textbf{A}\\
    	\includegraphics[width = \linewidth]{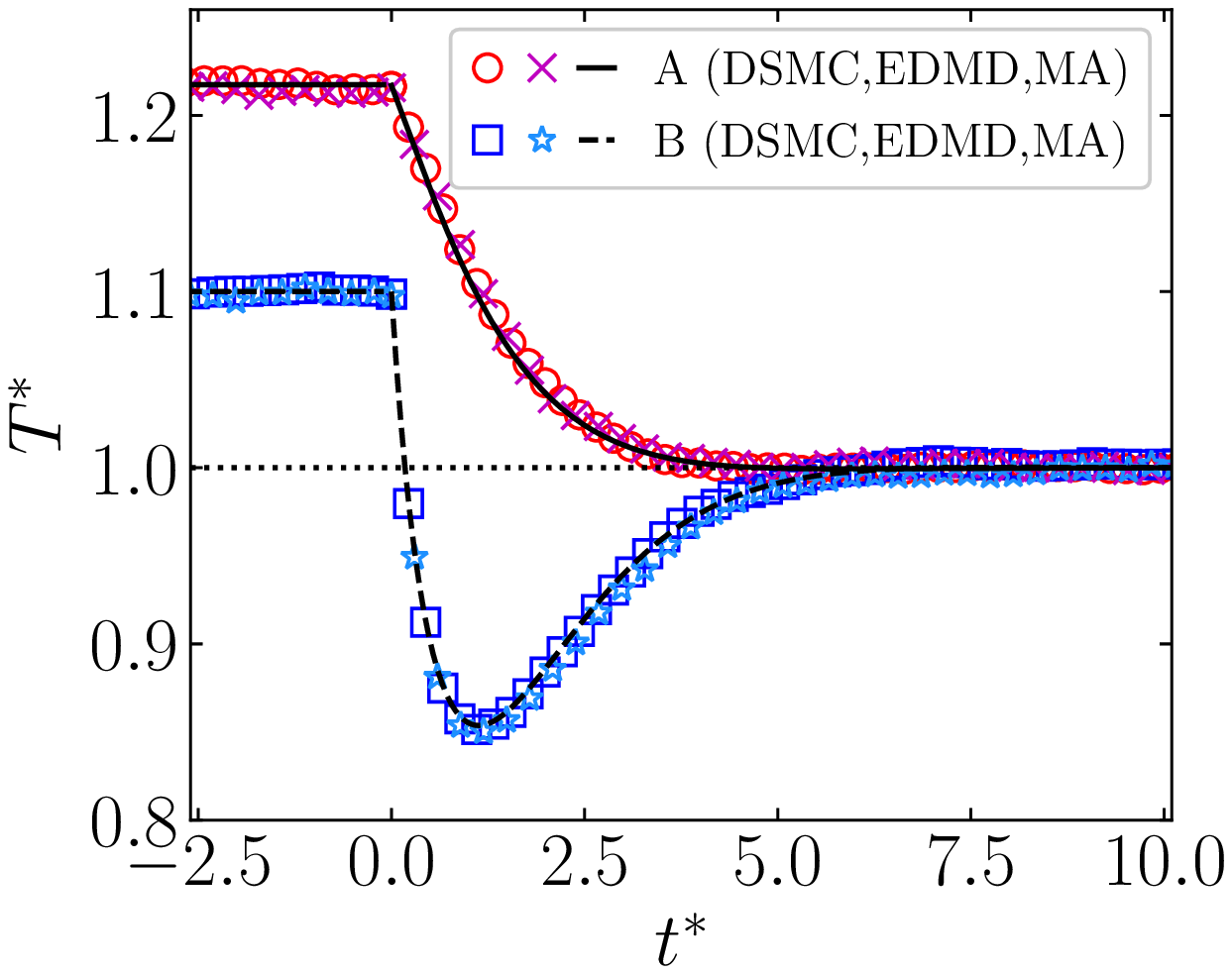}
    \end{minipage}  
    \begin{minipage}[t]{.3\linewidth}
    	\textbf{B}\\
    	\includegraphics[width = \linewidth]{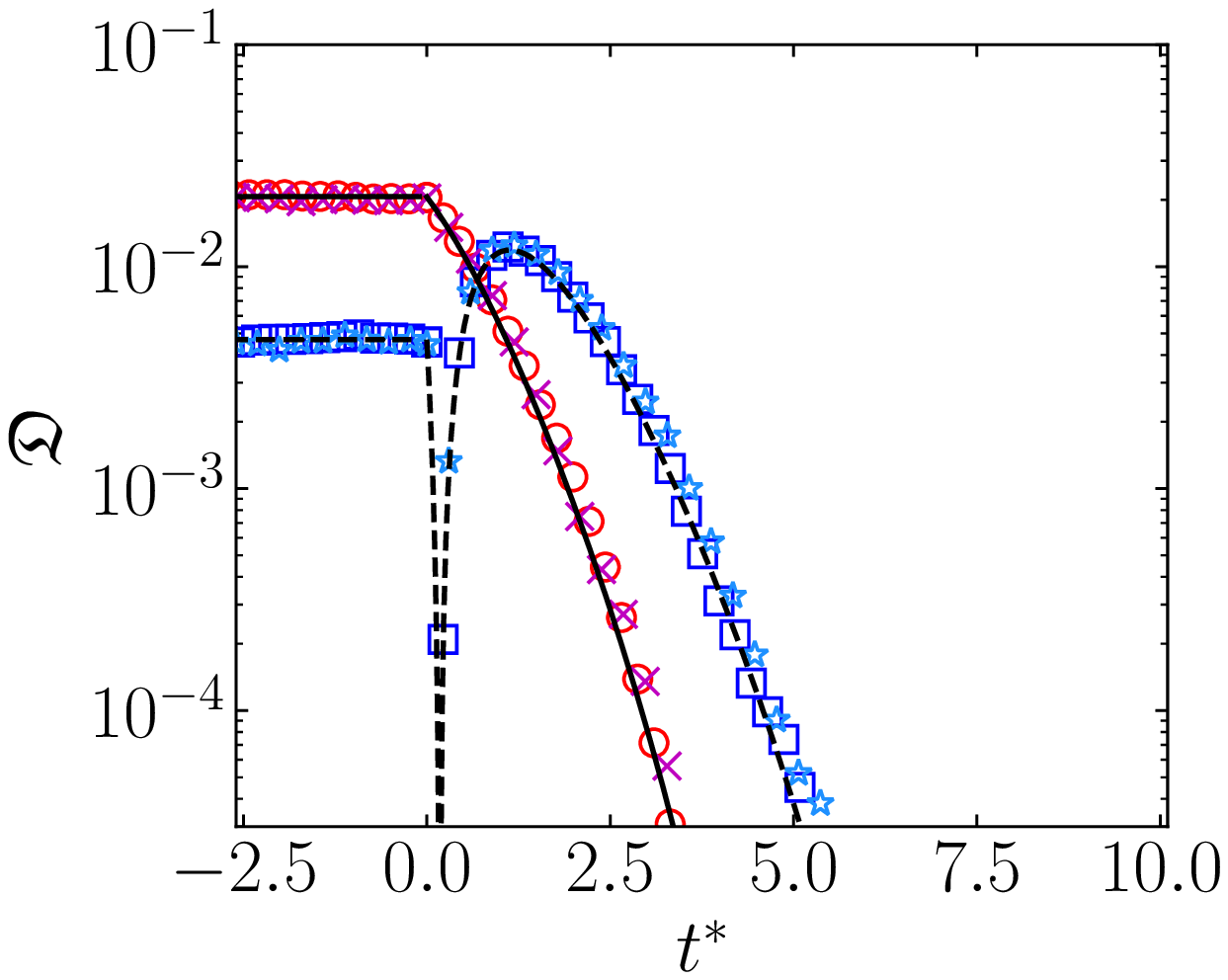} 
    	\end{minipage}  
    	\begin{minipage}[t]{.3\linewidth}
    	\textbf{C}\\
    	\includegraphics[width = \linewidth]{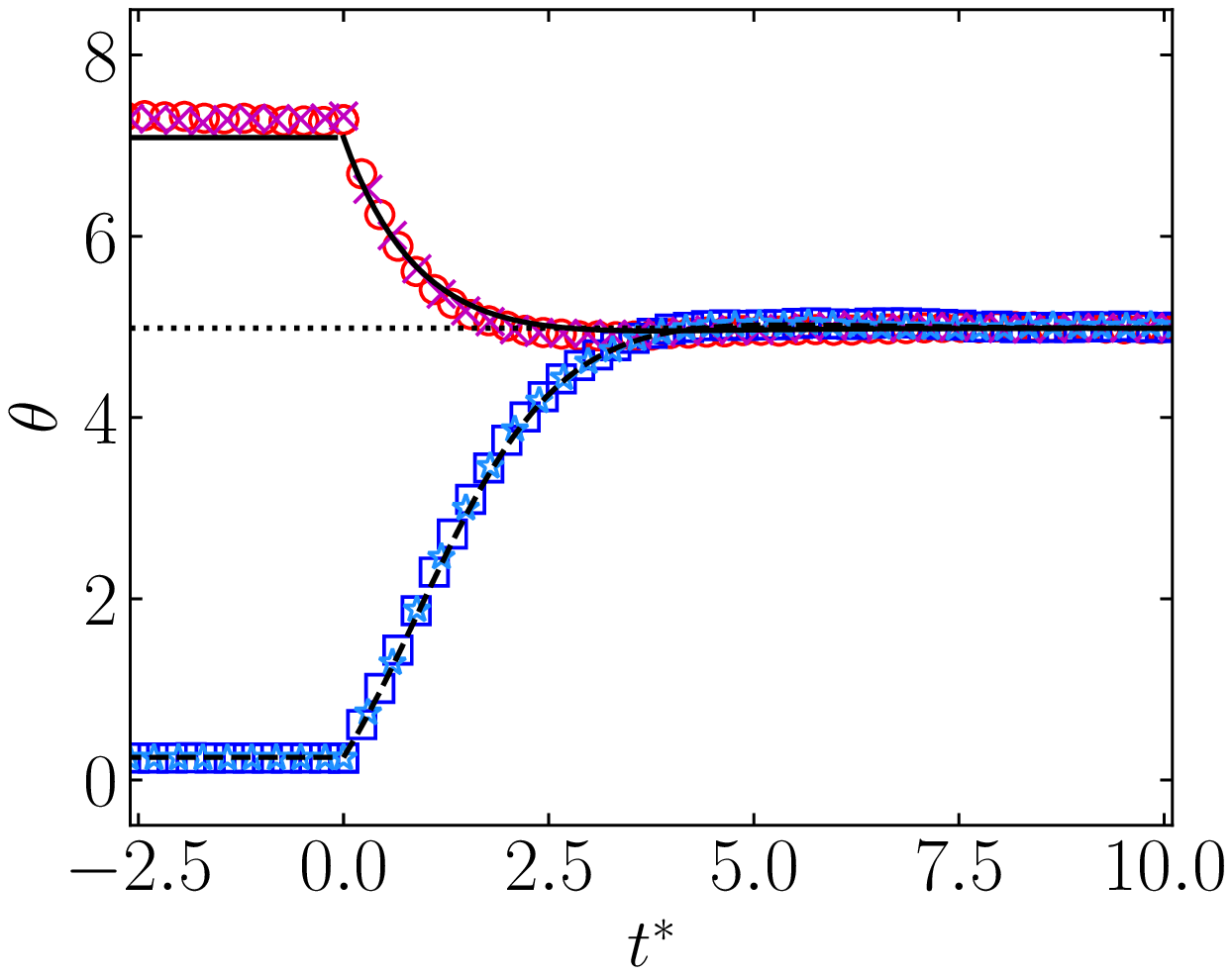}
        \end{minipage}\\ 
    	\begin{minipage}[t]{.3\linewidth}
    	\textbf{D}\\
    	\includegraphics[width = \linewidth]{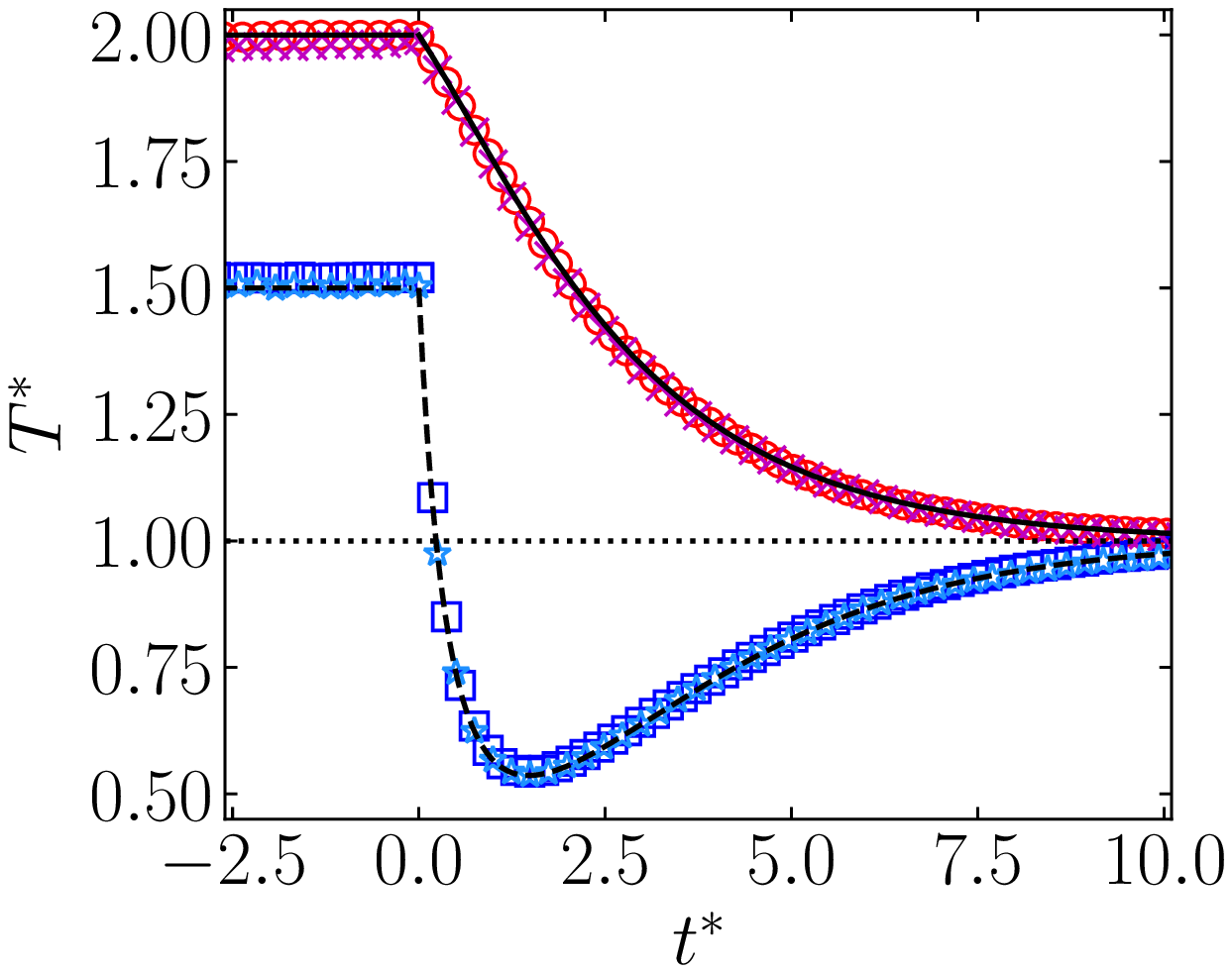}
        \end{minipage}
    	\begin{minipage}[t]{.3\linewidth}
    	\textbf{E}\\
    	\includegraphics[width = \linewidth]{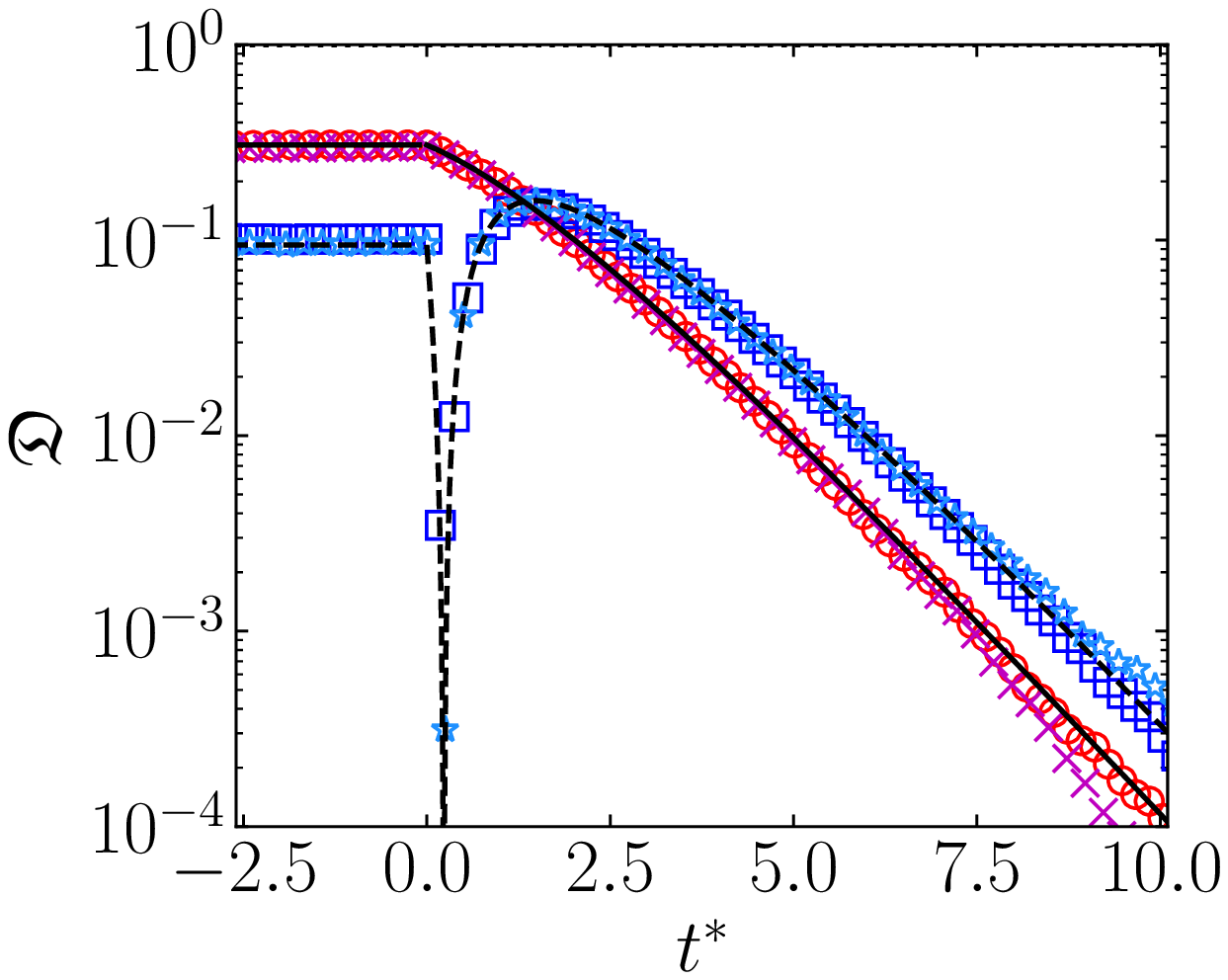}
        \end{minipage}
    	\begin{minipage}[t]{.3\linewidth}
    	\textbf{F}\\
    	\includegraphics[width = \linewidth]{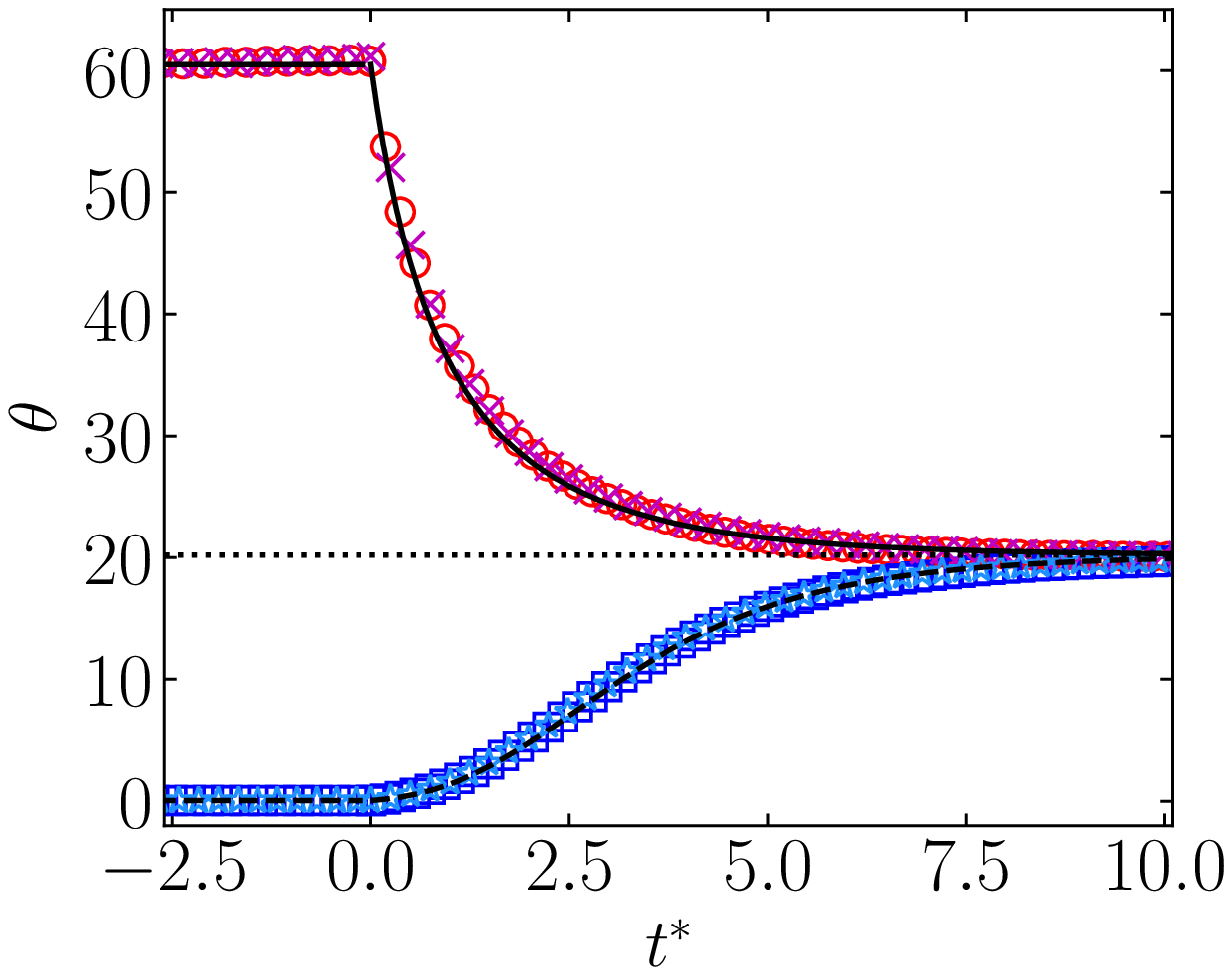}\\
        \end{minipage}\\
    	\begin{minipage}[t]{.3\linewidth}
    	\textbf{G}\\
    	\includegraphics[width = \linewidth]{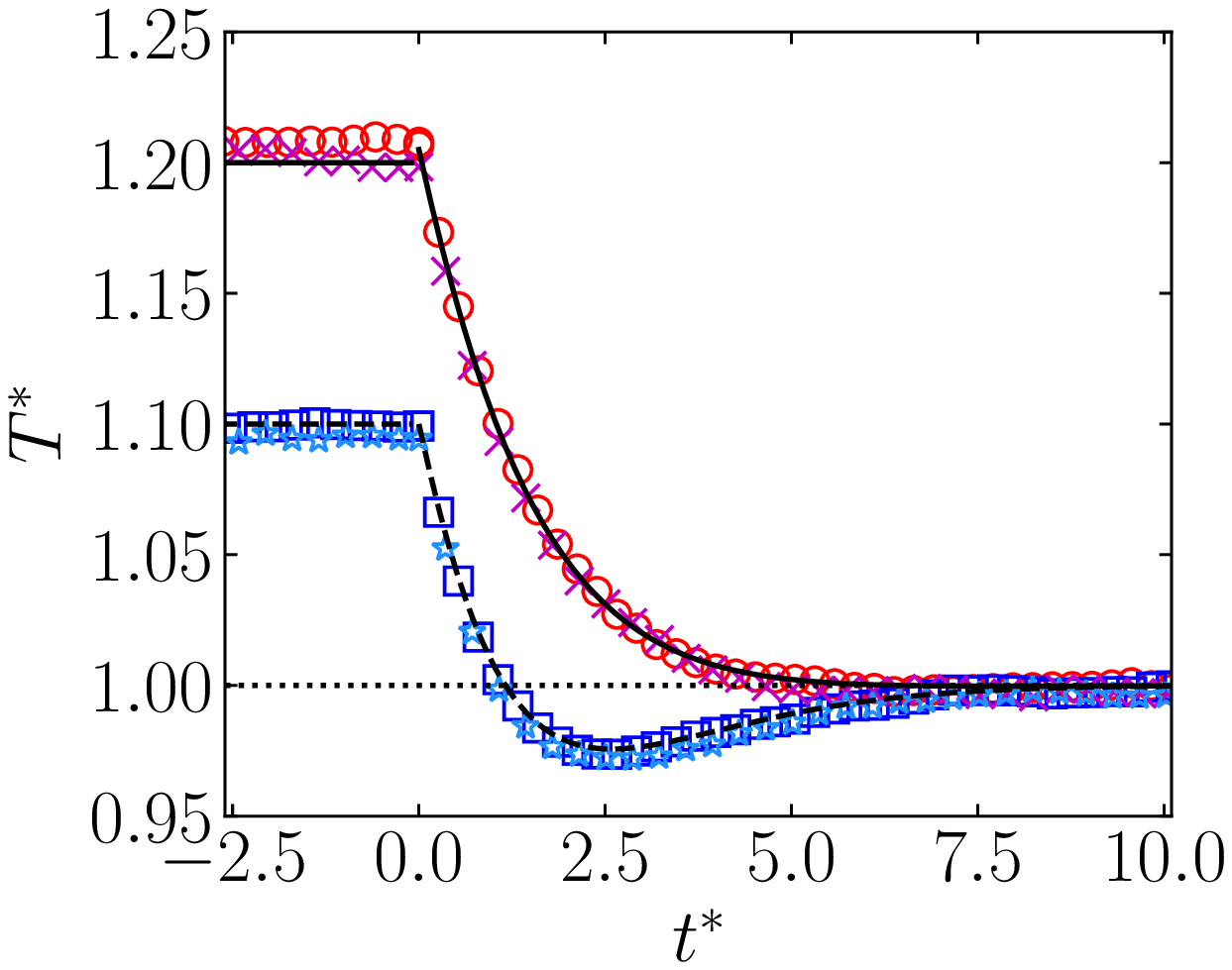}
        \end{minipage}
    	\begin{minipage}[t]{.3\linewidth}
    	\textbf{H}\\
    	\includegraphics[width = \linewidth]{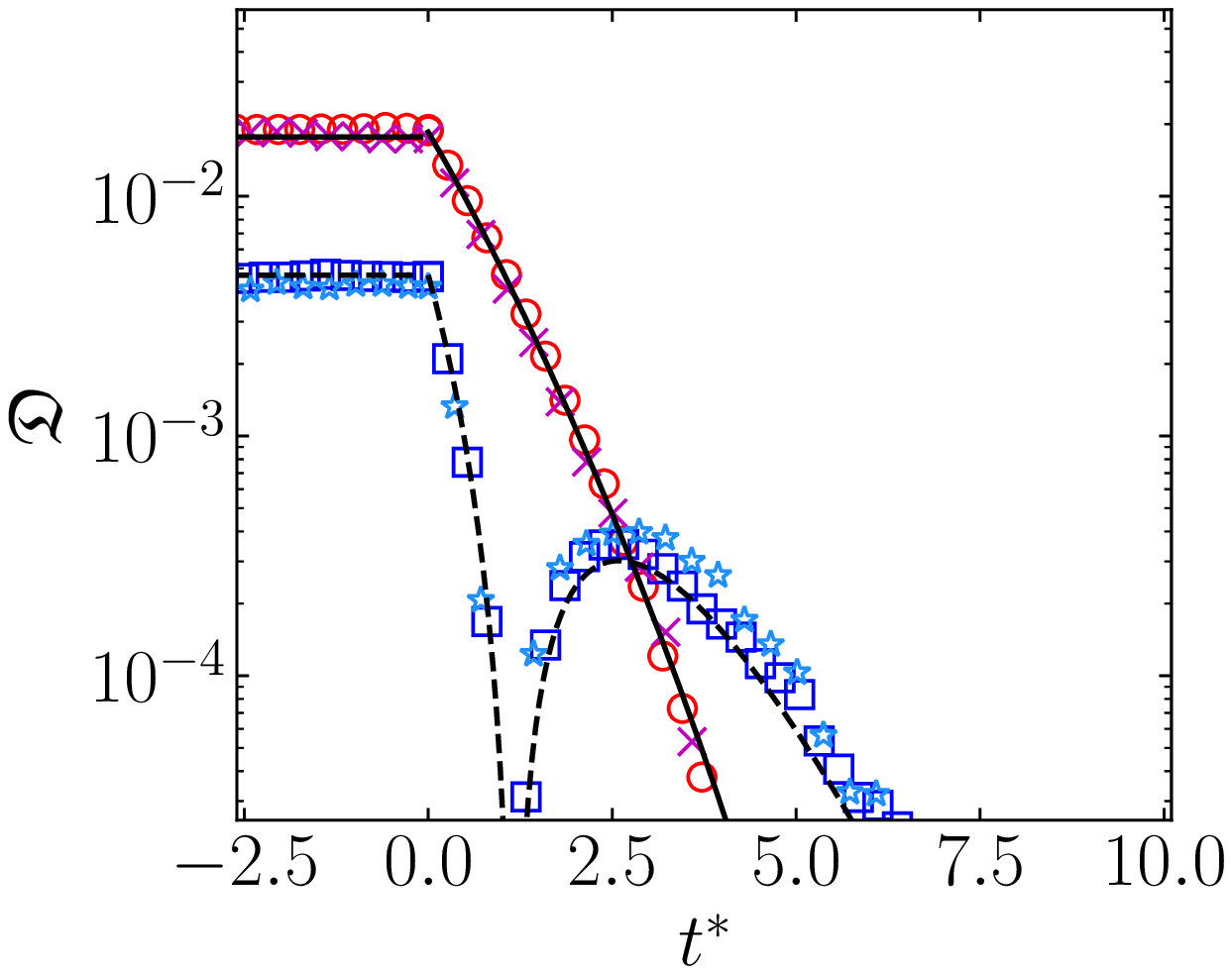}
        \end{minipage}
    	\begin{minipage}[t]{.3\linewidth}
    	\textbf{I}\\
    	\includegraphics[width = \linewidth]{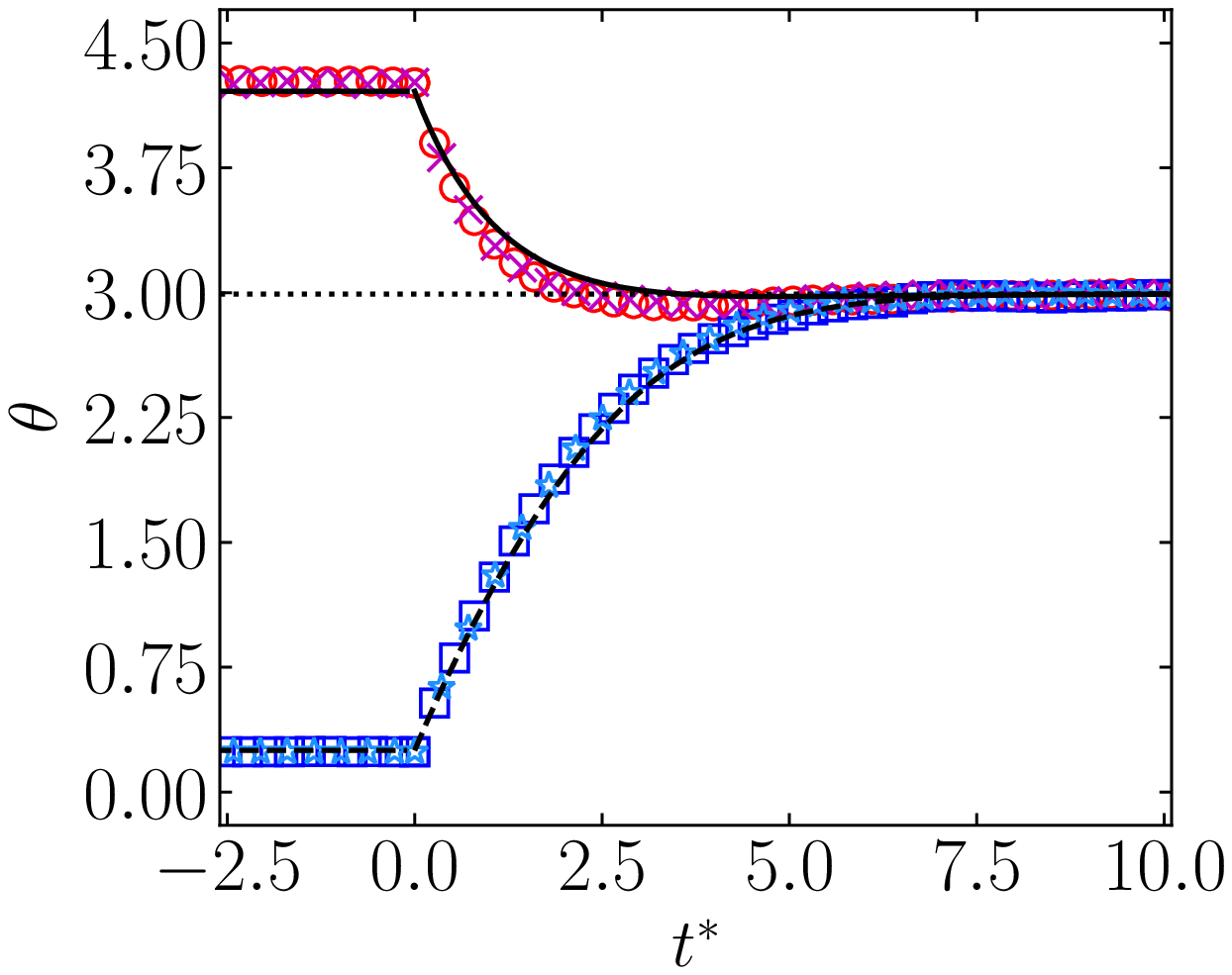}\\
        \end{minipage}\\ 
    	\begin{minipage}[t]{.3\linewidth}
    	\textbf{J}\\
    	\includegraphics[width = \linewidth]{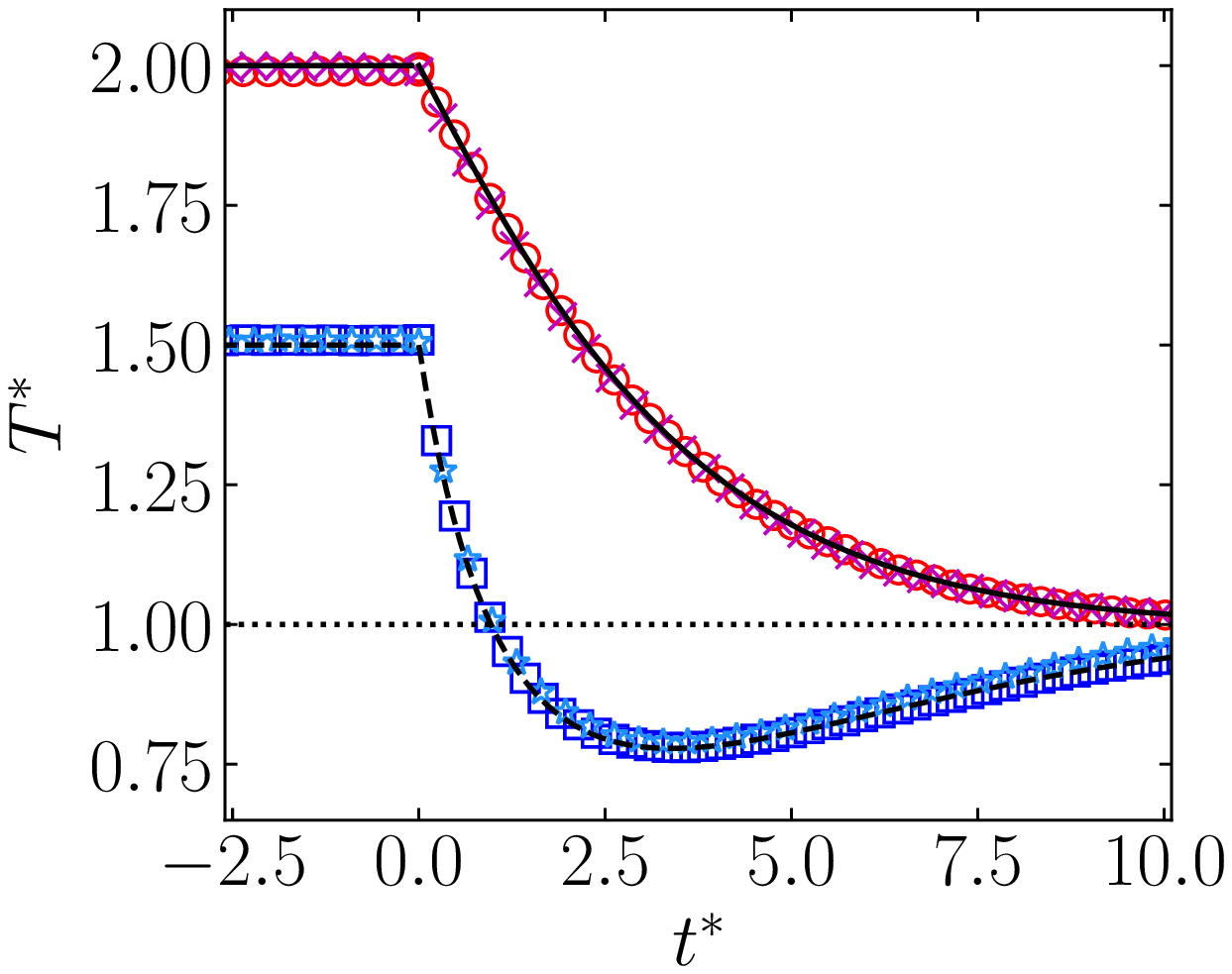}
        \end{minipage}
    	\begin{minipage}[t]{.3\linewidth}
    	\textbf{K}\\
    	\includegraphics[width = \linewidth]{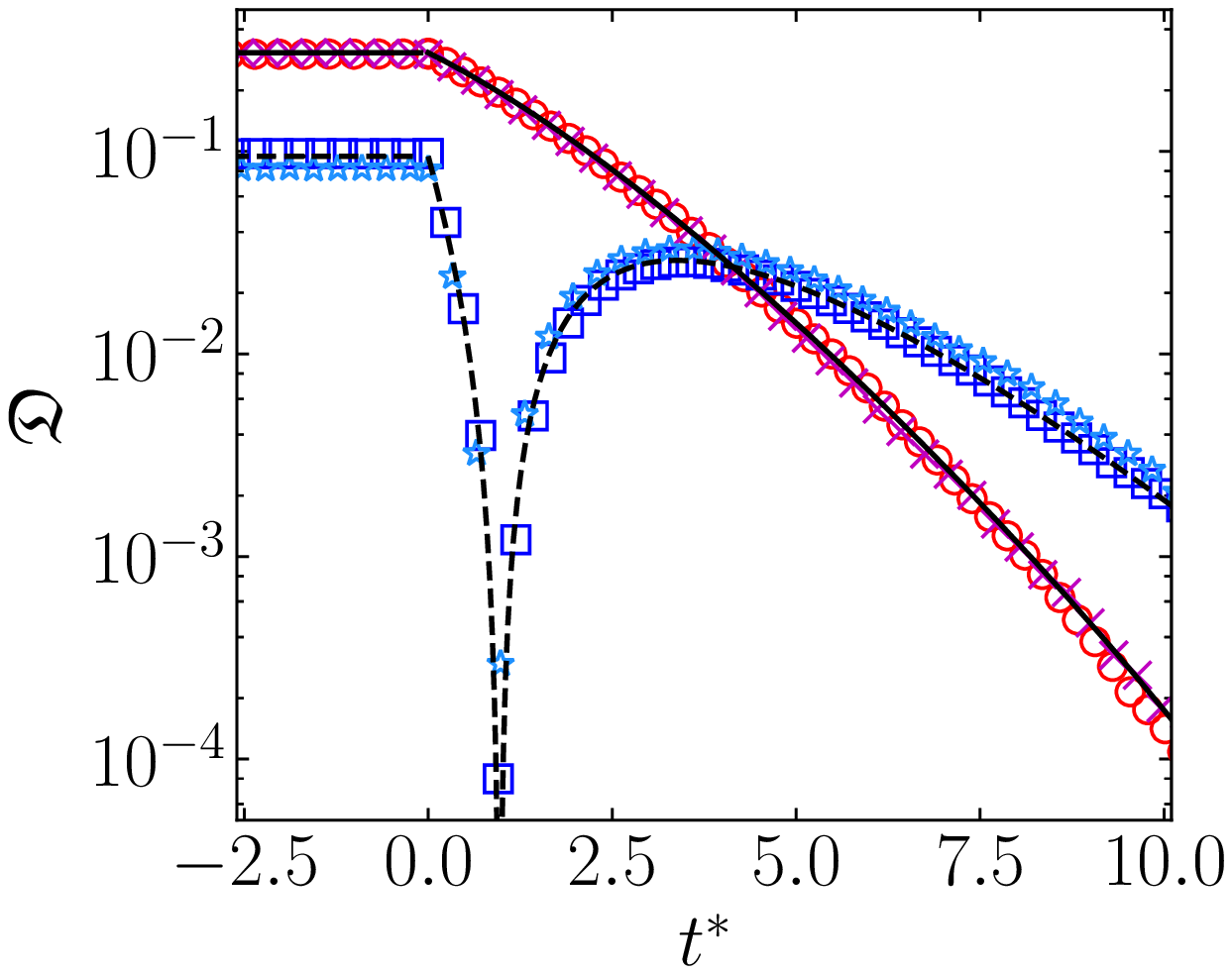}
        \end{minipage}
    	\begin{minipage}[t]{.3\linewidth}
    	\textbf{L}\\
    	\includegraphics[width = \linewidth]{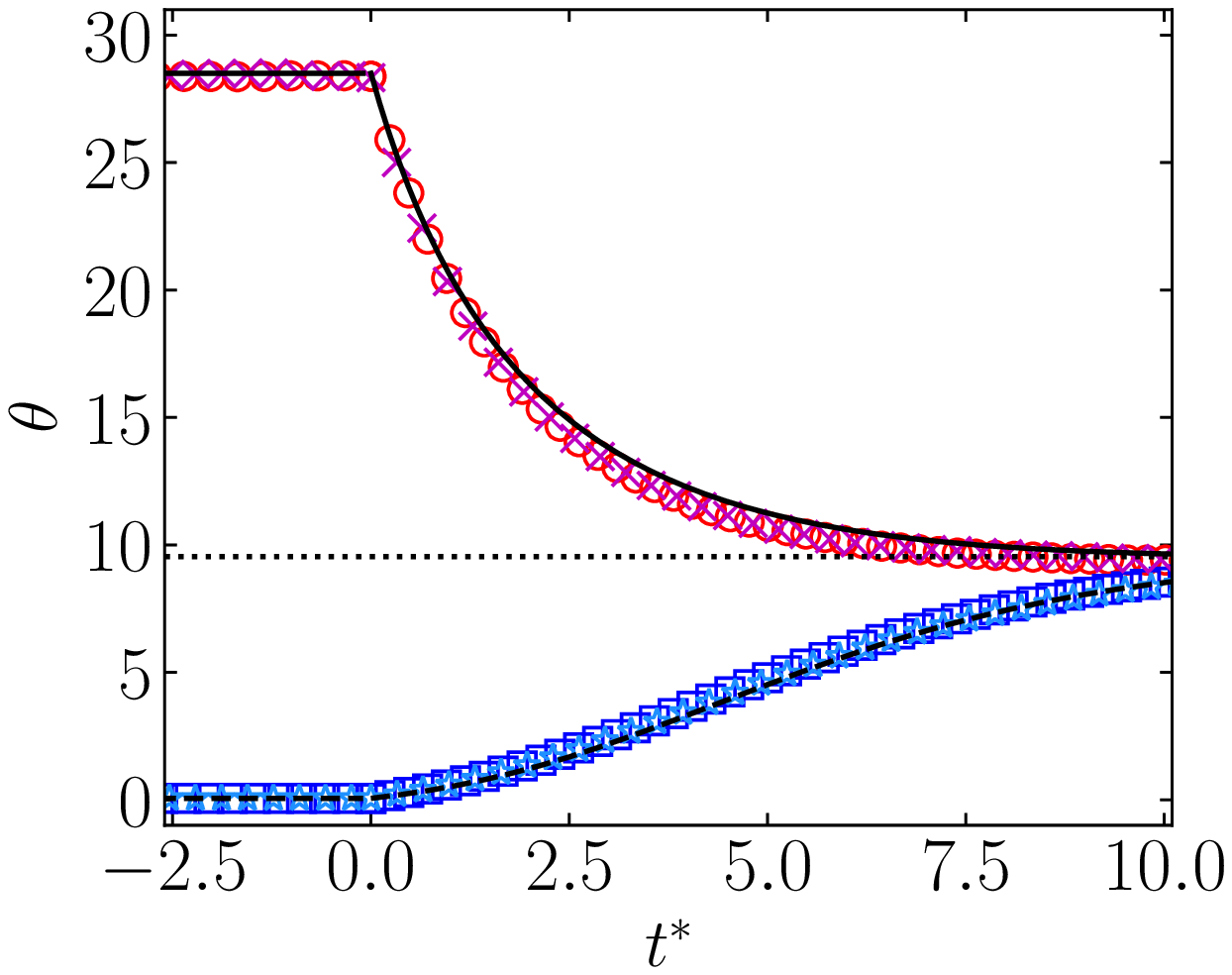}
    	\end{minipage}
        \caption{Time evolution of $T^*$, $\mathfrak{D}$, and $\theta$ for OME initialization protocol described in section.~\ref{sec:3.3.2} with $\varepsilon_\rf=0.9$.  \textbf{(A)}--\textbf{(C)}: $(\alpha,\beta)=(0.7,0)$; \textbf{(D)}--\textbf{(F)}: $(\alpha,\beta)=(0.7,-0.7)$; \textbf{(G)}--\textbf{(I)}; $(\alpha,\beta)=(0.9,0)$; and \textbf{(J)}--\textbf{(L)}: $(\alpha,\beta)=(0.9,-0.7)$. Thick and dashed lines correspond to the theoretical prediction from  Equations~\eqref{eq:Tstar-theta-ev}, dotted lines represented the steady state value, and symbols refer to DSMC and EDMD simulation results. }
    \label{fig:OME}
\end{figure}

In order to check the initialization protocols for detecting SME and OME, we have run DSMC and EDMD simulations. The simulation details are the same as introduced in section~\ref{sec:2.3}. Simulation points correspond to the average over ensembles of $100$ replicas. Again, no instabilities were observed.

\subsubsection{Standard Mpemba effect}

Figure~\textbf{\ref{fig:SME}} presents results for the SME protocol introduced in section~\ref{sec:3.3.1}.  As we can observe, the theoretical predictions agree very well with DSMC and EDMD simulation data.

The chosen initial temperature conditions are  $(T_{0A}^*,T_{0B}^*)=(3,2)$ for Figure~\textbf{\ref{fig:SME}A}, $(T_{0A}^*,T_{0B}^*)=(4,2)$ for  Figures~\textbf{\ref{fig:SME}C} and \textbf{\ref{fig:SME}G}, and $(T_{0A}^*,T_{0B}^*)=(3,2.92)$ for Figure~\textbf{\ref{fig:SME}E}. Moreover, $\varepsilon_\rf=0.1$ for all cases, except for the case of Figures~\textbf{\ref{fig:SME}E} and \textbf{\ref{fig:SME}F}, that is chosen to be $\varepsilon_\rf=0.6$. The values of $\Tn_{A}/\Tn_\rf$ and $\Tn_{B}/\Tn_\rf$  are given in each case by Equation~\eqref{eq:prior/posterior} to ensure the desired initial temperatures. All these cases avoid overshoot, in agreement with the case $\theta_0=\theta^\st(\varepsilon=0$) shown in Figure~\textbf{\ref{fig:PD_overshoot}}. Moreover, the cases with $\varepsilon_\rf=0.1$ are inside the region where SME is predicted to be present in Figure~\textbf{\ref{fig:PD_MP}A}.

The value $\varepsilon_\rf=0.6$ for the system $(\alpha=0.9,\beta=0)$ was chosen instead of $\varepsilon_\rf=0.1$ to avoid the need of taking very high initial temperatures (see Figure~\textbf{\ref{fig:PD_MP}A}) and also to prevent  overshoot (see Figure~\textbf{\ref{fig:PD_overshoot}}). The price paid for this choice of coefficients of restitution is that the difference $\theta_{0A}-\theta_{0B}$ is not too high, see Figures~\textbf{\ref{fig:theta_st_diff}} and \textbf{\ref{fig:SME}F}. Then, initial temperature values are restricted to be very similar and the crossing characterizing the SME is much less pronounced than in the other cases. However, the inset in Figure~\textbf{\ref{fig:SME}E} shows a well defined change of sign of the the difference $T_A^*-T_B^*$.

\subsubsection{Overshoot Mpemba effect}
The theoretical results stemming from the  OME protocol introduced in section~\ref{sec:3.3.2} are compared with simulations in Figure~\textbf{\ref{fig:OME}}, again with an excellent agreement. 

The initial temperatures are  $(T_{0A}^*,T_{0B}^*)=(1.22,1.1)$ for Figure~\textbf{\ref{fig:OME}A}, $(T_{0A}^*,T_{0B}^*)=(2,1.5)$ for Figures~ \textbf{\ref{fig:OME}D} and \textbf{\ref{fig:OME}J}, and $(T_{0A}^*,T_{0B}^*)=(1.2,1.1)$ for Figure~\textbf{\ref{fig:OME}G}. Moreover, a common value $\varepsilon_\rf=0.9$ is chosen to ensure that overshoot appears for the colder samples, as depicted in Figure~\textbf{\ref{fig:PD_overshoot}}, and  that OME is expected from the phase diagram shown in Figure~\textbf{\ref{fig:PD_MP}B}.

Whereas no crossing between $T^*_A$ and $T^*_B$ is observed in Figure~\ref{fig:OME}, such a crossing is present between the entropy-like quantities $\mathfrak{D}_A$ and $\mathfrak{D}_B$ [see Equation~\eqref{eq:KLD}]. As explained in section~\ref{sec:3.2}, this characterizes the  OME, where the initially colder system relaxes later to the steady state.

\section{Concluding Remarks}
\label{sec:4}

In this work, we have studied the homogeneous states of a dilute granular gas made of inelastic and rough hard disks lying on a two-dimensional plane. The inelasticity and roughness are mathematically described by constant    coefficients of normal ($\alpha$) and tangential ($\beta$) restitution, respectively. In order to avoid frozen long-time limiting states, the disks are assumed to be heated by the stochastic force- and torque-based ST. This novel stochastic thermostat injects energy to both translational and rotational degrees of freedom. Each specific thermostat of this type is univocally determined by its associated noise temperature, $\Tn\geq 0$ [see Equation~\eqref{eq:Tn}], and the rotational-to-total noise intensity, $0\leq \varepsilon\leq 1$ [see Equation~\eqref{eq:noise_paramsB}].

The system is assumed to be fully described by the instantaneous one-particle VDF,  its dynamics being then given  by the BFPE, Equation~\ref{eq:BFPE_v2}. It is known that the steady-state VDF is not a Maxwellian, as occurs for $\varepsilon=0$ \cite{VS15}, and this is even more the case with the  instantaneous transient VDF. However, these nonGaussianities  are expected to be small enough as to approximate the  VDF by a Maxwellian, as previously done for $\varepsilon=0$  \cite{TLLVPS19}. Therefore, we have worked under the two-temperature MA introduced in section~\ref{sec:2.2}, which allows us to account for the dynamics of the systems just in terms of the reduced total granular temperature, $T^*$, and the rotational-to-translational temperature ratio, $\theta$, according to Equations~\eqref{eq:Tstar-theta-ev}. The steady-state values can be explicitly expressed in terms of the mechanical properties of the disks, see Equation~\eqref{eq:steady_values}. Steady and transient states predicted by the MA are tested via DSMC ans EDMD, with very good agreement, as observed in Figures~\textbf{\ref{fig:steady_state}} and \textbf{\ref{fig:evol}}. This reinforces the validity of our approach. As expected, $\theta^\st$ is an increasing function of $\varepsilon$ and independent of $\Tn$, whereas $\widetilde{T}^\st \equiv T^\st/\Tn$ can be a nonmonotonic function of $\varepsilon$ (see Figure~\textbf{\ref{fig:steady_state}}).

The main core of this work has been the description of ME  in cooling processes in this system, with special emphasis on  the elaboration of preparation protocols for the generation of the initial states. We noted that, if $T^*-1$ does not change its sign during the evolution, the usual form of ME, SME, can emerge if $\theta_{0A}\ll \theta_{0B}$, A being the initially hotter sample. This SME is characterized by a single crossing (or, in general, an odd number of them) between the temperature curves, thus inducing that the initially colder system, B, relaxes more slowly toward the final steady state. However, we have realized that an overshoot (or change of sign of $T^*-1$) might appear if the rotational-to-total noise intensity is larger than  a certain critical value   $\varepsilon_\ct(T^*_0,\theta_0)$, as shown for some cases in Figure~\textbf{\ref{fig:PD_overshoot}}. Therefore, the SME is no longer ensured and, following a recent work \cite{MSP22}, ME might appear without a crossing between $T^*_A$ and $T^*_B$ but with a crossing between the entropy-like quantities $\mathfrak{D}_A$ and $\mathfrak{D}_B$ [see Equation~\eqref{eq:KLD}]. This peculiar type of ME is termed OME and is described in section~\ref{sec:3.2}. According to the MA, the most favorable condition for the OME changes from the SME one to $\theta_{0A}\gg \theta_{0B}$.

Protocols for generating initial conditions to observe both SME and OME have been presented in sections~\ref{sec:3.3.1} and \ref{sec:3.3.2}, respectively. While reminiscent of the protocols previously considered in the case of sheared inertial suspensions \cite{THS21},  the protocols proposed here represent novel instructions to elaborate a ME experiment in homogeneous states of granular gaseous systems. We have based those protocols on the steady states of the ST, taking advantage of the increase of $\theta^\st$ as  $\varepsilon$ increases. Thus, to guarantee the biggest possible difference $|\theta_{0B}-\theta_{0A}|$ we have fixed prior thermalization processes with $(\varepsilon_A,\varepsilon_B)=(0,1)$ for SME and $(\varepsilon_A,\varepsilon_B)=(1,0)$  for OME. Moreover, in this prior thermalization stage, $\Tn_A/\Tn_\rf$ and $\Tn_B/\Tn_\rf$, where  $(\Tn_\rf,\varepsilon_\rf)$ characterizes the posterior thermostat, are chosen to fine-tune both $T_{0A}^*$ and $T_{0B}^*$.  It is crucial to choose $\varepsilon_\rf< \min\{\varepsilon_\ct(T_{0A}^*,\theta_{0A}),\varepsilon_\ct(T_{0B}^*,\theta_{0B})\}$ for SME to avoid overshoot, and $\varepsilon_\rf>\varepsilon_\ct(T_{0B}^*,\theta_{0B})$ for OME to ensure the overshoot of the initially colder sample. These protocols, together with the MA predictions, allowed us to elaborate phase diagrams for SME and OME emergences, as depicted in Figure~\textbf{\ref{fig:PD_MP}}.

The theoretical descriptions of the SME and OME initialization protocols have been tested by DMSC and EDMD simulations, finding a very good agreement between theory and simulations, as observed in Figures~\textbf{\ref{fig:SME}} and \textbf{\ref{fig:OME}}, respectively. 

One can then conclude that, despite its simplicity,  the MA captures very well the dynamics for this system. In turn, this implies that the SME and OME protocols for the initial-state preparation described in this paper are trustworthy.


Finally, we expect that this work will be useful to the ME community in the search for practical protocols able to generate the adequate initial states. In addition, given the simplicity of the hard-disk system studied in this paper, we hope it can be experimentally realizable, thus opening up the possibility of reproducing the ME by an adequate control of the external forcing mechanisms.

\section*{Conflict of Interest Statement}

The authors declare that the research was conducted in the absence of any commercial or financial relationships that could be construed as a potential conflict of interest.

\section*{Author Contributions}
AM and AS contributed to the conception and design of the study. AM performed the computer simulations and  wrote the first draft of the manuscript.  Both authors contributed to manuscript revision, read, and approved the submitted version.
\section*{Funding}

The authors acknowledge financial support from Grant No.~PID2020-112936GB-I00 funded by MCIN/AEI/10.13039/501100011033, and from Grants No.~IB20079 and No.~GR21014 funded by Junta de Extremadura (Spain) and by ERDF
``A way of  making Europe.''
AM is grateful to the Spanish Ministerio de Ciencia, Innovaci\'on y Universidades for a predoctoral fellowship FPU2018-3503.

\section*{Acknowledgments}

The authors are grateful to the computing facilities of the Instituto de Computaci\'on Cient\'ifica Avanzada of the University of Extremadura (ICCAEx), where the simulations were run.

\section*{Data Availability Statement}
The datasets presented in this study can be found in online repositories. The names of the repository/repositories and accession number(s) can be found below: [\href{https://github.com/amegiasf/MpembaSplitting}{https://github.com/amegiasf/MpembaSplitting}].

\bibliographystyle{Frontiers-Vancouver} 

\end{document}